\documentclass[a4paper]{extarticle}
\usepackage[T1]{fontenc} 
\usepackage[english]{babel} 
\usepackage{indentfirst}

\usepackage[hmarginratio=1:1,left=20mm,right=20mm,bottom=25mm,top=25mm,columnsep=20pt]{geometry}

\usepackage[hang,normalsize,labelfont=bf,up,textfont=it,up,skip=0pt,justification=justified]{caption} 

\usepackage{abstract}

\usepackage{titlesec}
\titleformat{\section}[block]{\Large\bfseries\centering}{\thesection.}{1em}{}
\titleformat{\subsection}[block]{\large\bfseries}{\thesubsection.}{1em}{}
\titleformat{\subsubsection}[block]{\normalsize\bfseries}{\thesubsubsection.}{1em}{}

\usepackage{fancyhdr}
\renewcommand{\headrulewidth}{0pt}
\renewcommand{\footrulewidth}{0pt}

\usepackage{titling} 

\usepackage[breaklinks=true]{hyperref} 

\usepackage{amsmath,amssymb}
\usepackage{graphicx}
\usepackage{multirow}
\usepackage{algorithm}
\usepackage[noend]{algpseudocode}
\usepackage{color}
\usepackage{bm}
\usepackage{lastpage}

\makeatletter

\newcommand{\ArtTitle}[2]{\def\@ArtTitle{#1}\def\@ArtTitleShort{#2}}
\newcommand{\ArtJournalInfo}[1]{\def\@ArtJournalInfo{#1}}
\newcommand{\ArtJournalRef}[1]{\def\@ArtJournalRef{#1}}
\newcommand{\ArtAuthors}[9]{
\def\@ArtAuthorA{#1}\def\@ArtAuthorAmail{#2}\def\@ArtAuthorAAff{#3}
\def\@ArtAuthorB{#4}\def\@ArtAuthorBmail{#5}\def\@ArtAuthorBAff{#6}
\def\@ArtAuthorC{#7}\def\@ArtAuthorCmail{#8}\def\@ArtAuthorCAff{#9}
}

\newcommand{\ArtAuthorsBis}[9]{
\def\@ArtAuthorD{#1}\def\@ArtAuthorDmail{#2}\def\@ArtAuthorDAff{#3}
\def\@ArtAuthorE{#4}\def\@ArtAuthorEmail{#5}\def\@ArtAuthorEAff{#6}
\def\@ArtAuthorF{#7}\def\@ArtAuthorFmail{#8}\def\@ArtAuthorFAff{#9}
}

\newcommand{\ArtAbstract}[1]{\def\@ArtAbstract{#1}}

\newcommand{\makeArtTitle}{

{
\includegraphics[height=0.06\textheight,trim={7cm 13cm 7cm 13cm},clip]{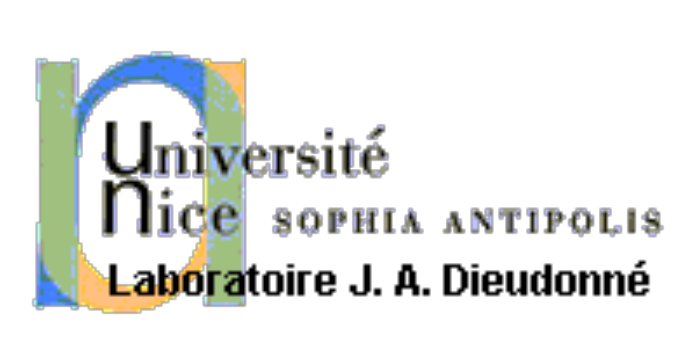}
\hspace{0.15\textwidth}
\includegraphics[height=0.06\textheight,trim={4cm 12cm 4cm 12cm},clip]{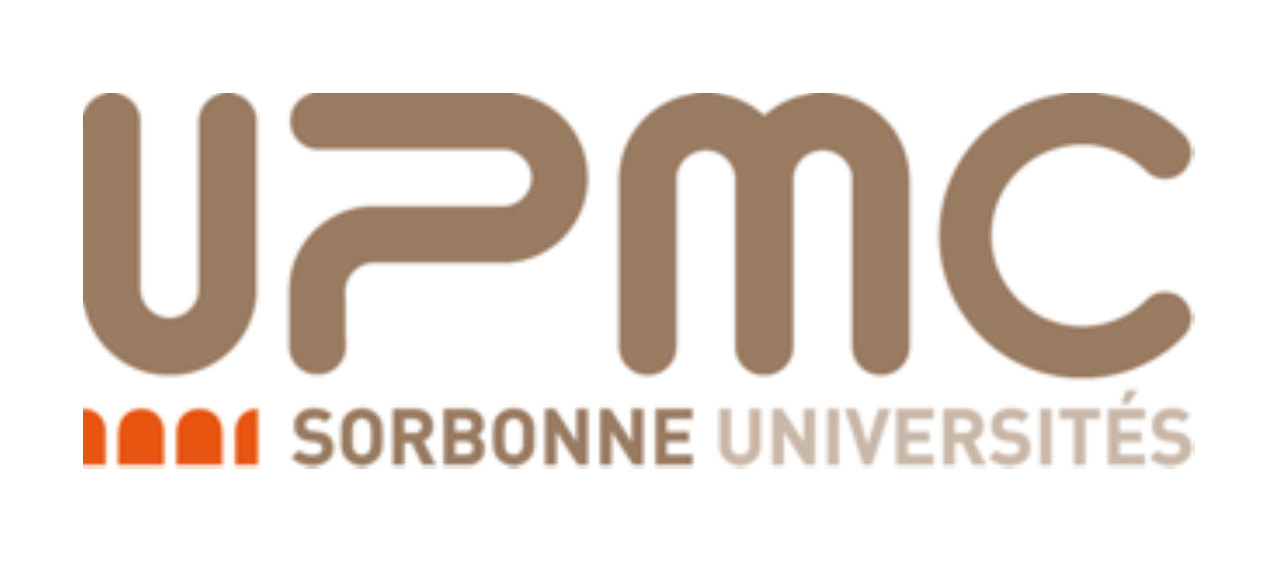}
\hspace{0.15\textwidth}
\includegraphics[height=0.06\textheight,trim={9cm 13cm 9cm 13cm},clip]{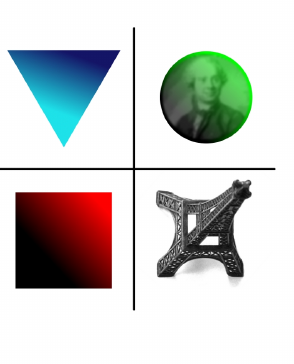}
\hspace{0.15\textwidth}
\includegraphics[height=0.06\textheight,trim={6cm 10cm 6cm 10cm},clip]{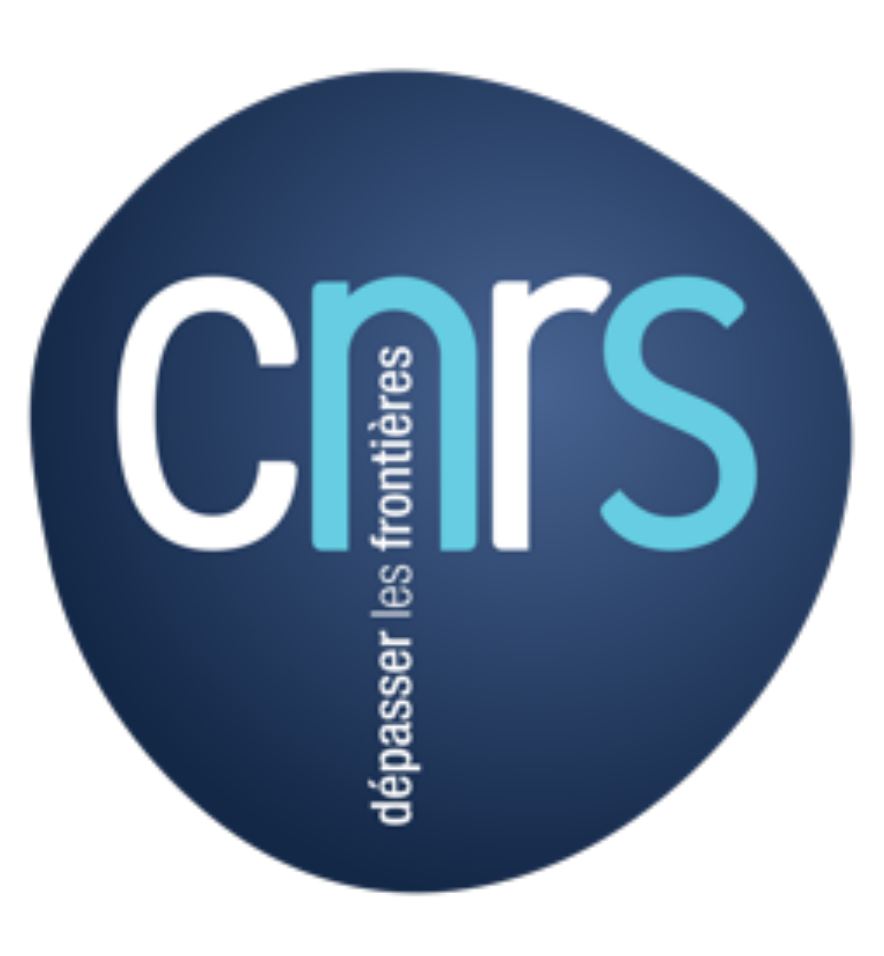}
}

\par\vspace{0.05\textheight}

{
\begin{minipage}{0.99\textwidth}
\centering
\Large\@ArtAuthorA
\end{minipage}
\hfill
\par\vspace{0.0\textheight}
{\centering 
\small\@ArtAuthorAAff\\
}
\par\vspace{0.015\textheight}
\begin{minipage}{0.99\textwidth}
\centering
\Large\@ArtAuthorD
\end{minipage}
\hfill
\par\vspace{0.0\textheight}
{\centering 
\small\@ArtAuthorDAff\\
}
\par\vspace{0.015\textheight}
\begin{minipage}{0.49\textwidth}
\centering
\Large\@ArtAuthorB
\end{minipage}
\begin{minipage}{0.49\textwidth}
\centering
\Large\@ArtAuthorC
\end{minipage}
\hfill
\par\vspace{0.0\textheight}
{\centering 
\small\@ArtAuthorCAff\\
}
\par\vspace{0.05\textheight}
}

\begin{center}
\begin{minipage}{0.9\textwidth}
\centering
\huge
\@ArtTitle
\end{minipage}
\end{center}
\par\vspace{0.05\textheight}

{
{\Large\centering\textbf{Abstract\\}}
\par\vspace{0.02\textheight}
{\noindent
\textit{\@ArtAbstract}
}
}
\par

\par\vspace*{\fill}
{\huge
\href{http://www.arXiv.org}{\nolinkurl arXiv} - \href{http://hal.upmc.fr}{\nolinkurl HAL}
}

\renewcommand{\footrulewidth}{1pt}
\fancyfoot{
\raggedright
\@ArtJournalInfo
\@ArtJournalRef
\\
\underline{Contact}: \@ArtAuthorAmail
}

\pagebreak
}

\newcommand{\makeArtTableOfContent}{
\tableofcontents
\pagebreak
}

\makeatother

\ArtTitle{
Low-Shapiro hydrostatic reconstruction technique
for blood flow simulation in large arteries
with varying geometrical and mechanical properties
}{
Low-Shapiro hydrostatic reconstruction for blood flow
}
\ArtJournalInfo{
Published in the \textit{Journal of Computational Physics}.
}
\ArtJournalRef{\url{http://dx.doi.org/10.1016/j.jcp.2016.11.032}}

\ArtAuthors
{Arthur R. Ghigo}{arthur.ghigo@dalembert.upmc.fr}
{Sorbonne Universit\'es, CNRS and UPMC Universit\'e Paris 06, UMR 7190, Institut Jean Le Rond $\partial$'Alembert}
{Jose-Maria Fullana}{}
{Sorbonne Universit\'es, CNRS and UPMC Universit\'e Paris 06, UMR 7190, Institut Jean Le Rond $\partial$'Alembert}
{Pierre-Yves Lagr\'ee}{}
{Sorbonne Universit\'es, CNRS and UPMC Universit\'e Paris 06, UMR 7190, Institut Jean Le Rond $\partial$'Alembert}

\ArtAuthorsBis
{Olivier Delestre}{delestre@unice.fr}
{Laboratoire J.A. Dieudonn\'e, UMR CNRS 7351 \& Polytech' Nice Sophia, University of Nice Sophia Antipolis (UNS)}
{}{}{}
{}{}{}

\ArtAbstract
{The purpose of this work is to construct a simple, efficient and accurate well-balanced numerical scheme for one-dimensional (1D) blood flow in large arteries with varying geometrical and mechanical properties. As the steady states at rest are not relevant for blood flow, we construct two well-balanced hydrostatic reconstruction techniques designed to preserve low-Shapiro number steady states that may occur in large network simulations. The Shapiro number $S_h=u/c$ is the equivalent of the Froude number for shallow water equations and the Mach number for compressible Euler equations. The first is the low-Shapiro hydrostatic reconstruction (HR-LS), which is a simple and efficient method, inspired from the hydrostatic reconstruction technique (HR). The second is the subsonic hydrostatic reconstruction (HR-S), adapted here to blood flow and designed to exactly preserve all subcritical steady states. We systematically compare HR, HR-LS and HR-S in a series of single artery and arterial network numerical tests designed to evaluate their well-balanced and wave-capturing properties. The results indicate that HR is not adapted to compute blood flow in large arteries as it is unable to capture wave reflections and transmissions when large variations of the arteries' geometrical and mechanical properties are considered. On the contrary, HR-S is exactly well-balanced and is the most accurate hydrostatic reconstruction technique. However, HR-LS is able to compute low-Shapiro number steady states as well as wave reflections and transmissions with satisfying accuracy and is simpler and computationally less expensive than HR-S. We therefore recommend using HR-LS for 1D blood flow simulations in large arterial network simulations.
}

\newcommand{\myreferences}{mybibfile.bib}

\begin{document}


\makeArtTitle


\makeatletter
\fancyhf{}

\renewcommand{\headrulewidth}{1pt}
\fancyhead[L]{\@ArtTitleShort \: - \: \@ArtAuthorA \, et al.}
\fancyhead[R]{\thepage /\pageref{LastPage}} 

\renewcommand{\footrulewidth}{0pt}
\fancyfoot{}
\makeatother


\makeArtTableOfContent


\section{Introduction}

Since the early work of Euler \cite{Euler1844}, one-dimensional (1D) models have been successfully used to describe the flow of blood in the large arteries of the systemic network \cite{PedleyBook1980,Alastruey2011,Wang2015,Muller2014,Boileau2015}. They have proved to be valuable and efficient tools to capture pulse wave propagation in large network simulations and obtain satisfactory evaluations of average quantities such as the cross-sectional area ($A$), the flow rate ($Q$) or the pressure ($P$) \cite{Alastruey2009,Politi2016}. In recent works, 1D models have also been used to compute inverse problems to obtain patient specific parameters \cite{Lagree2000,Martin2005,Dumas2012}. Due to their simplicity, efficiency and the reduced number of parameters they require, we hope that in the near future these 1D models will be intensively used by medical practitioners to make pre-and-post operative diagnosis and perform patient specific simulations of surgeries.

\begin{figure}[!h]
\makebox[1.\textwidth][c]{
\begin{minipage}[t]{0.33\textwidth}
  \centering
  \includegraphics[scale=0.15,angle=0]{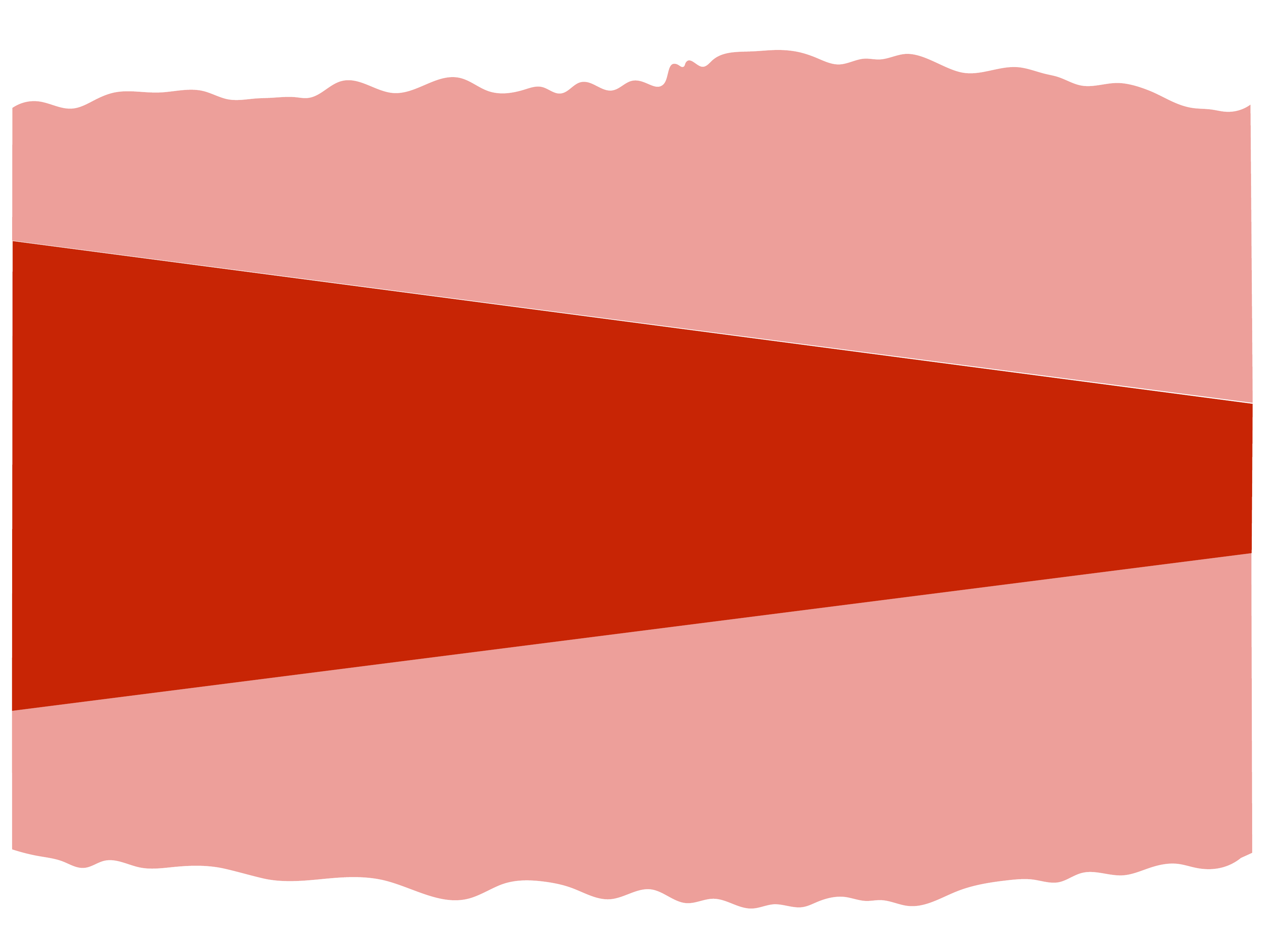}\\
\end{minipage} 
\begin{minipage}[t]{0.33\textwidth}
  \centering
  \includegraphics[scale=0.15,angle=0]{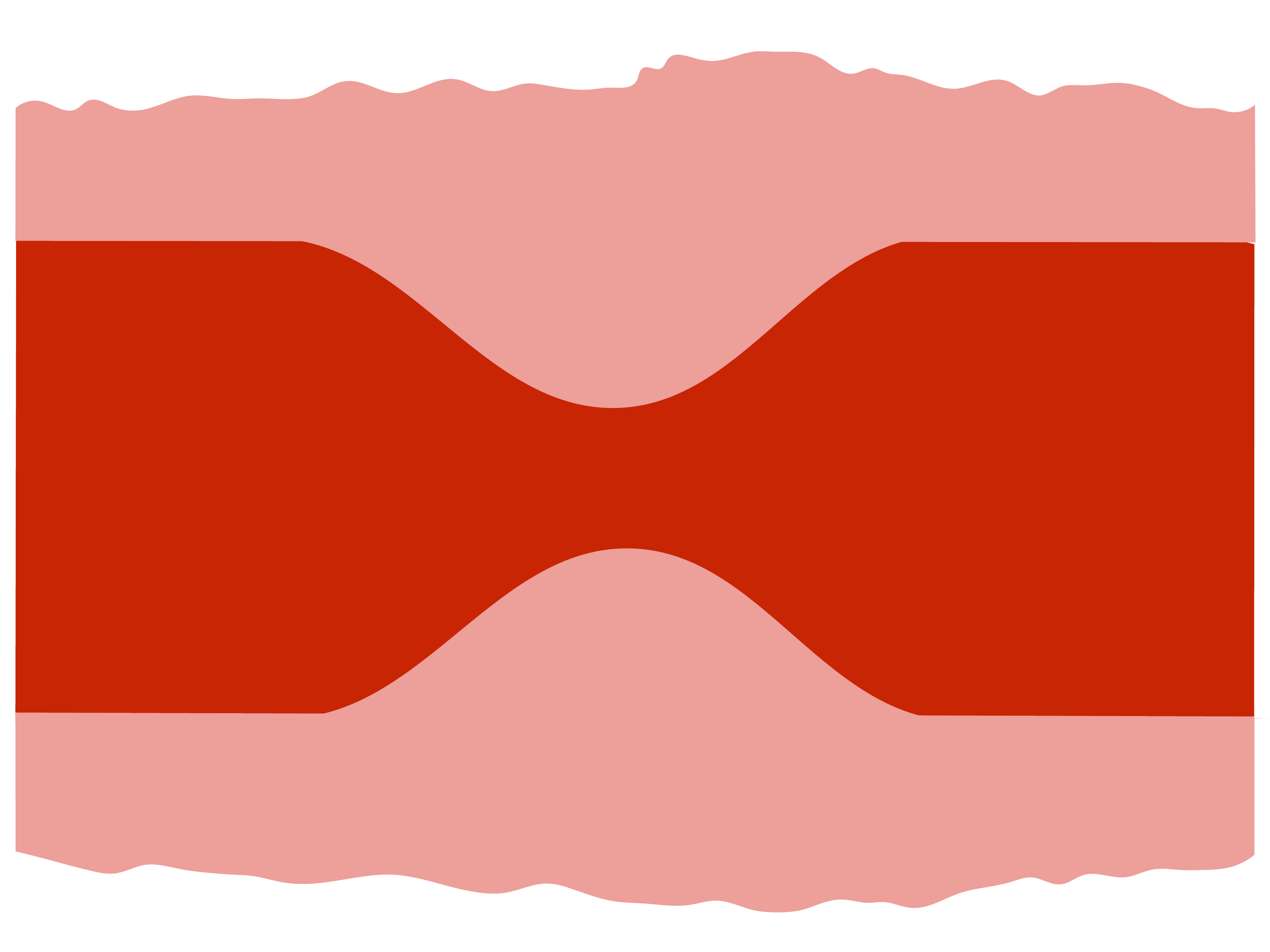}\\
\end{minipage} 
\begin{minipage}[t]{0.33\textwidth}
  \centering
  \includegraphics[scale=0.15,angle=0]{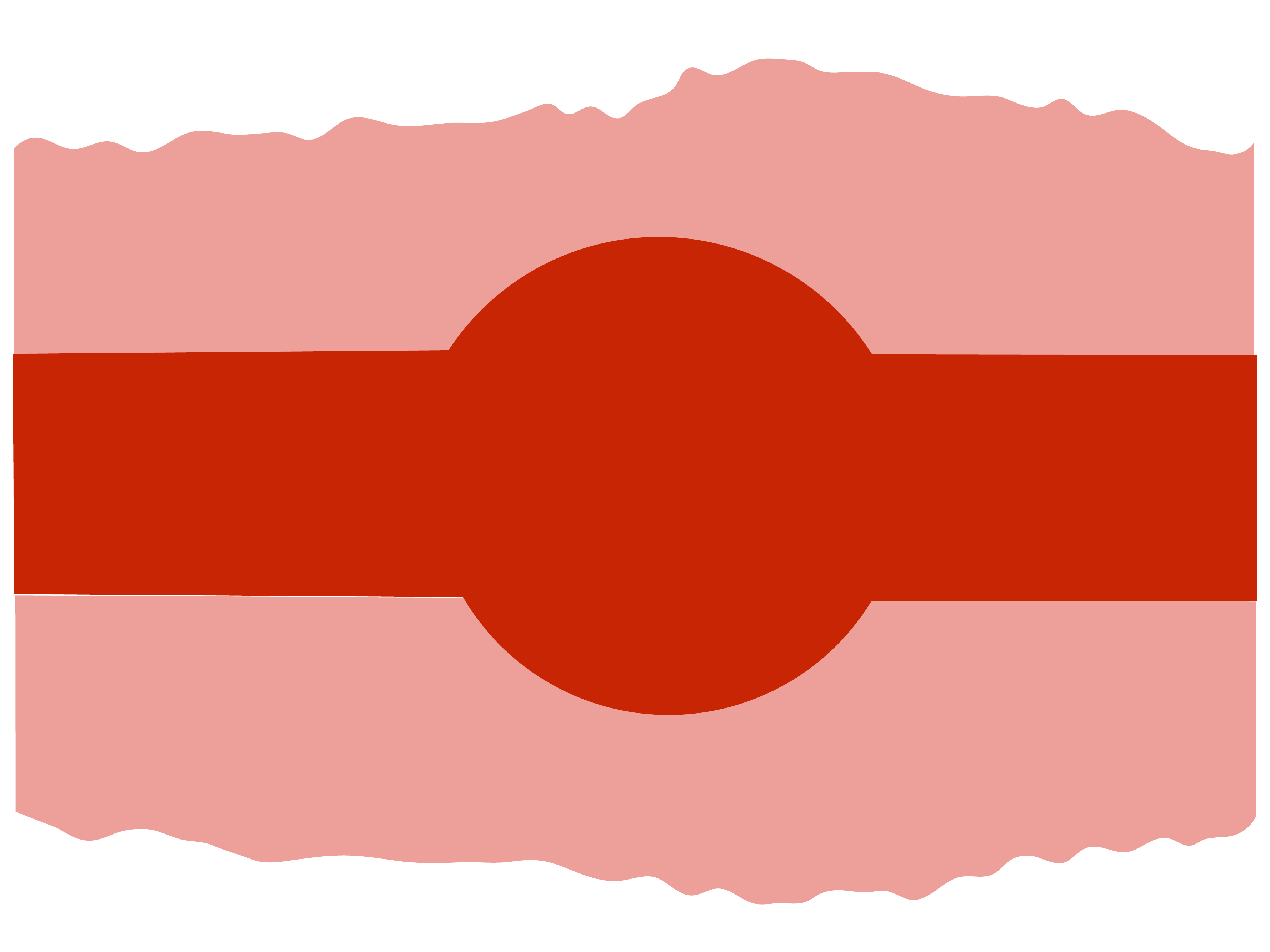}\\
\end{minipage} 
}
\caption{Schematic representations of possible arterial geometrical configurations.\\
\underline{\textit{Left}}: Taper; \underline{\textit{Center}}: Stenosis; \underline{\textit{Right}}: Aneurysm.}
\label{fig:Taper-Stenosis-Aneurysm}
\end{figure}

In physiological situations, the mechanical and geometrical properties of the arterial wall can vary locally. These variations can be caused by tapering (figure \ref{fig:Taper-Stenosis-Aneurysm} left), pathologies such as stenoses (figure \ref{fig:Taper-Stenosis-Aneurysm} center) or aneurysms (figure \ref{fig:Taper-Stenosis-Aneurysm} right) and endovascular prosthesis (stent). Mathematically, they result in a source term in the momentum conservation equation that prevents from writing the system in a conservation-law form. A naive discretization of this nonconservative source term can lead to spurious oscillations of the numerical solution and the failure of the numerical method, especially close to steady states \cite{Delestre2012}. This problem was originally pointed out by Roe \cite{Roe1987} for the scalar equation with source terms and reflects a truncation error between the discretization of the conservative flux gradient and the nonconservative source term that does not vanish close to steady states. Since the works of Berm\'udez and V\'azquez \cite{Bermudez1994} and LeRoux \cite{Gosse1996,Greenberg1996} in the context of shallow-water equations, numerical schemes that preserve some steady states at a discrete level are called well-balanced.\\

The aim of this study is to propose a simple, robust and efficient well-balanced numerical method for blood flow in an artery with variations of its mechanical and geometrical properties. As blood flow equations are mathematically similar to shallow water equations, several well-balanced numerical schemes have been derived for 1D blood flow equations with varying geometrical and mechanical properties. A popular approach consists in expressing the system in terms of primitive variables, namely the cross-sectional area ($A$) and the flow velocity $u$. The resulting system can be written in a conservation-law form, even in the presence of varying geometrical and mechanical properties. However, it has been proved for shallow water equations that this formulation is not mass-conservative and can lead to erroneous estimations of the wave celerity \cite{ToroBook2001}. This analysis is also valid for blood flow equations and the numerical solutions obtained with a nonconservative system will be incorrect in the presence of elastic jumps. Indeed, the Rankine-Hugoniot jump relation of the nonconservative form is different from the one of the conservative form. \u{C}ani\'c \cite{Canic2002} and Sherwin \cite{Sherwin2003} were among the first to address the issue of the nonconservative source term for blood flow simulation. \u{C}ani\'c proposed to treat the nonconservative product in this source term through jump conditions, while Sherwin used a two-rarefaction Riemann solver when the material properties varied abruptly. More recently, Toro and Siviglia \cite{Toro2013} reformulated the 1D conservative system with varying geometrical and mechanical properties as a homogeneous quasi-linear system and solved the associated Riemann problem. To do so, they introduced an auxiliary steady variable containing the geometrical and mechanical properties of the artery, and also included variations of the external pressure. In the framework of path-conservative methods \cite{Pares2006}, M\"uller and Toro \cite{Muller2013} used this augmented quasi-linear system to propose an exactly well-balanced numerical scheme for all steady states (subcritical, transcritical and supercritical). Murillo and Garc\'ia-Navarro \cite{Murillo2015} derived an energy balanced numerical scheme in the framework of augmented solvers for arteries with varying mechanical and geometrical properties, and also variations of the external pressure. In \cite{Delestre2012}, Delestre and Lagr\'ee successfully applied the hydrostatic reconstruction (HR), proposed in \cite{Audusse2004} for shallow water equations, to compute blood flow in arteries with varying cross-sectional area. In more recent work \cite{Delestre2016}, Delestre extended the hydrostatic reconstruction (HR) to arteries with varying cross-sectional area and arterial wall rigidity. \\

The hydrostatic reconstruction (HR) meets the simplicity and efficiency requirements for 1D blood flow simulation and will be the reference well-balanced method used in this study. The hydrostatic reconstruction (HR) can be used with any finite-volume numerical flux for a conservative problem and guarantees the following natural properties of shallow water flows:
\begin{itemize}
\item well-balanced for the steady states at rest, or hydrostatic equilibria;
\item the conservation of mass;
\item the non-negativity of the water-height $h$;
\item the ability to compute dry states and transcritical flows;
\item a discrete or semi-discrete entropy inequality, which enables to compute the entropic solution in the presence of a discontinuity. 
\end{itemize}
Unfortunately, the steady states at rest preserved by the hydrostatic reconstruction (HR) are not relevant for blood flow as they only occur in "dead men" \cite{Delestre2012}. We propose two extensions of the hydrostatic reconstruction adapted to blood flow simulation in large arteries.

By relaxing some of the properties of the hydrostatic reconstruction (HR) such as the ability to compute dry states, we derive an extension of the hydrostatic reconstruction, that we refer to as the "low-Shapiro" hydrostatic reconstruction (HR-LS). The low-Shapiro hydrostatic reconstruction (HR-LS) accurately preserves low-Shapiro number steady states that may occur in large network simulations. The Shapiro number $S=u/c$ is the equivalent of the Froude number for shallow water equations and the Mach number for compressible Euler equations. We also adapt the subsonic hydrostatic reconstruction (HR-S), proposed by Bouchut \cite{Bouchut2010}, to blood flow equations with variable geometrical and mechanical properties. The subsonic hydrostatic reconstruction (HR-S) exactly preserves all subcritical steady states, including low-Shapiro number steady states. By construction, both the low-Shapiro hydrostatic reconstruction (HR-LS) and the subsonic hydrostatic reconstruction (HR-S) are able to accurately compute wave reflections and transmissions. The different numerical methods are then tested and compared in a series of steady and unsteady physiological flow configurations, where both the geometrical and mechanical wall properties vary. \\

This work is organized as follows. In section \ref{sec:Math-Model} we derive the hyperbolic system of equations that describes the flow of blood in large arteries and recall its main mathematical properties. In section \ref{sec:Num}, we present a kinetic numerical scheme for the homogeneous problem and the boundary conditions used in the examples presented in this study. In section \ref{sec:HR}, we introduce the low-Shapiro hydrostatic reconstruction (HR-LS) and the subsonic hydrostatic reconstruction (HR-S) for blood flow in arteries with varying mechanical and geometrical wall properties. In sections \ref{sec:Ex-Single} and \ref{sec:Ex-55}, we present a series a steady and unsteady test cases for a single artery and a 55 arteries network, in which we evaluate the performances of the different hydrostatic reconstruction techniques.\\

\section{Mathematical model}
\label{sec:Math-Model}

\subsection{Model derivation}

The 1D models for blood flow are derived by averaging over the cross-sectional area of each artery a simplified Navier-Stokes system of equations. These simplified equations are obtained using the long wave approximation (${D / \lambda} \ll 1$, ratio between the averaged diameter of an artery $D$ and the average wave length of the pulse wave $\lambda$) and supposing the axial symmetry of blood flow ($\partial_{\theta} = 0$). We recall that in arteries the ratio ${D / \lambda}$ is of the order of $1 \times 10^{-2}$, therefore the long wave hypothesis is asymptotically valid. Because blood and wall viscosities will damp the effects we want to highlight, namely pulse wave propagation, we neglect them in the rest of this study. We use the inviscid system of equations describing the flow of blood in an elastic artery at the axial position $x$ and time $t$
\begin{equation}
\left\{
\begin{split}
\frac{\partial A }{\partial t }& + \frac{\partial Q }{\partial x } = 0\\
\frac{\partial Q }{\partial t }& +  \frac{\partial  }{\partial x }\left( \frac{Q^2}{A}\right) =  -\frac{A}{\rho}\frac{\partial P }{\partial x }.
\end{split}
\right.
\label{eq:BF-Sys-Pressure}
\end{equation}
The variables $Q$, $A$ and $P$ are respectively the flow rate, the cross-sectional area and the blood pressure. We also introduce the flow velocity $u = \frac{Q}{A}$. The parameter $\rho$ is the density of blood and is supposed constant. For a description of the derivation of system \eqref{eq:BF-Sys-Pressure} we refer the reader to \cite{Lambert1958,Barnard1966,Hughes1973}. To close the system of equations, the variation of pressure is linked to the deformation of the artery. Assuming that the arterial wall is a homogeneous, incompressible Hookean solid and that the artery is represented by a thin-cylinder whose sections move independently of one another, the following wall law is obtained, describing the elastic, or spring-like, behavior of the arterial wall
\begin{equation}
P\left( x,t \right) = P_0 + K\left( x\right) \left( \sqrt{A\left( x,t \right)}-\sqrt{A_0\left( x\right)} \right),
\label{eq:Pressure-Elastic}
\end{equation}
where $A_0$ is the cross-sectional area at rest of the artery and $K$ is the arterial wall rigidity. Both quantities can vary with the axial position $x$. More complex and general pressure laws can be used (for example in veins \cite{Pedley1996}), yet equation \eqref{eq:Pressure-Elastic} contains sufficient information to describe the main features of blood flow in large arteries \cite{Wang2015,Politi2016}. Combining both system \eqref{eq:BF-Sys-Pressure} and equation \eqref{eq:Pressure-Elastic} we obtain the final 1D nonconservative system of equations
\begin{equation}
\left\{
\begin{split}
&\frac{\partial A }{\partial t } + \frac{\partial Q }{\partial x } = 0\\
&\frac{\partial Q }{\partial t } +  \frac{\partial F }{\partial x } = S_T,
\end{split}
\right.
\label{eq:BF-Sys-Topo}
\end{equation}
where $F$ is the momentum flux
\begin{equation}
F = \frac{Q^2}{A} + \frac{K}{3\rho}A^{\frac{3}{2}},
\label{eq:BF-Flux}
\end{equation}
and $S_T$ is a source term taking into account the possible variations of the geometrical and mechanical properties of the arterial wall
\begin{equation}
S_T = \frac{A}{\rho} \left( \frac{\partial  }{\partial x }\left( K\sqrt{A_0} \right) - \frac{2}{3}\sqrt{A}\frac{\partial K }{\partial x } \right).
\label{eq:BF-Source-Topo}
\end{equation}

\subsection{Hyperbolic system}

System \eqref{eq:BF-Sys-Topo} can be written as a system of balance laws
\begin{equation}
\frac{\partial \mathbf{U} }{\partial t } 
+ \frac{\partial  }{\partial x }\left[ \mathbf{F}\left( \mathbf{U},K \right) \right]
= \mathbf{S}\left( \mathbf{U},K \right) \frac{\partial \bm{\sigma} }{\partial x }.
\label{eq:BF-Sys}
\end{equation}
$\mathbf{U}$ and $\mathbf{F}$ are respectively the vector of conservative variables and the vector of mass and momentum flux
\begin{equation}
\mathbf{U}=  
\begin{bmatrix}
	A\\
	Q\\
\end{bmatrix},
\qquad
\mathbf{F}\left( \mathbf{U},K \right) = 
\begin{bmatrix}
	Q\\
	F\\
\end{bmatrix}
,
\end{equation}
and the vector $\bm{\sigma}$ and the matrix $\mathbf{S}$ are defined as:
\begin{equation}
\bm{\sigma}= 
\begin{bmatrix}
  K\\
  Z\\
\end{bmatrix}
=
\begin{bmatrix}
  K\\
  K\sqrt{A_0}\\
\end{bmatrix},
\qquad
  \mathbf{S}\left( \mathbf{U} \right) = 
  \begin{bmatrix}
  0 & 0 \\
-\frac{2}{3}\frac{A^{\frac{3}{2}}}{\rho} & \frac{A}{\rho} \\
  \end{bmatrix}.
\end{equation}
The main difficulty of system \eqref{eq:BF-Sys} lies in the presence of the nonconservative source term $\mathbf{S}\frac{\partial \bm{\sigma} }{\partial x}$. This nonconservative term vanishes when the cross-sectional area at rest $A_0$ and the arterial wall rigidity $K$ are constant, and system \eqref{eq:BF-Sys} is reduced to the following system of conservation laws
\begin{equation}
\frac{\partial \mathbf{U} }{\partial t } + \frac{\partial  }{\partial x }\left[\mathbf{F}\left( \mathbf{U},K \right)\right] = 0.
\label{eq:BF-Sys-Conservative}
\end{equation}
The conservative system \eqref{eq:BF-Sys-Conservative} has been thoroughly studied by many authors and we only briefly recall its properties. Additional details can be found in \cite{Formaggia2003}. To analyze the mathematical properties of the system \eqref{eq:BF-Sys-Conservative}, we compute the Jacobian matrix of the flux vector $\mathbf{F}$
\begin{equation}
\mathbf{J}\left( \mathbf{U},K \right) = \frac{\partial \mathbf{F} }{\partial \mathbf{U} }=
\begin{bmatrix}
0 & 1\\
\frac{K\sqrt{A}}{2 \rho}-\frac{Q^2}{A^2}	&	\frac{2Q}{A}	\\
\end{bmatrix}.
\end{equation}
$\mathbf{J}\left( \mathbf{U,K} \right)$ has two real eigenvalues $\lambda_1$ and $\lambda_2$, respectively associated to two right eigenvectors $\mathbf{R_1}$ and $\mathbf{R_2}$
\begin{equation}
\lambda_1 = \frac{Q}{A} - c
, \quad
\lambda_2 = \frac{Q}{A} + c ,
\qquad
\mathbf{R_1} = 
\begin{bmatrix}
	1\\
	\lambda_1 \\
\end{bmatrix} 
, \quad
\mathbf{R_2} = 
\begin{bmatrix}
	1\\
	\lambda_2 \\
\end{bmatrix} .
\end{equation}
The variable $c$ is the Moens-Korteweg wave speed \cite{Moens1878,Korteweg1878} and corresponds to the speed of pulse waves in an artery
\begin{equation}
c = \sqrt{\frac{K}{2\rho}\sqrt{A}}.
\label{eq:Moens-Korteweg-c}
\end{equation}
The hyperbolicity of the system is characterized by the Shapiro number $S_h$, introduced by Shapiro in \cite{Shapiro1977}
\begin{equation}
S_h = \frac{u}{ c} = \frac{1}{c} \frac{Q}{A} .
\end{equation}
$S_h$ as the analogue of the Froude number $F_r$ for the shallow-water equations or of the Mach number $M_a$ for compressible flows. Depending on the value of $S_h$, we distinguish two flow regimes, represented respectively by the subcritical velocity domain $\mathbb{U}_{sub}$ and the supercritical velocity domain $\mathbb{U}_{sup}$
\begin{equation}
\left\{
\begin{split}
\mathbb{U}_{sub} = & \left\{ \frac{Q}{A} \in \mathbb{R} \: | \: A > 0 , \: K > 0, \: Z > 0, \: S_h < 1 \right\} \\
\mathbb{U}_{sup} = & \left\{ \frac{Q}{A} \in \mathbb{R} \: | \: A > 0 , \: K > 0, \: Z > 0, \: S_h > 1 \right\} . \\
\end{split}
\right.
\end{equation}
In both regions $\mathbb{U}_{sub}$ and $\mathbb{U}_{sup}$, system \eqref{eq:BF-Sys-Conservative} is strictly hyperbolic as $\lambda_1 \neq \lambda_2$ and the right eigenvectors $\mathbf{R_1}$ and $\mathbf{R_2}$ are linearly independent. However, when $S_h=1$ the flow is critical and the system looses its strict hyperbolicity. In this case resonance phenomena can occur, leading to a possible loss of uniqueness of the solution \cite{Liu1987,Isaacson1992,LevequeBook2002,Han2012}. 

In physiological conditions, blood flow is almost always subcritical. Nevertheless, very specific pathologies may lead to supercritical flows but will not be the subject of this study. Only subcritical solutions of system \eqref{eq:BF-Sys-Conservative} and more generally of system \eqref{eq:BF-Sys} in $\mathbb{U}_{sub}$ will be considered here.\\

For solutions of system \eqref{eq:BF-Sys-Conservative} in $\mathbb{U}_{sub}$, linear algebra shows that the Jacobian matrix $\mathbf{J}$ is diagonalizable in the form $\mathbf{J} =\mathbf{R} \mathbf{\Delta} \mathbf{R}^{-1}$, where the columns of $\mathbf{R}$ are the right eigenvectors $\mathbf{R_1}$ and $\mathbf{R_2}$ and $\mathbf{\Delta}$ is a diagonal matrix containing the eigenvalues of $\mathbf{J}$. Introducing a new vector $\mathbf{W} = \left[W_1,W_2\right]^{T}$ such that $\partial_{\mathbf{U}} \mathbf{W} = \mathbf{R}^{-1}$, system \eqref{eq:BF-Sys-Conservative} can be written as:
\begin{equation}
\frac{\partial \mathbf{W} }{\partial t } + \mathbf{\Delta} \frac{\partial \mathbf{W} }{\partial x } = 0.
\label{eq:bf-characteristic}
\end{equation} 
Finally, by integrating the equation $\partial_{\mathbf{U}} \mathbf{W} = \mathbf{R}^{-1}$, the following expression for $\mathbf{W}$ is obtained
\begin{equation}
\mathbf{W} = 
\begin{bmatrix}
W_1\\
W_2\\
\end{bmatrix} = 
\begin{bmatrix}
\frac{Q}{A} - 4 c\\
\\
\frac{Q}{A} + 4 c\\
\end{bmatrix}.
\label{eq:bf-Riemann-Invariants}
\end{equation}
The vector $\mathbf{W}$ is often referred to as the Riemann invariant vector and is linked to the conservative variables
\begin{equation}
\left\{
\begin{split}
& A = \left( \frac{2 \rho}{K} \right)^2 \left( \frac{W_2 - W_1}{8} \right)^4 \\
& Q = A \frac{W_1 + W_2}{2} .\\
\end{split}
\right. 
\label{eq:bf-AQW}
\end{equation}
The relations \eqref{eq:bf-AQW} are useful to define the boundary conditions at the inlet and outlet of the computational domain.\\

The vector $\mathbf{U}$, solution of system \eqref{eq:BF-Sys-Conservative}, satisfies an entropy inequality linked to the entropy pair $\left( \eta,G \right)$
\begin{equation}
\frac{\partial \eta }{\partial t} + \frac{\partial G }{\partial x} \leq  0,
\label{BF:eq-Entropy-Inequality}
\end{equation}
where $\eta$ is the entropy and $G$ is the entropy flux
\begin{equation}
\left\{
\begin{split}
&\eta\left( \mathbf{U},K \right) =  \frac{Q^2}{2A} + \frac{2}{3}\frac{K}{\rho} A^{\frac{3}{2}} \\
&G\left( \mathbf{U} ,K\right) = \left( \frac{Q^2}{2A} + \frac{K}{\rho}A^{\frac{3}{2}} \right)\frac{Q}{A}.  \\
\end{split}
\right.
\end{equation}
This entropy inequality is extended to solutions of system \eqref{eq:BF-Sys} through a new entropy pair $\left( \tilde{\eta},\tilde{G} \right)$ taking into account the vector $\bm{\sigma}$
\begin{equation}
\left\{
\begin{split}
&\tilde{\eta}\left( \mathbf{U},\bm{\sigma} \right) =  \eta\left( \mathbf{U},K \right) - \frac{Z}{\rho} A \\
&\tilde{G}\left( \mathbf{U} ,\bm{\sigma}\right) = G\left( \mathbf{U},K \right) - \frac{Z}{\rho} Q. \\
\end{split}
\right.
\end{equation}
This entropy inequality is closely linked to the variation of the physical energy of the system. The existence of such an inequality is essential in order to select the correct physical solution across discontinuities \cite{GosseBook2013}.\\

System \eqref{eq:BF-Sys} admits non-trivial steady solutions, verifying the following steady state system of equations
\begin{equation}
\left\{
\begin{split}
& Q = C_1\\
& \frac{1}{2} \frac{Q^2}{A^2} + \frac{1}{\rho} \left( K\sqrt{A}-Z \right) = C_2 , \\
\end{split}
\right.
\label{eq:BF-Steady-State-Equation}
\end{equation}
where $C_1$ and $C_2$ are two constants. In the following, we note $E = \frac{1}{2} \frac{Q^2}{A^2} + \frac{1}{\rho} \left( K\sqrt{A}-Z \right)$ the energy discharge. A particular family of steady states are the steady states at rest, or "man at eternal rest" equilibria, defined by
\begin{equation}
\left\{
\begin{split}
& Q = 0\\
&  K\sqrt{A}-Z = C_2 . \\
\end{split}
\right.
\label{eq:BF-Steady-State-Equation-Rest}
\end{equation}
For shallow water flows, steady states mainly occur in lakes and verify the "man at eternal rest" equilibria \eqref{eq:BF-Steady-State-Equation-Rest}. In arteries, steady or quasi-steady flow regimes are observed in small segments when the frequency of the pulse wave is greatly reduced due to a high resistance of the flow, for example after severe stenoses or in smaller arteries. In these cases, the relevant equilibria are no longer the steady states at rest but the non-zero flow steady states described by system \eqref{eq:BF-Steady-State-Equation}.

\section{Numerical scheme for the homogeneous conservative system}
\label{sec:Num}

In this section we describe the finite volume numerical scheme used to solve the homogeneous conservative system \eqref{eq:BF-Sys-Conservative}. The spatial domain is discretized in a series of cells $C_i$ defined as
\begin{equation}
C_i = \left[x_{i-\frac{1}{2}}, \: x_{i+\frac{1}{2}}  \right] = \left[x_i - \frac{\Delta x}{2}, \: x_i + \frac{\Delta x}{2}  \right] , \quad i \in \left[1,N\right],
\end{equation}
where $\Delta x$ is the cell size, supposed constant for simplicity. The time domain is also discretized using a constant time step $\Delta t$ and the discrete times are defined as
\begin{equation}
t^n = n \Delta t , \quad n \in \mathbb{N}.
\end{equation}

\subsection{Finite volume numerical scheme}

We first derive the integral form of the conservative system \eqref{eq:BF-Sys-Conservative} by integrating it with respect to $t$ and $x$ over $\left] t^n, \: t^{n+1}\right[ \times C_i$ \cite{LevequeBook2002}
\begin{equation}
\left.
\begin{split}
& \int_{C_i} \left[ \mathbf{U}\left( x,t^{n+1} \right) - \mathbf{U}\left( x,t^{n} \right) \right] \mathrm{d}x + \\
& \int_{t^n}^{t^{n+1}} \left[ \mathbf{F}\left( \mathbf{U}\left( x_{i+\frac{1}{2}},t \right), K\left( x_{i+\frac{1}{2}} \right) \right) - \mathbf{F}\left( \mathbf{U}\left( x_{i-\frac{1}{2}},t \right), K\left( x_{i-\frac{1}{2}} \right) \right) \right] \mathrm{d}t = 0
.
\end{split}
\right.
\label{eq:BF-Finite-Volume-Scheme}
\end{equation}
We then approximate the integrals in \eqref{eq:BF-Finite-Volume-Scheme} using the discrete variable $\mathbf{U_i^n}$ and the numerical flux $\mathbf{F^n_{i+\frac{1}{2}}}$, corresponding respectively to an approximation of the space average of the exact solution $\mathbf{U}$ over the cell $C_i$ at time $t^n$
\begin{equation}
\mathbf{U_i^n} \approx \frac{1}{\Delta x} \int_{C_i} \mathbf{U}\left( x,t^n \right)\text{d}x ,
\end{equation}
and to an approximation of the time average of $\mathbf{F}$ at the cell interface $I_{i+\frac{1}{2}}$
\begin{equation}
\mathbf{F^n_{i+\frac{1}{2}}} \approx  \frac{1}{\Delta t} \int_{t^n}^{t^{n+1}}  \mathbf{F}\left( \mathbf{U}\left( x_{i+\frac{1}{2}},t \right), K\left( x_{i+\frac{1}{2}} \right) \right) \mathrm{d}t .
\end{equation}
Using these definitions, we obtain the following explicit finite volume numerical scheme
\begin{equation}
\mathbf{U_i^{n+1}} = \mathbf{U_i^n} - \frac{\Delta t}{\Delta x} \left[ \mathbf{F^n_{i+\frac{1}{2}}} - \mathbf{F^n_{i-\frac{1}{2}}} \right].
\label{eq:BF-First-Order-Scheme}
\end{equation} 
We define $\mathbf{F^n_{i+\frac{1}{2}}}$ as a two-points numerical flux vector, namely
\begin{equation}
\mathbf{F^n_{i+\frac{1}{2}}} = \mathcal{F}\left( \mathbf{U_L},\mathbf{U_R} \right)
=
\begin{bmatrix}
\mathcal{F_A}\left( \mathbf{U_L},\mathbf{U_R} \right)\\
\mathcal{F_Q}\left( \mathbf{U_L},\mathbf{U_R} \right)\\
\end{bmatrix}
.
\end{equation}
As we focus only on first-order finite volume numerical schemes, the vectors $\mathbf{U_L}$ and $\mathbf{U_R}$ at the cell interface $I_{i+\frac{1}{2}}$ at time $t^n$ are defined as 
\begin{equation}
\mathbf{U_L} = \mathbf{U_i^n}, 
\qquad
\mathbf{U_R} = \mathbf{U_{i+1}^{n}}.
\label{eq:BF-UL-UR}
\end{equation}
The choice of the function $\mathcal{F}$ defines the numerical flux and thus the finite volume scheme. Several possibilities exist, and a review of the most common ones applied to blood flow equations can be found in \cite{Delestre2012,Muller2013,Muller2015,Wang2015,Murillo2015,Audebert2016}. 

\subsection{Kinetic numerical flux}

We choose to compute the function $\mathcal{F}$ using a kinetic numerical flux, and a review of this method applied to different systems of equations can be found in \cite{Bouchut1999}. The kinetic method was first introduced for shallow water equations in \cite{Perthame2001}, combined with the hydrostatic reconstruction (HR) in \cite{Audusse2005} and adapted to the blood flow in \cite{Delestre2012,Audebert2016}. The principal motivations for choosing a kinetic numerical flux are that it preserves the positivity of the cross-sectional area and its numerical diffusion is better suited to compute resonant solutions \cite{BouchutBook2004, Andrianov2005}. We briefly recall the classical kinetic approach.\\

Following \cite{Perthame2001,Audusse2005}, we introduce the real, positive, even and compactly supported function $\chi\left( w \right)$, verifying the following properties
\begin{equation}
\left\{
\begin{split}
& \chi\left( -w \right) = \chi \left( w \right) \\
& \int_\mathbb{R} \chi \left( w \right)\mathrm{d}w = \int_\mathbb{R}w^2 \chi \left( w \right) \mathrm{d}w = 1 .\\
\end{split}
\right.
\label{eq:kin-chi-Properties}
\end{equation}
We choose the following expression for the function $\chi\left( w \right)$
\begin{equation}
\chi\left( w \right) = 
\left\{
\begin{split}
& \frac{1}{2 \sqrt{3}} &\text{ if } |w| \leq \sqrt{3}\\
& 0 &\text{ else}.\\
\end{split}
\right.
\label{eq:kin-chi}
\end{equation}
Using this function, we define the kinetic Maxwellian, or so-called \textit{Gibbs equilibrium}, which represents the density of microscopic particles moving at the velocity $\xi \in \mathbb{R}$
\begin{equation}
M\left( x,t,\xi \right)=M\left( A,\xi-u \right) = \frac{A\left( x,t \right)}{\tilde{c}} \chi \left( \frac{\xi - u}{\tilde{c}} \right),
\label{eq:kin-Maxwellian}
\end{equation}
where
\begin{equation}
\tilde{c} = \sqrt{\frac{K}{3 \rho} \sqrt{A}}.
\label{eq:kin-c}
\end{equation}
Noticing that the integral and the first and second moments on $\mathbb{R}$ of $M$ respectively allow to recover $A$, $Q$ and $F$, it can be proved \cite{Perthame2001} that $\mathbf{U}$ is solution of system \eqref{eq:BF-Sys-Conservative} if and only if $M$ satisfies the following linear kinetic equation
\begin{equation}
\frac{\partial M }{\partial t } + \xi \frac{\partial M }{\partial x } = \mathcal{Q}\left( x,t,\xi \right),
\label{eq:BF-kin-eq}
\end{equation}
where $\mathcal{Q}\left( x,t,\xi \right)$ is a collision term that satisfies
\begin{equation}
\int_{\mathbb{R}} \mathcal{Q}\mathrm{d}\xi = \int_{\mathbb{R}} \xi \mathcal{Q}\mathrm{d}\xi = 0.
\label{eq:BF-kin-Collision}
\end{equation}
As the equation \eqref{eq:BF-kin-eq} is linear, it can be approximated by a simple upwind scheme. The flux function $\mathcal{F}$ is then obtained using the integral and the first moment of the upwind numerical flux used to solve the linear kinetic equation \eqref{eq:BF-kin-eq}, and writes
\begin{equation}
\mathcal{F}\left( \mathbf{U_L},\mathbf{U_R} \right) = \mathcal{F}^+\left( \mathbf{U_L} \right) + \mathcal{F}^-\left( \mathbf{U_R} \right),
\end{equation}
where $\mathbf{U_L}$ and $\mathbf{U_R}$ are defined as in \eqref{eq:BF-UL-UR}. The fluxes $\mathcal{F}^+\left( \mathbf{U} \right)$ and $\mathcal{F}^- \left( \mathbf{U} \right)$ are defined as 
\begin{equation}
\left\{
\begin{split}
& \mathcal{F}^+\left( \mathbf{U} \right) &=& \int_{\xi \geq 0} \xi 
\begin{bmatrix}
1\\
\xi\\
\end{bmatrix}
 M \left( A,\xi-u \right)\text{d} \xi\\
& \mathcal{F}^-\left( \mathbf{U} \right) &=& \int_{\xi \leq 0} \xi 
\begin{bmatrix}
1\\
\xi\\
\end{bmatrix}
 M \left( A,\xi-u \right)\text{d} \xi .\\
\end{split}
\right.
\end{equation}
After some computation, we find that
\begin{equation}
\left\{
\begin{split}
&\mathcal{F}^+\left( \mathbf{U} \right) =
\frac{A}{2 \sqrt{3} \tilde{c}}
\begin{bmatrix}
\frac{1}{2}\left(\left(\xi_{p}^+\right)^2 - \left(\xi_{m}^+\right)^2 \right)\\
\frac{1}{3}\left( \left(\xi_{p}^+\right)^3- \left(\xi_{m}^+\right)^3 \right)\\
\end{bmatrix}
\\
&\mathcal{F}^-\left( \mathbf{U} \right) =
\frac{A}{2 \sqrt{3} \tilde{c}} 
\begin{bmatrix}
\frac{1}{2}\left(\left(\xi_{p}^-\right)^2 - \left(\xi_{m}^-\right)^2 \right)\\
\frac{1}{3}\left( \left(\xi_{p}^-\right)^3- \left(\xi_{m}^-\right)^3 \right)\\
\end{bmatrix} ,\\
\end{split}
\right.
\label{eq:kin-num-flux}
\end{equation}
with
\begin{equation}
\left\{
\begin{split}
&\xi_p^+ = \max\left(0,u + \sqrt{3}\tilde{c}\right), \qquad &\xi_m^+ = \max\left(0,u - \sqrt{3}\tilde{c}\right)\\
&\xi_p^- = \min\left(0,u + \sqrt{3}\tilde{c}\right), \qquad &\xi_m^- = \min\left(0,u - \sqrt{3}\tilde{c}\right).\\
\end{split}
\right.
\label{eq:kin-xi}
\end{equation}
The stability of the scheme is ensured if at each time $t^{n}$, the time step $\Delta t$ verifies the following CFL (Courant, Friedrichs and Lewy) \cite{Courant1967} condition
\begin{equation}
\Delta t \leq \min_{i=1}^N \frac{\Delta x}{|u_i^n| + \tilde{c}_i^n}.
\label{eq:CFL-kin}
\end{equation}

\subsection{Initial condition}

All numerical simulations presented in this study are initialized by the following solution of the steady state at rest system \eqref{eq:BF-Steady-State-Equation-Rest}
\begin{equation}
Q = 0 \quad \mathrm{ and } \quad  A = A_0 ,\\
\label{eq:Initial-Condition}
\end{equation}
and the initial vector of conservative variable in the cell $C_i$ is then
\begin{equation}
\mathbf{U_i^0} = 
\begin{bmatrix}
A_{0,i}\\
0\\
\end{bmatrix} .
\label{eq:Initial-Vector}
\end{equation}

\subsection{Subcritical boundary condition}
\label{sec:BC}

In each artery at time $t^{n}$, boundary conditions are imposed in inlet and outlet ghost cells, respectively noted $C_{in}$ and $C_{out}$, by setting the value of their associated vector of conservative variable $\mathbf{U_{in}^n}$ and $\mathbf{U_{out}^n}$. As we compute subcritical solutions of system \eqref{eq:BF-Sys} in $\mathbb{U}_{sub}$, one boundary condition is imposed in the inlet ghost cell $C_{in}$ and one boundary condition is imposed in the outlet ghost cell $C_{out}$, respectively allowing to determine one component of $\mathbf{U_{in}^n}$ and one component of $\mathbf{U_{out}^n}$. To compute the remaining unknown components of $\mathbf{U_{in}^n}$ and $\mathbf{U_{out}^n}$, we follow the methodology proposed by Bristeau and Coussin \cite{Bristeau2001} and Alastruey \cite{Alastruey2008}. In the following, we assume that in each cell $C_i$ at time $t^{n}$, the discrete vector of conservative variables $\mathbf{U_i^n}$ is known.

\subsubsection{Inlet boundary condition: imposed flow rate $Q_{in}$}

We describe here a methodology to impose the flow rate $Q_{in}\left( t^{n} \right) = Q_{in}^n$ at the interface between the first cell of the computational domain $C_1$ and the inlet ghost cell $C_{in}$, namely
\begin{equation}
\mathcal{F_A}\left( \mathbf{U_{in}^n} , \mathbf{U_1^n} \right) = Q_{in}^n .
\label{eq:bc-inlet-flux}
\end{equation}
Taking advantage of the fact that the kinetic flux function $\mathcal{F}$ can be split in two, equation \eqref{eq:bc-inlet-flux} can be expressed as
\begin{equation}
\mathcal{F_A}^+\left( \mathbf{U_{in}^n} \right) +\mathcal{F_A}^-\left( \mathbf{U_1^n} \right) = Q_{in}^n.
\label{eq:bc-inlet-flux-kin}
\end{equation}
To ensure the stability of the scheme, this condition is imposed in an upwind manner. Following \cite{Bristeau2001}, we define the quantity 
\begin{equation}
a_1 = Q_{in}^n - \mathcal{F_A}^-\left( \mathbf{U_1^n} \right).
\end{equation}
Two possible cases exist:
\begin{itemize}
\item If $a_1 \leq 0$, the dominant part of the information is coming from inside the computational domain. As we are performing an upwind evaluation of the inlet boundary condition, we impose
\begin{equation}
\left\{
\begin{split}
& \mathcal{F_A}^+\left( \mathbf{U_{in}^n} \right) = 0\\
& \mathcal{F_Q}^+\left( \mathbf{U_{in}^n} \right) = 0 .\\
\end{split}
\right.
\label{eq:bc-inlet-flux-kin-a1-neg}
\end{equation}
\item If $a_1 > 0$, the dominant part of the information is coming from outside the computational domain. In this case, we impose
\[
\mathcal{F_A}^+\left( \mathbf{U_{in}^n} \right) = a_1.
\]
An additional equation is required to completely determine $\mathbf{U_{in}^n}$. We take advantage of the characteristic structure of the problem: as we are in the subcritical case, there exists an outgoing characteristic on which the Riemann invariant $W_1$ is constant. Using this property, we assume that a correct estimation of the cell average value of the outgoing Riemann invariant $W_1\left(\mathbf{U_{in}^n}\right)$ is $W_1\left( \mathbf{U_1^n} \right)$. Finally, we impose
\begin{equation}
\left\{
\begin{split}
& \mathcal{F_A}^+\left( \mathbf{U_{in}^n} \right) = a_1\\
&W_1\left(\mathbf{U_{in}^n}\right) = W_1\left( \mathbf{U_1^n} \right) . \\
\end{split}
\right.
\label{eq:bc-inlet-flux-kin-a1-pos}
\end{equation}
\end{itemize}
$\mathbf{U_{in}^n}$ is obtained by solving either system \eqref{eq:bc-inlet-flux-kin-a1-neg} or system \eqref{eq:bc-inlet-flux-kin-a1-pos}. This can be done using a classic Newton's method in a limited number of iterations ($\sim 5$).

\subsubsection{Unsteady outlet boundary condition: reflection of the outgoing characteristic}
\label{sec:BC-Rt}

We propose here a methodology to characterize the incoming information at the outlet of the computational domain. Indeed, as we are in the subcritical regime, there exists an outgoing characteristic on which the Riemann invariant $W_2$ is constant and an incoming characteristic on which propagates the Riemann invariant $W_1$. As in the previous case, the cell average value of the outgoing Riemann invariant $W_2\left(\mathbf{U_{out}^n}\right)$ can be estimated by $W_2\left( \mathbf{U_N^n} \right)$ and we impose
\begin{equation}
W_2\left(\mathbf{U_{out}^n}\right) = W_2\left( \mathbf{U_N^n} \right).
\end{equation}
The value of the incoming Riemann invariant $W_1\left(\mathbf{U_{out}^n}\right)$ is unknown as it propagates on a characteristic coming from outside the computational domain. In large artery simulations, it is common to estimate the incoming Riemann invariant $W_1\left(\mathbf{U_{out}^n}\right)$ as a fraction of the outgoing Riemann invariant $W_2\left(\mathbf{U_{out}^n}\right)$ \cite{Wang2015,Alastruey2008,Alastruey2009,Murillo2015}. This fraction is quantified by a reflection coefficient $R_t$ such that
\begin{equation}
W_1\left(\mathbf{U_{out}^n}\right) - W_1\left(\mathbf{U_{out}^0}\right) = - R_t \left[ W_2\left(\mathbf{U_{out}^n}\right) - W_2\left(\mathbf{U_{out}^0}\right) \right],
\label{eq:bc-outlet-Rt}
\end{equation}
where $W_1\left(\mathbf{U_{out}^0}\right)$ and $W_2\left(\mathbf{U_{out}^0}\right)$ are the initial Riemann invariants of the ghost cell $C_{out}$. The reflection coefficient $R_t$, whose value ranges between $0$ and $1$, models the reflective and resistive behavior of the network that is not taken into account in the numerical simulation and lies distal (anatomically located far from the point of reference) to the outlet of the computational domain. Finally, using the relations \eqref{eq:bf-AQW}, we solve the following system of equations to obtain $\mathbf{U_{out}^n}$
\begin{equation}
\left\{
\begin{split}
&W_2\left(\mathbf{U_{out}^n}\right) = W_2\left( \mathbf{U_N^n} \right)\\
&W_1\left(\mathbf{U_{out}^n}\right) - W_1\left(\mathbf{U_{out}^0}\right) = - R_t \left[ W_2\left(\mathbf{U_{out}^n}\right) - W_2\left(\mathbf{U_{out}^0}\right) \right] .\\
\end{split}
\right.
\label{eq:bc-outlet-sys}
\end{equation}
When we wish to remove any incoming information, or equivalently any distal reflection, we set $R_t=0$.

\subsubsection{Steady outlet boundary condition: imposed cross-sectional area $A_{out}$}
\label{sec:BC-At}

We describe here a methodology to impose the cross-sectional area $A_{out}$ at the outlet of the computational domain. Indeed, when the flow rate is imposed at the inlet of the computational domain and the vector $\bm{\sigma}$ is known, the steady states verifying system \eqref{eq:BF-Steady-State-Equation} are completely determined if we impose the value of the cross-sectional area at the outlet of the computational domain. We set in the outlet ghost cell $C_{out}$ a constant cross-sectional area $A_{out}$
\[
A_{out}^n = A_{out}.
\]
We need only to compute $Q_{out}^n$ to completely determine the outlet vector of conservative variables $\mathbf{U_{out}^n}$. To do so, we estimate as in the previous section $W_2\left( \mathbf{U_{out}^n} \right)$ by
\[
W_2\left(\mathbf{U_{out}^n}\right) = W_2\left( \mathbf{U_N^n} \right),
\] 
and using the relations \eqref{eq:bf-AQW}, we compute $W_1\left(\mathbf{U_{out}^n}\right)$ and then $Q_{out}^n$
\begin{equation}
\left\{
\begin{split}
& W_1\left(\mathbf{U_{out}^n}\right) = W_2\left(\mathbf{U_{out}^n}\right) - 8 c_{out}^n\\
& Q_{out}^n = A_{out}^n \frac{W_1\left(\mathbf{U_{out}^n}\right)+W_2\left(\mathbf{U_{out}^n}\right)}{2} . \\
\end{split}
\right.
\label{eq:bc-out-sys}
\end{equation}

\subsubsection{Junction boundary condition: conservation of mass and continuity of total pressure}

In large network simulations, boundary conditions must also be provided at every junction point, where arteries that are neither the inlet segment nor the terminal segments are connected together. At these junctions points, the outlets of the parent arteries are linked to the inlets of the daughter arteries through the conservation of mass and the continuity of total pressure, or equivalently the energy discharge \cite{Sherwin2003-2}. We consider here a junction point where a single parent artery $A_P$ is connected to $N_D$ daughter arteries $\left( A_{Di} \right)_{i=1}^{N_D}$. The values of $\mathbf{U_{out}^n} |_{A_P}$ and of $\left(\mathbf{U_{in}^n} |_{A_{Di}}\right)_{i=1}^{N_D}$ must be computed, with a total of $2\left( N_D+1 \right)$ unknowns. $N_D+1$ equations are obtained by estimating the outgoing Riemann invariant of the parent and daughter arteries as
\begin{equation}
\left\{
\begin{split}
&W_2\left(\mathbf{U_{out}^n}\right)|_{A_P} = W_2\left( \mathbf{U_N^n} \right)|_{A_P}\\
&W_1\left(\mathbf{U_{in}^n}\right)|_{A_{Di}} = W_1\left( \mathbf{U_1^n} \right)|_{A_{Di}} , \quad i\in \left[1,N_D\right] .\\
\end{split}
\right.
\label{eq:bc-conj-Riemann}
\end{equation}
The missing equations are provided by the conservation of mass and total pressure, or equivalently the energy discharge, at each junction point \cite{Raines1974,Sherwin2003}
\begin{equation}
\left\{
\begin{split}
&Q_{out}^n|_{A_P} = \sum_{i=1}^{N_D}Q_{in}^n |_{A_{Di}} \\
&\left[\frac{1}{2} \rho \left(\frac{Q_{out}^n}{A_{out}^n}\right)^2 + K \sqrt{A_{out}^n} - Z\right]|_{A_P} = \left[\frac{1}{2} \rho \left(\frac{Q_{in}^n}{A_{in}^n}\right)^2 + \right. \\
& \qquad \qquad \qquad \qquad \left.K \sqrt{A_{in}^n} - Z \right]|_{A_{Di}}  , \quad i=1,...,N_D .\\
\end{split}
\right.
\label{eq:bc-conj-Q-P}
\end{equation}
In practice, since in physiological conditions the flow is always subcritical, we can simplify the problem and impose only the continuity of pressure \cite{Alastruey2009}, neglecting the advection terms in the second equation of system \eqref{eq:bc-conj-Q-P}
\begin{equation}
\left\{
\begin{split}
&Q_{out}^n|_{A_P} = \sum_{i=1}^{N_D}Q_{in}^n |_{A_{Di}} \\
&\left[K \sqrt{A_{out}^n} - Z\right]|_{A_P} = \left[ K \sqrt{A_{in}^n} - Z \right]|_{A_{Di}}, \quad i=1,...,N_D .\\
\end{split}
\right.
\label{eq:bc-conj-Q-P-Low-Fr}
\end{equation}
This set of equations allows to accurately compute wave reflections and transmissions if a change of impedance occurs between the parent and the daughter arteries. Accurately computing reflected waves is crucial to obtain physiological wave forms in the simulated network, as the observed pressure and flow waves are the sum of the incoming and the multiple reflected waves \cite{Alastruey2009,Alastruey2011,Politi2016}.

\section{Hydrostatic reconstruction}
\label{sec:HR}

In many physiological configurations, the geometrical and mechanical properties of an artery vary significantly with its length. In the scope of this paper, these geometrical and mechanical gradients are limited to variations of the cross-sectional area at rest $A_0$ and the arterial wall rigidity $K$. To prevent spurious oscillations of the numerical solution of system \eqref{eq:BF-Sys} close to steady states, a well-balanced numerical scheme is required to properly balance the source term $S_T$ and the flux gradient $\partial_x \mathbf{F}$.\\

In order to make an explicit analogy with the well-balanced methods derived for shallow water equations, we introduce the following notations
\begin{equation}
\mathcal{P}(A,K)= \frac{K}{3 \rho} A^{\frac{3}{2}}
, \qquad
\mathcal{E}(A,K) = \frac{2K}{3 \rho} \sqrt{A}
, \qquad
H = K \sqrt{A}.
\label{eq:SW-Notations}
\end{equation}
With these notations, we have $\frac{\partial \mathcal{E} }{\partial A }= \frac{\mathcal{P}}{A^2}$ and the flux vector $\mathbf{F}$ can be expressed as
\begin{equation}
\mathbf{F}\left( \mathbf{U},K \right) = 
\begin{bmatrix}
	Q\\
	\frac{Q^2}{A} + \mathcal{P}(A,K)\\
\end{bmatrix}.
\label{eq:bf-Flux-Vector}
\end{equation}
Moreover, the steady state systems \eqref{eq:BF-Steady-State-Equation} and \eqref{eq:BF-Steady-State-Equation-Rest} can respectively be written as
\begin{equation}
\left\{
\begin{split}
& Q = C_1\\
& \frac{1}{2} \frac{Q^2}{A^2} + \mathcal{E}(A,K) + \frac{\mathcal{P}(A,K)}{A}  -\frac{Z}{\rho} = C_2 ,\\
\end{split}
\right.
\label{eq:BF-Steady-State-Equation-SW}
\end{equation}
and
\begin{equation}
\left\{
\begin{split}
& Q = 0\\
& H-Z = C_2 .\\
\end{split}
\right.
\label{eq:BF-Rest-Steady-State-Equation-SW}
\end{equation}

\subsection{The hydrostatic reconstruction: HR}

The hydrostatic reconstruction (HR) was introduced by Audusse \cite{Audusse2004} for shallow water equations and applied to blood flow equations by Delestre \cite{Delestre2012,Delestre2016}. Through a reconstruction of the conservative variables, HR allows to obtain a simple and efficient well-balanced numerical scheme given any finite volume numerical flux for the homogeneous problem \eqref{eq:BF-Sys-Conservative}. It is simple to implement and can easily be adapted to different pressure laws with multiple varying parameters, which is useful when considering veins, collapsible tubes and external pressure variations \cite{Pedley1996,Cavallini2010,Muller2014}. This technique allows to preserve at a discrete level the steady states at rest \eqref{eq:BF-Rest-Steady-State-Equation-SW} and guarantees that the scheme verifies some natural properties of the shallow water equations (listed as bullets in the introduction), such as the positivity of the water height (equivalent of the cross-sectional area $A$), the ability to compute dry states and transcritical flows  and a semi-discrete entropy inequality. This last property is necessary to select the admissible entropy solution across a discontinuity, as explained in \cite{GosseBook2013}.

On both sides of each cell interface $I_{i+\frac{1}{2}}$, reconstructed conservative variables are defined to preserve the following system of equations, which coincides with the steady states at rest system \eqref{eq:BF-Rest-Steady-State-Equation-SW} when the flow rate $Q$ or the velocity $u$ are zero
\begin{equation}
\left\{
\begin{split}
& u = \frac{Q}{A} = C_1 \\
& H-Z = C_2 . \\
\end{split}
\right.
\label{eq:WB-HRU-sys-SW}
\end{equation}
Details on the derivation of HR for blood flow in an artery with variable cross-sectional area $A$ and variable arterial wall rigidity $K$ can be found in \cite{Delestre2016}.\\

In large arteries, the steady states at rest preserved by HR only occur for "dead men" or distal to an obliterated segment and are of little interest when simulating blood flow in the systemic network. However, in regions of large flow resistance such as small arteries, arterioles or arteries severely constricted by a stenosis, the flow looses its pulsatility and reaches steady or near-steady states with a non-zero flow rate. These quasi-steady flow configurations can occur in large network simulations when the level of arterial precision extends to small arteries and arterioles or in the presence of a very severe stenosis. They are described by the steady state system \eqref{eq:BF-Steady-State-Equation-SW}. Therefore, a modification of HR is necessary to capture the relevant steady states for blood flow in large arteries, described by system \eqref{eq:BF-Steady-State-Equation-SW}.

\subsection{The low-Shapiro hydrostatic reconstruction: HR-LS}

System \eqref{eq:BF-Steady-State-Equation-SW} is nonlinear and difficult to solve in practice. However, in physiological conditions, blood flow is subcritical with a Shapiro number of the order of $S_h \approx 1 \times 10^{-2}$. Therefore, the nonlinear advection term $\frac{1}{2}\frac{Q^2}{A^2}$ in system \eqref{eq:BF-Steady-State-Equation-SW} can be neglected at first order with respect to the term $\mathcal{E}(A,K) + \frac{\mathcal{P}(A,K)}{A}  -\frac{Z}{\rho}$ that scales as $c^2$. Doing so, we obtain the following simplified low-Shapiro number steady state system of equations
\begin{equation}
\left\{
\begin{split}
& Q = C_1\\
& H-Z  = C_2 . \\
\end{split}
\right.
\label{eq:WB-HRQ-sys-SW}
\end{equation}
System \eqref{eq:WB-HRQ-sys-SW} coincides with the steady state at rest system \eqref{eq:BF-Rest-Steady-State-Equation-SW} when $Q$ or $u$ are zero and is an asymptotically correct approximation of the steady state system \eqref{eq:BF-Steady-State-Equation-SW} in low-Shapiro number flow regimes. It also contains the correct conservation properties to obtain low-Shapiro number wave reflections if a change of impedance occurs at the interface between two cells of the computational domain. Indeed, the conservation properties of system \eqref{eq:WB-HRQ-sys-SW} are identical to those of system \eqref{eq:bc-conj-Q-P-Low-Fr}, which have proved to be adequate to compute wave reflections and transmissions at junction points \cite{Alastruey2009,Wang2015}. System \eqref{eq:WB-HRQ-sys-SW} is the basis for the derivation of the modification of HR we propose in this study, referred to as the low-Shapiro hydrostatic reconstruction (HR-LS) and better suited to compute blood flow in physiological conditions.

HR-LS aims at preserving low-Shapiro number steady states system \eqref{eq:WB-HRQ-sys-SW} in an artery with a varying cross-sectional area at rest $A_0$ and arterial wall rigidity $K$. Similarly to HR, the well-balanced property is enforced by defining reconstructed variables on both sides of each cell interface $I_{i+\frac{1}{2}}$ according to the reconstruction procedure \eqref{eq:WB-HRQ-sys-SW}. In the following, variables noted with $"^*"$ will refer to the reconstructed variables. Given the vectors of conservative variables $\mathbf{U_L}$ and $\mathbf{U_R}$ and the vectors $\bm{\sigma_L}$ and $\bm{\sigma_R}$ at the left and right of the interface $I_{i+\frac{1}{2}}$ between cells $C_i$ and $C_{i+1}$, the discrete analogue of system \eqref{eq:WB-HRQ-sys-SW} writes
\begin{equation}
\left\{
\begin{split}
&Q_{L}^* = Q_{L}\\
&H_{L}^*- Z^* = H_L - Z_L ,\\
\end{split}
\right.
 \quad
\left\{
\begin{split}
&Q_{R}^* = Q_{R}\\
&H_{R}^*- Z^* = H_R - Z_R .\\
\end{split}
\right.
\label{eq:WB-HRQ-sys-SW-Discrete}
\end{equation}
By solving system \eqref{eq:WB-HRQ-sys-SW-Discrete} and preserving the positivity of $H$, we obtain the following reconstructed variables
\begin{equation}
\left\{
\begin{split}
&H_L^* = \max \left( 0,\:Z^* + H_L - Z_L \right)\\
&Q_L^* = Q_{L} ,\\
\end{split}
\right.
\quad
\left\{
\begin{split}
&H_R^* = \max \left( 0,\:Z^* + H_R - Z_R \right)\\
& Q_R^* = Q_{R} . \\
\end{split}
\right.
\end{equation}
The reconstructed variable $Z^*$ is chosen considering nonlinear stability arguments that require that
\[
\left\{
\begin{split}
&0 \leq H_{L}^* \leq H_{L}\\
&0 \leq H_{R}^* \leq H_{R} ,\\
\end{split}
\right.
\]
to preserve the positivity of $H$. A simple choice is the downwind value
\begin{equation}
\begin{split}
Z^* = \min \left( Z_L,Z_R \right) .
\end{split}
\end{equation}
In order to obtain the reconstructed values $A_L^*$ and $A_R^*$, we must select a reconstruction for $K^*$. Following \cite{BouchutBook2004,Delestre2016} we choose
\begin{equation}
K^* = \max(K_L,K_R).
\end{equation}
Therefore, we directly have
\begin{equation}
A_{L}^* = \left( \frac{H_{L}^*}{K^*} \right)^2
, \qquad
A_{R}^* = \left( \frac{H_{R}^*}{K^*} \right)^2 .
\end{equation}
Finally, at each side of the interface $I_{i+\frac{1}{2}}$, we obtain the reconstructed conservative vectors 
\begin{equation}
\mathbf{U_L}^* = 
\begin{bmatrix}
A_L^*\\
Q_L^*\\
\end{bmatrix}
, \qquad
\mathbf{U_R}^* = 
\begin{bmatrix}
A_R^*\\
Q_R^*\\
\end{bmatrix} ,
\label{eq:bf-WB-Conservative-Vector}
\end{equation}
that will be used to compute the numerical flux $\mathcal{F}\left( \mathbf{U_L^*} , \mathbf{U_R}^* \right)$.\\

A conservative formulation for the source term $S_T$ is obtained by integrating over the cell $C_i$ the steady flux gradient in which the nonlinear advection term is neglected. This approximation is valid in low-Shapiro number flow regimes, and therefore particularly appropriate for blood flow in large arteries. The following conservative expression for $S_T$ is obtained, expressed in terms of the reconstructed conservative vector $U^*$
\begin{equation}
\begin{split}
S_{Ti}^n = & \frac{1}{\Delta x} \int_{C_i} S_T\left( \mathbf{U},\bm{\sigma} \right)  \text{d}x  =  \mathcal{P}\left( {A_{L,i+\frac{1}{2}}^*},K_{i+\frac{1}{2}}^* \right) - \mathcal{P}\left( {A_{R,i-\frac{1}{2}}^*},K_{i-\frac{1}{2}}^* \right) ,\\
\end{split}
\end{equation}
where $\left( {A_{L,i+\frac{1}{2}}^*},{A_{R,i-\frac{1}{2}}^*} \right)$ are the reconstructed cross-sectional areas at the left of the cell interface $I_{i+\frac{1}{2}}$ and at the right of the cell interface $I_{i-\frac{1}{2}}$ respectively and $\left( {K_{i+\frac{1}{2}}^*} , {K_{i-\frac{1}{2}}^*} \right)$ are the reconstructed arterial wall rigidities at the cell interfaces $I_{i+\frac{1}{2}}$ and $I_{i-\frac{1}{2}}$ respectively. For consistency reasons, we modify the previous expression and write
\begin{equation}
\left.
\begin{split}
&S_{Ti} =&&
\left[\mathcal{P}\left( {A_{L,i+\frac{1}{2}}^*},K_{i+\frac{1}{2}}^* \right) - \mathcal{P}\left( {A_{L,i+\frac{1}{2}}},K_{L,i+\frac{1}{2}} \right) \right] - \\
&&& \left[\mathcal{P}\left( {A_{R,i-\frac{1}{2}}^*},K_{R,i-\frac{1}{2}}^* \right) - \mathcal{P}\left( {A_{R,i-\frac{1}{2}}},K_{R,i-\frac{1}{2}} \right) \right] . \\
\end{split}
\right.
\end{equation}
To simplify the expression, we introduce the notation
\begin{equation}
\mathcal{P}\left( A,A^*,K,K^* \right)=\mathcal{P}\left( A^*,K^* \right) - \mathcal{P}\left( A,K \right).
\end{equation}
With these notations, the first order well-balanced finite-volume numerical scheme for system \eqref{eq:BF-Sys} is simply
\begin{equation}
\mathbf{U_i^{n+1}} = \mathbf{U_i^n} - \frac{\Delta t}{\Delta x} \left[ \mathbf{F^{n*}_{i+\frac{1}{2}}} - \mathbf{F^{n*}_{i-\frac{1}{2}}} \right],
\label{
}
\end{equation}
with 
\begin{equation}
\left\{
\begin{split}
&\mathbf{F^{n*}_{i+\frac{1}{2}}} = && \mathcal{F}\left( \mathbf{U_{L,i+\frac{1}{2}}}^*,\mathbf{U_{R,i+\frac{1}{2}}}^*,K_{i+\frac{1}{2}}^* \right) +\\
& & &
\begin{bmatrix}
0 \\
 \mathcal{P}\left( A_{L,i+\frac{1}{2}},A_{L,i+\frac{1}{2}}^*,K_{L,i+\frac{1}{2}},K_{i+\frac{1}{2}}^* \right)\\
\end{bmatrix}
\\
&\mathbf{F^{n*}_{i-\frac{1}{2}}} = & & \mathcal{F}\left( \mathbf{U_{L,i-\frac{1}{2}}}^*,\mathbf{U_{R,i-\frac{1}{2}}}^*,K_{i-\frac{1}{2}}^* \right) + \\
& & &
\begin{bmatrix}
0 \\
 \mathcal{P}\left( A_{R,i-\frac{1}{2}},A_{R,i-\frac{1}{2}}^*,K_{R,i-\frac{1}{2}},K_{i-\frac{1}{2}}^* \right) \\
\end{bmatrix}
 .\\
\end{split}
\right.
\end{equation}
It is straightforward to see that HR-LS is well-balanced for the steady states at rest system \eqref{eq:BF-Rest-Steady-State-Equation-SW} and provides a good evaluation of low-Shapiro number steady states system \eqref{eq:WB-HRQ-sys-SW}. It also guarantees the following natural properties of blood flow equations:
\begin{itemize}
\item the conservation of mass;
\item the non-negativity of the cross-sectional area $A$;
\item correct reflection and transmission conditions when variations of vessel impedance occur.
\end{itemize}
In physiological conditions, the arteries never completely collapse, therefore the numerical scheme no longer needs to be able to compute dry states. Furthermore, as the flow is subcritical and the heart input signal is not discontinuous, transcritical or supercritical regimes and discontinuities of the conservative variables do no occur. Hence the semi-discrete entropy inequality as well as the ability to compute transcritical flows are no longer crucial requirements of the numerical scheme. Finally, the viscosity of the blood and of the arterial wall, that are not taken into account in the theoretical part of this study, are of great importance in arteries and have diffusive and dissipative effects that remove high frequency components and therefore any discontinuity in the conservative variables. This last point will be addressed in the last example section \ref{sec:Ex-55}.

\subsection{The subsonic hydrostatic reconstruction: HR-S}

In \cite{Bouchut2010}, Bouchut proposed an extension of HR, referred to as the subsonic hydrostatic reconstruction (HR-S), ideal for blood flow simulations in large arteries. HR-S is well-balanced for all subcritical steady states \eqref{eq:BF-Steady-State-Equation-SW} and also preserves the good properties of HR (listed as bullets in the introduction), that is the positivity of the water height (equivalent of the cross-sectional area $A$), the ability to compute dry states and transcritical flows and a semi-discrete entropy inequality. HR-S is also able to correctly capture wave reflections and transmissions in regions where the impedance of the arterial wall changes. Indeed, the subcritical steady states system \eqref{eq:BF-Steady-State-Equation-SW} coincides with the junction conservation properties \eqref{eq:bc-conj-Q-P}. However, HR-S requires the resolution of the nonlinear steady state system \eqref{eq:BF-Steady-State-Equation-SW} at each time step at every cell interface presenting a gradient of the artery's geometrical or mechanical properties. This increases the computational cost compared to HR and HR-LS, especially if the region requiring a well-balanced treatment is not limited to a few cells.

In this section, we present the derivation of HR-S adapted to blood flow in an artery where both variations of cross-sectional area at rest $A_0$ and variations of the arterial wall rigidity $K$ are taken into account. HR-S will serve as a reference exactly well-balanced method to be compared to HR and HR-LS. In particular, HR-S will allow us to assess if relaxing the dry-state property and the semi-discrete entropy inequality in HR-LS impacts solutions of blood flow in physiological conditions. With the notations \eqref{eq:SW-Notations} and \eqref{eq:BF-Steady-State-Equation-SW}, we are in the framework introduced by Bouchut \cite{Bouchut2010}. Therefore, we will only briefly recall the main steps of the derivation of HR-S and additional details can be found in the cited publication.

\subsubsection{Well-balanced subsonic positivity-preserving reconstruction procedure for the cross-sectional area $A$}

Similarly to HR and HR-LS, the well-balanced property is enforced by defining reconstructed variables on both sides of each cell interface $I_{i+\frac{1}{2}}$ according to the reconstruction procedure \eqref{eq:BF-Steady-State-Equation-SW}. Variables noted with $"^*"$ will refer to the reconstructed variables. Following \cite{Bouchut2010}, we introduce the function $f$
\begin{equation}
\left.
\begin{split}
f: \: \:& \mathbb{R}\times \left(\mathbb{R}^{+*}\right)^2 &\rightarrow& \mathbb{R} \\
& \left(Q,A,K \right) &\rightarrow& \frac{1}{2}\frac{Q^2}{A^2} + \left[ \mathcal{E}\left( A,K \right) + \frac{\mathcal{P}\left( A,K \right)}{A} \right] ,\\
\end{split}
\right.
\label{eq:WB-HRS-f}
\end{equation}
and given the vectors of conservative variables $\mathbf{U_L}$ and $\mathbf{U_R}$ and the vectors $\bm{\sigma_L}$ and $\bm{\sigma_R}$ at the left and right of the interface $I_{i+\frac{1}{2}}$ between cells $C_i$ and $C_{i+1}$, the discrete analogue of system \eqref{eq:BF-Steady-State-Equation-SW} writes
\begin{equation}
\left.
\begin{split}
&\left\{
\begin{split}
& Q_L^* = Q_L \\
& f\left( Q_L^*,A_L^*,K^* \right) = f\left( Q_L,A_L,K_L \right)   + \delta_L\\
\end{split}
\right.\\
&\left\{
\begin{split}
& Q_R^* = Q_R \\
& f\left( Q_R^*,A_R^*,K^* \right) = f\left( Q_R,A_R,K_R \right)   + \delta_R ,\\
\end{split}
\right.\\
\end{split}
\right.
\label{eq:BF-Steady-State-Equation-f}
\end{equation}
with
\begin{equation}
\left\{
\begin{split}
&\delta_L =  \frac{1}{\rho}\left( Z^* - Z_L \right)\\
&\delta_R = \frac{1}{\rho}\left( Z^* - Z_R \right) .\\
\end{split}
\right.
\label{eq:BF-HRS-delta}
\end{equation}
Similarly to HR-LS, the reconstruction of the flow rate $Q^*$ is straightforward. However, contrary to HR and HR-LS, system \eqref{eq:BF-Steady-State-Equation-f} is nonlinear in $A^*$ and is difficult to solve analytically. To help the resolution of system \eqref{eq:BF-Steady-State-Equation-f}, we recall the following properties (see \cite{Bouchut2010} for details).\\

\begin{figure}[!h]
\begin{center}
\makebox[1.\textwidth][c]{
\begin{minipage}[t]{1\textwidth}
  \centering
  \includegraphics[scale=0.35,angle=0,trim={4cm 4cm 2.4cm 4cm},clip]{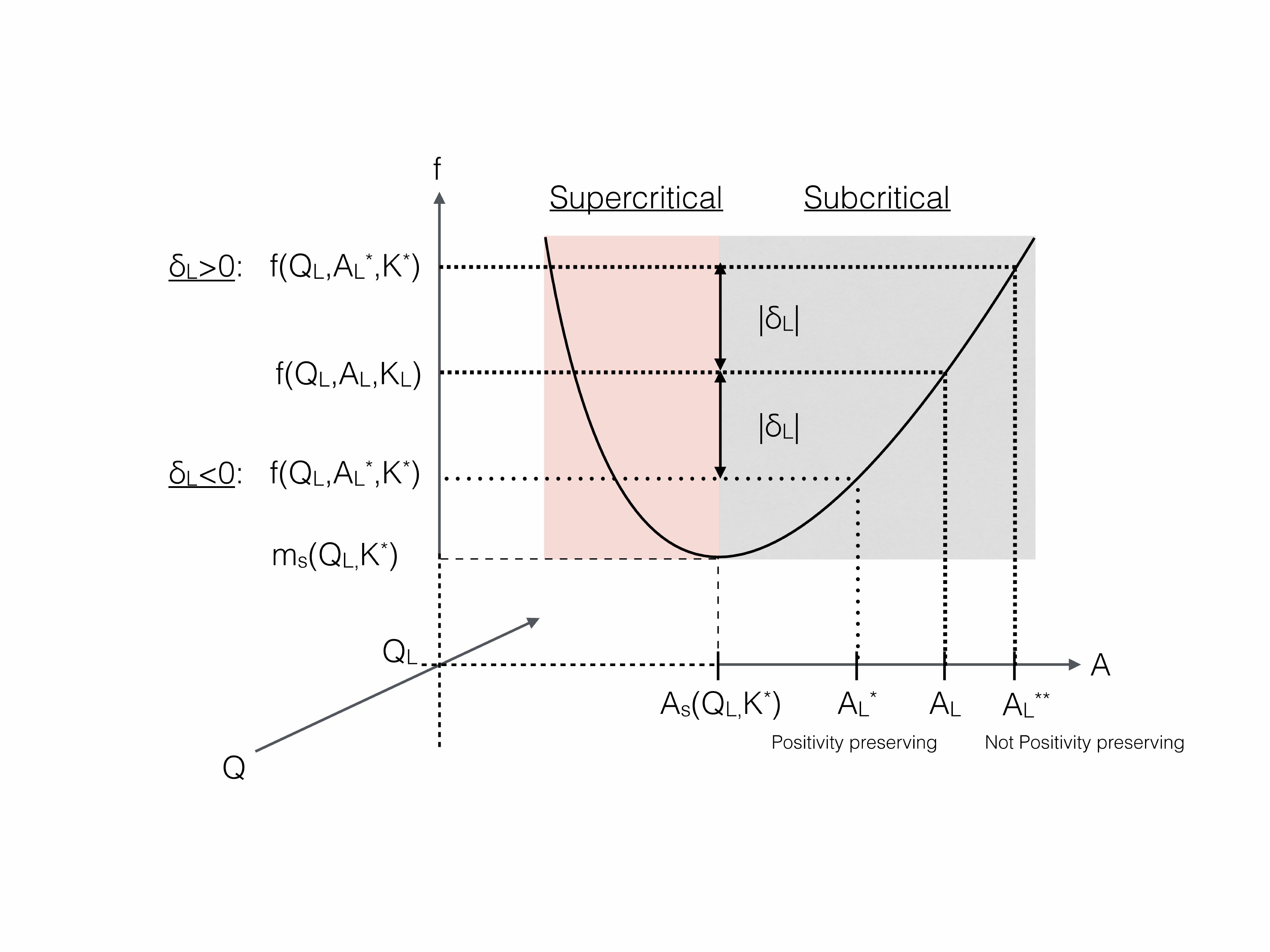}\\
\end{minipage}%
}
\caption{Representation of the function $f\left( Q_L,\cdot,\cdot\right)$. The abscissa of the intersections between function $f$ and the straight lines representing the different values of $f\left( Q_L^*,A_L^*,k^* \right)$ give the possible values of $A_L^*$. A graphical analysis shows that conditions \eqref{eq:WB-HRS-A-Inequality} and \eqref{eq:WB-HRS-A-Inequality-Sub} are met only for $\delta_L < 0$.}
\label{fig:HRS-f}
\end{center}
\end{figure}

For fixed values of $Q$ and $K$, function $f$ admits a minimum in $A_s\left( Q,K \right)$ and $m_s\left( Q,K \right)$ is the minimum value of $f$
\begin{equation}
A_s\left( Q,K \right) = \left( \frac{2 \rho }{K} Q^2 \right)^{\frac{2}{5}}
, \qquad
m_s\left( Q,K \right) = \frac{5}{4}\frac{K}{\rho}\left[ \frac{2 \rho }{K} Q^2  \right]^{\frac{1}{5}} .
\label{eq:WB-HRS-As-ms}
\end{equation}
For fixed values of $Q$ and $K$ and since the function $f$ is convex, system \eqref{eq:BF-Steady-State-Equation-f} admits a subcritical and a supercritical solution for the cross-sectional area $A$ if $f\left( Q,A,K \right)> m_s\left( Q,K \right)$. Furthermore, if $A > A_s\left( Q,K \right)$ the flow is subcritical with $U \in \mathbb{U}_{sub}$ and inversely if $A < A_s\left( Q,K \right)$ the flow is supercritical with $U \in \mathbb{U}_{sup}$ (see figure \ref{fig:HRS-f}). \\

Using these properties, Bouchut \cite{Bouchut2010} proposed a reconstruction procedure for the cross-sectional area $A^*$. The first step is to select reconstructions of the variables $Z^*$ and $K^*$ that preserve the positivity of $A$ and select the subcritical solution of system \eqref{eq:BF-Steady-State-Equation-f}. The following inequalities must be verified to respectively preserve the positivity of $A$ and select the subcritical solution of system \eqref{eq:BF-Steady-State-Equation-f}
\begin{equation}
\left\{
\begin{split}
&A_{L}^* \leq A_{L}\\
&A_{R}^* \leq A_{R} ,\\
\end{split}
\right.
\label{eq:WB-HRS-A-Inequality}
\end{equation}
and 
\begin{equation}
\left\{
\begin{split}
&A_s \leq A_{L}^* \\
&A_s \leq A_{R}^* .\\
\end{split}
\right.
\label{eq:WB-HRS-A-Inequality-Sub}
\end{equation}
The inequalities \eqref{eq:WB-HRS-A-Inequality-Sub} are naturally verified as we consider only subcritical flow configurations. On the contrary, the inequalities \eqref{eq:WB-HRS-A-Inequality} are verified if inequalities \eqref{eq:WB-HRS-A-Inequality-Sub} are true and if $Z^*$ and $K^*$ are chosen such that $\delta_{L,R}\leq 0$. A simple choice for $Z^*$ and $K^*$ is
\begin{equation}
Z^* = \min \left( Z_L,Z_R \right)
, \qquad
K^* = \max(K_L,K_R).
\label{eq:WB-HRS-Z-K}
\end{equation}
Given the expressions \eqref{eq:WB-HRS-Z-K} for $Z^*$ and $K^*$, we adapted the reconstruction procedure for the cross-sectional area $A^*$ proposed by Bouchut \cite{Bouchut2010} to blood flow in arteries with variable cross-sectional area $A_0$ and variable arterial wall rigidity $K$. It is summarized in figure \ref{fig:HRS-f} and is presented in the algorithm \ref{alg:HRS-A-Reconstruction}. The algorithm \ref{alg:HRS-A-Reconstruction} describes the steps that need to be followed to obtain the reconstructed cross-sectional area $A_{L}^*$, solution of system \eqref{eq:BF-Steady-State-Equation-f}. The same algorithm can be applied to reconstruct $A_R^*$.\\

\begin{algorithm}[H]
\caption{Algorithm to compute the reconstructed cross-sectional area $A_L^*$ to enforce the well-balanced property by interface for the steady state system \eqref{eq:BF-Steady-State-Equation-SW}.}
\label{alg:HRS-A-Reconstruction}
\begin{algorithmic}

\If{$\delta_L =0$} 
	\State $A_{L}^* \gets A_L$
\Else 
	\If{$u_L \geq c_L$}
		\State $A_{L}^* \gets A_L$
	\Else
		\If{$f\left( Q_L,A_L,K_L \right) + \delta_L > m_s\left( Q_L,K^* \right)$}
			\State \[
			\left\{
			\begin{split}
			&Q_{L}^* = Q_{L}\\
			&f\left( Q_{L}^*,A_L^*,K^* \right) = f\left( Q_L,A_L,K_L \right) + \delta_L\\
			\end{split}
			\right.
			\]
			\State The solution of the system can be obtained numerically using a recursive procedure.
		\Else
			\State $A_{L}^* \gets A_s(Q_L,K^*)$
		\EndIf	
	\EndIf
\EndIf

\end{algorithmic}
\end{algorithm}

\subsubsection{Well-balanced subsonic first-order numerical scheme}

Similarly to HR and HR-LS, a conservative formulation for the source term $S_T$ is obtained by integrating over the cell $C_i$ the steady flux gradient. However, the nonlinear advection term is no longer neglected and an additional flux term is introduced to take it into account
\begin{equation}
\begin{split}
S_{Ti} =& 
\mathcal{P}\left( A_{L,i+\frac{1}{2}},A_{L,i+\frac{1}{2}}^*,K_{L,i+\frac{1}{2}},K_{i+\frac{1}{2}}^* \right) + \\
&\mathcal{T}_L\left( \mathbf{U_{L,i+\frac{1}{2}}},\mathbf{U_{L,i+\frac{1}{2}}}^*,\mathbf{U_{R,i+\frac{1}{2}}}^*,K_{i+\frac{1}{2}}^* \right) - \\
& \mathcal{P}\left( A_{R,i-\frac{1}{2}},A_{R,i-\frac{1}{2}}^*,K_{R,i-\frac{1}{2}},K_{i-\frac{1}{2}}^* \right) - \mathcal{T}_R\left( \mathbf{U_{R,i-\frac{1}{2}}}, \mathbf{U_{L,i-\frac{1}{2}}}^*,\mathbf{U_{R,i-\frac{1}{2}}}^*,K_{i-\frac{1}{2}}^* \right),
\end{split}
\end{equation}
where $\left( {A_{L,i+\frac{1}{2}}^*},{A_{R,i-\frac{1}{2}}^*} \right)$ are the reconstructed cross-sectional areas at the left the cell interface $I_{i+\frac{1}{2}}$ and at the right of the cell interface $I_{i-\frac{1}{2}}$ respectively and $\left( {K_{i+\frac{1}{2}}^*} , {K_{i-\frac{1}{2}}^*} \right)$ are the reconstructed arterial wall rigidities at the cell interfaces $I_{i+\frac{1}{2}}$ and $I_{i-\frac{1}{2}}$ respectively. The additional fluxes $\mathcal{T}_L$ and $\mathcal{T}_R$ are chosen such that the numerical scheme satisfies an entropy inequality by interface (see \cite{Bouchut2010} for details). The computation of $\mathcal{T}_L$ and $\mathcal{T}_R$ is presented in the algorithm \ref{alg:HRS-Flux-Computation}. Only the steps that need to be followed to obtain $\mathcal{T}_L$ are detailed in \ref{alg:HRS-Flux-Computation} but similar results are obtained for $\mathcal{T}_R$. 
\begin{algorithm}[H]
\caption{Algorithm to compute the flux $\mathcal{T}_L$ used in HR-S to balance the nonlinear advection term $\frac{Q^2}{A}$ and the source term $S_T$.}
\label{alg:HRS-Flux-Computation}
To simplify the expression of $\mathcal{T}_L$ we use the following notations
\[
\left\{
\begin{split}
&\mathcal{F_A} = \mathcal{F_A}\left( \mathbf{U_L^*},\mathbf{U_R^*},K^* \right)\\
&\mathcal{F_Q} = \mathcal{F_Q}\left( \mathbf{U_L^*},\mathbf{U_R^*},K^* \right)\\
&\mathcal{P} = \mathcal{P}\left( A_L,A_L^*,K_L,K^* \right)\\
& \Delta f = f\left( Q_L^*,A_L^*,K^* \right) - f\left( Q_L,A_L,K_L \right) - \delta_L, \\
\end{split}
\right.
\]
\begin{algorithmic}

\If{$\delta_L =0$} 
	\State $\mathcal{T}_L\left(\mathbf{U_L}, \mathbf{U_L^*},\mathbf{U_R^*},K^* \right) \gets 0$
\Else 
	\If{$u_L \geq c_L$}
		\State $ \mathcal{T}_L\left(\mathbf{U_L}, \mathbf{U_L^*},\mathbf{U_R^*},K^* \right) \gets  -\frac{A_L}{Q_L} \, \mathcal{F_A} \, \delta_L $
	\Else
		\If{$f\left( Q_L,A_L,K_L \right) + \delta_L > m_s\left( Q_L,K^* \right)$}
			\State \[
			\begin{split}
			\mathcal{T}_L\left(\mathbf{U_L}, \mathbf{U_L^*},\mathbf{U_R^*},K^* \right) \gets 
			&\frac{A_L - A_L^*}{A_L^*}  \left[ \mathcal{F_Q} - \mathcal{P} - \frac{Q_L^*}{A_L^*}\mathcal{F_A} \right] - \\
			&\mathcal{F_A} \left[ \frac{Q_L^*}{A_L^*} - \frac{Q_L}{A_L} \right]	
			\end{split}	
			\]
		\Else
			\State \[
			\begin{split}
			\mathcal{T}_L\left(\mathbf{U_L}, \mathbf{U_L^*},\mathbf{U_R^*},K^* \right) \gets 
			&\frac{A_L - A_L^*}{A_L^*}  \left[ \mathcal{F_Q} - \mathcal{P} - \frac{Q_L^*}{A_L^*}\mathcal{F_A} \right] - \\
			&\mathcal{F_A} \left[ \frac{Q_L^*}{A_L^*} - \frac{Q_L}{A_L} \right] +  \frac{A_L}{Q_L} \mathcal{F_A} \, \Delta f
			\end{split}	
			\]
		\EndIf	
	\EndIf
\EndIf

\end{algorithmic}
\end{algorithm}

Finally, the first-order well-balanced finite-volume numerical scheme for system \eqref{eq:BF-Sys} is still
\begin{equation}
\mathbf{U_i^{n+1}} = \mathbf{U_i^n} - \frac{\Delta t}{\Delta x} \left[ \mathbf{F^{n*}_{i+\frac{1}{2}}} - \mathbf{F^{n*}_{i-\frac{1}{2}}} \right],
\label{eq:BF-First-Order-Num-HRS}
\end{equation}
with 
\begin{equation}
\left\{
\begin{split}
&\mathbf{F^{n*}_{i+\frac{1}{2}}} & =& \mathcal{F}\left( \mathbf{U_{L,i+\frac{1}{2}}}^*,\mathbf{U_{R,i+\frac{1}{2}}}^*,K_{i+\frac{1}{2}}^* \right) +\\
& &&
\begin{bmatrix}
0 \\
 \mathcal{P}\left( A_{L,i+\frac{1}{2}},A_{L,i+\frac{1}{2}}^*,K_{L,i+\frac{1}{2}},K_{i+\frac{1}{2}}^* \right) + \mathcal{T}_L\left(\mathbf{U_{L,i+\frac{1}{2}}}, \mathbf{U_{L,i+\frac{1}{2}}}^*,\mathbf{U_{R,i+\frac{1}{2}}}^*,K_{i+\frac{1}{2}}^* \right)\\
\end{bmatrix}
\\
&\mathbf{F^{n*}_{i-\frac{1}{2}}} & =& \mathcal{F}\left( \mathbf{U_{L,i-\frac{1}{2}}}^*,\mathbf{U_{R,i-\frac{1}{2}}}^*,K_{i-\frac{1}{2}}^* \right) +\\
& &&
\begin{bmatrix}
0 \\
{P}\left( A_{R,i-\frac{1}{2}},A_{R,i-\frac{1}{2}}^*,K_{R,i-\frac{1}{2}},K_{i-\frac{1}{2}}^* \right) + \mathcal{T}_R\left(\mathbf{U_{R,i-\frac{1}{2}}}, \mathbf{U_{L,i-\frac{1}{2}}}^*,\mathbf{U_{R,i-\frac{1}{2}}}^*,K_{i-\frac{1}{2}}^* \right)\\
\end{bmatrix}
. \\
\end{split}
\right.
\end{equation}

In the following section, we present a series of numerical test-cases were we systematically compare HR, HR-LS and HR-S.

\section{Physiological examples in a single artery}
\label{sec:Ex-Single}

In this section we present a series of numerical computations designed to evaluate the performances in physiological conditions of the low-Shapiro hydrostatic reconstruction (HR-LS) in comparison with the hydrostatic reconstruction (HR) and the subsonic hydrostatic reconstruction (HR-S). All quantities are represented in centimeters, grams and seconds, or equivalently "cgs", which are the natural units to describe blood flow. Indeed, the density of blood is close to 1 in "cgs".\\

\begin{figure}[!h]
\begin{center}
\makebox[1.\textwidth][c]{
\begin{minipage}[t]{0.33\textwidth}
  \centering
  \includegraphics[scale=0.25,angle=0]{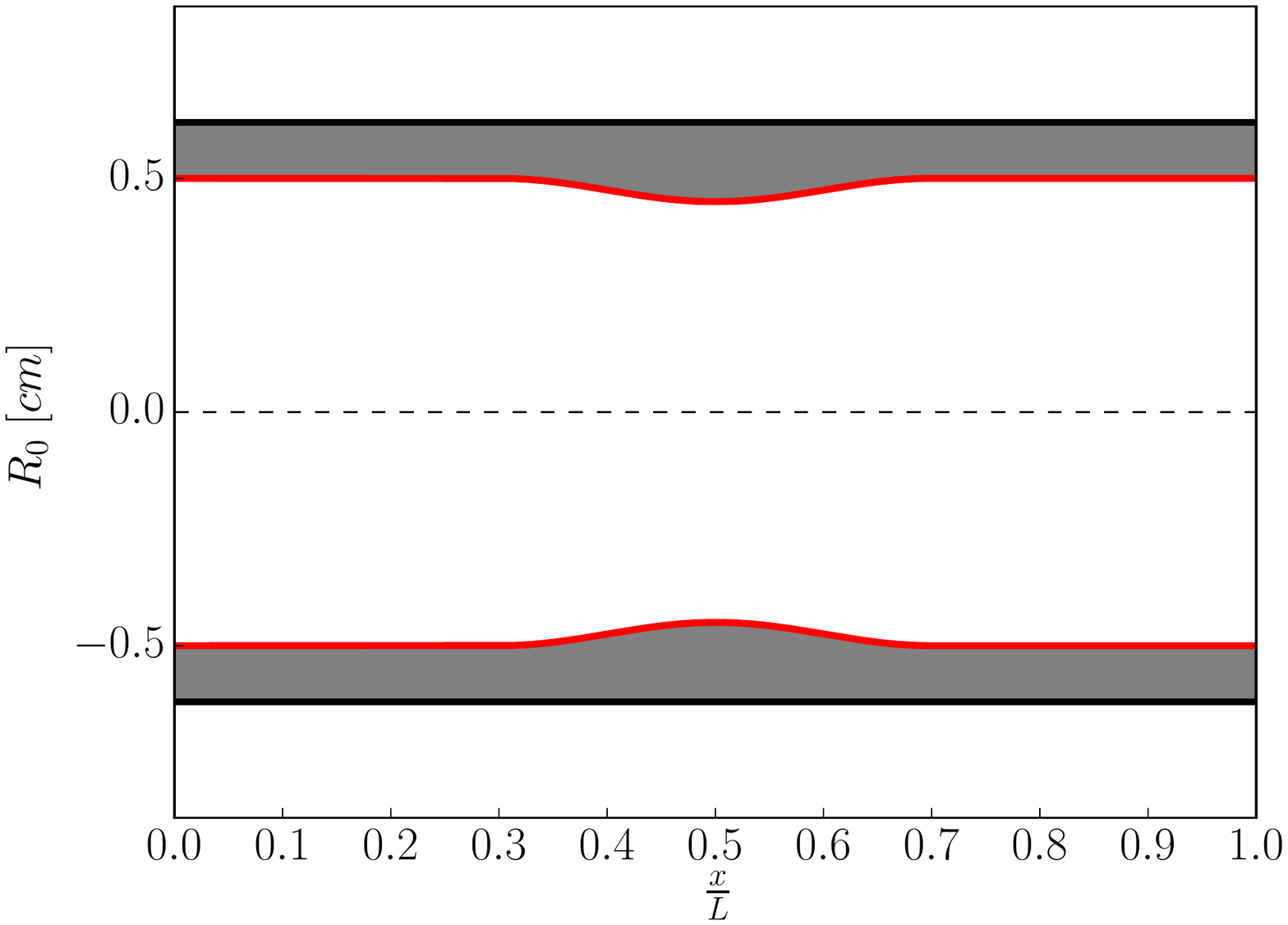}\\
\end{minipage}%
\begin{minipage}[t]{0.33\textwidth}
  \centering
  \includegraphics[scale=0.25,angle=0]{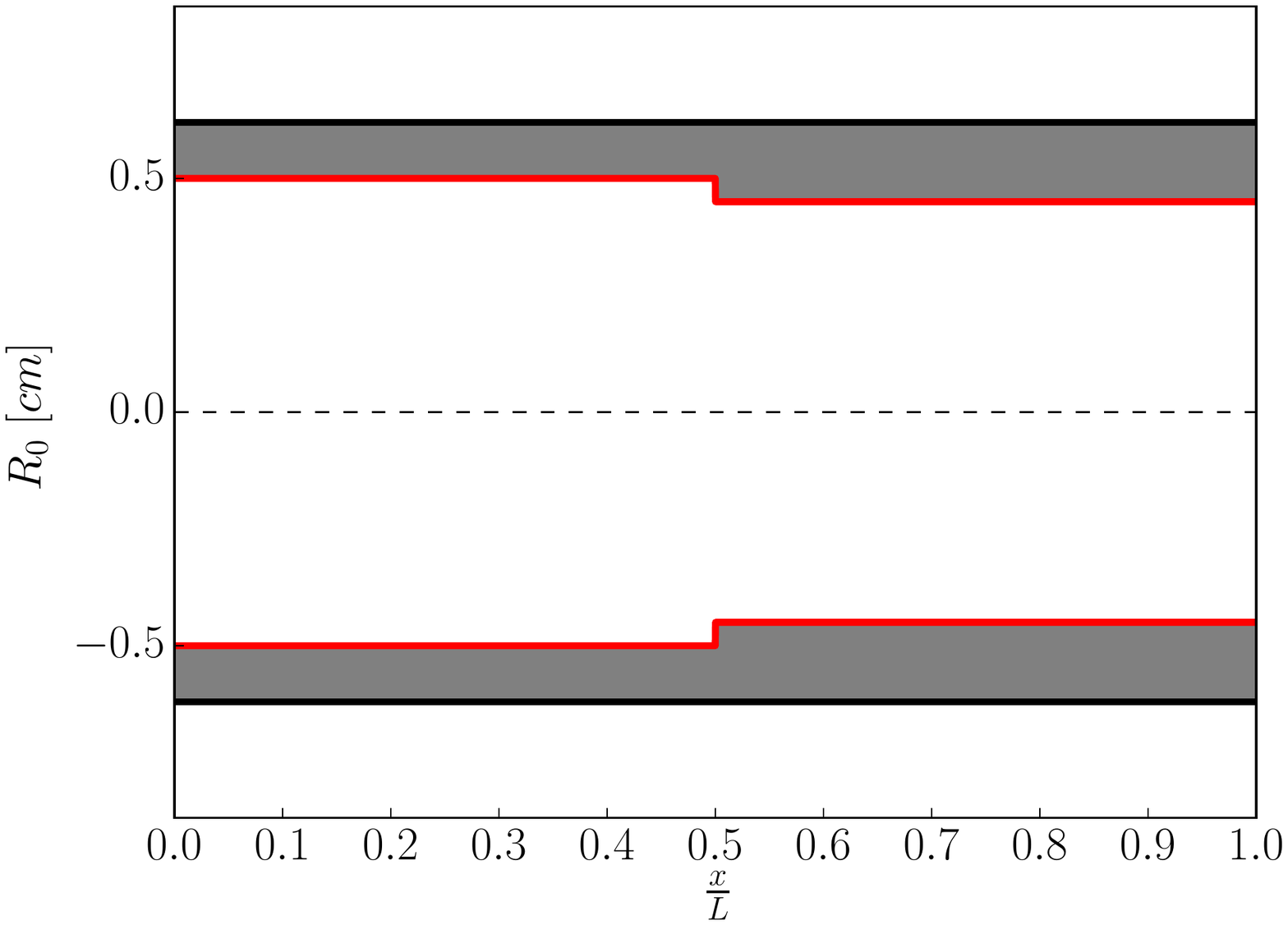}\\
\end{minipage}%
\begin{minipage}[t]{0.33\textwidth}
  \centering
  \includegraphics[scale=0.25,angle=0]{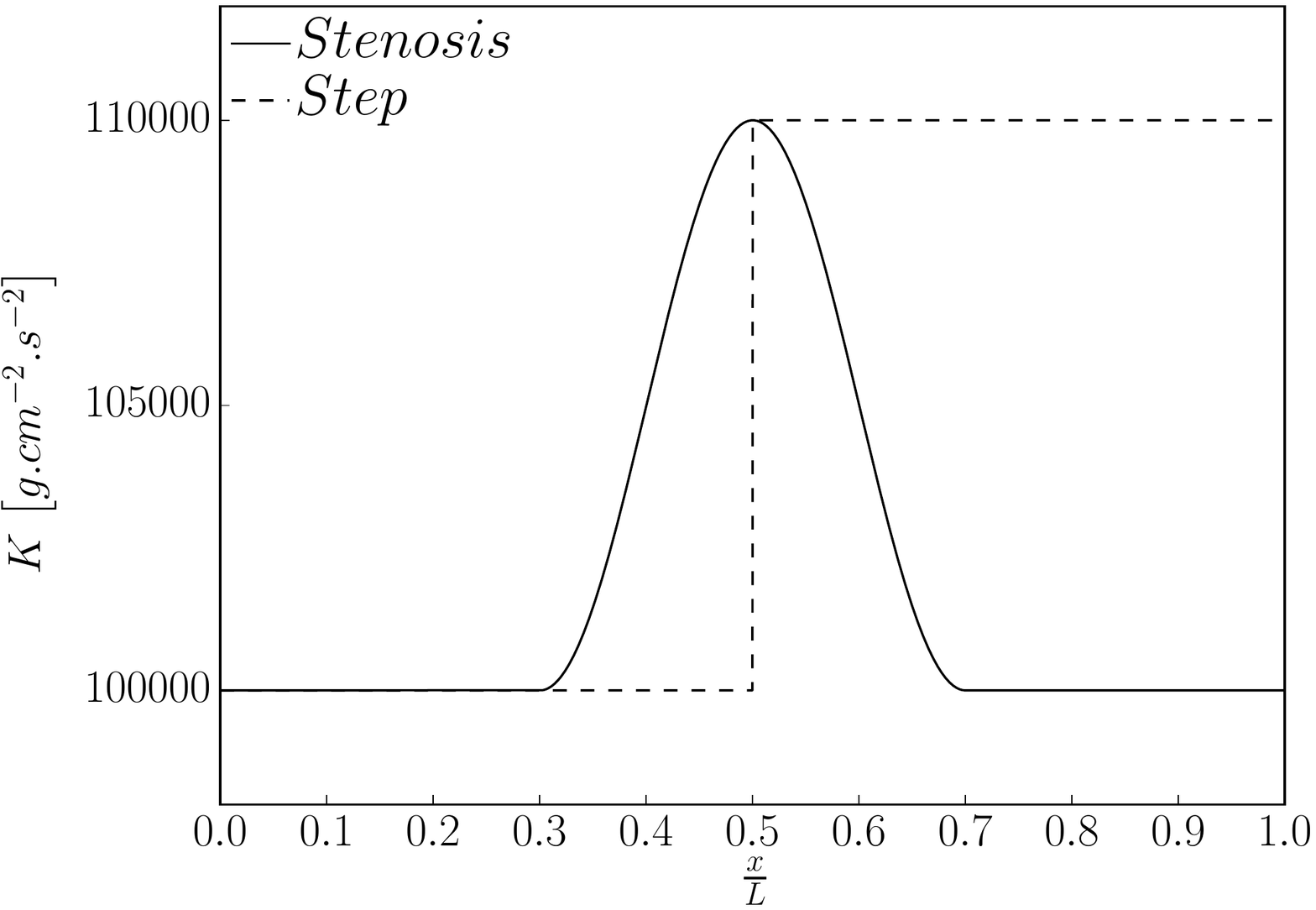}\\
\end{minipage}%
}
\caption{Representation of the radius at rest $R_0$ and the arterial wall rigidity $K$ for the smooth stenosis \eqref{eq:Ex-Stenosis-Geom} and the decreasing step \eqref{eq:Ex-Step-Geom} for $\Delta \mathcal{G}=10 \%$. \underline{\textit{Left}}: $R_0$ for the stenosis; \underline{\textit{Center}}: $R_0$ for the step; \underline{\textit{Right}}: $K$ for the stenosis (full line) and the step (dashed line).}
\label{fig:Stenosis-Step-R0-K}
\end{center}
\end{figure}

The following numerical simulations are performed in a single artery representative of a large artery such as the aorta. Table \ref{table:Artery-Properties} summarizes the values of the characteristic properties of blood and of the artery, namely the blood density $\rho$, the length $L$ of the artery and the inlet radius at rest and arterial wall rigidity $R_{in}$ and $K_{in}$, all written in "cgs".
\begin{table}[H]
\begin{center}
\def\arraystretch{1.2}
\begin{tabular}{c|c|c|c}
$\rho$ [$g.cm^{-3}$] & $L$ [$cm$] & $R_{in}$ [$cm$] & $K_{in}$ [$g.cm^{-2}.s^{-2}$] \\ 
\hline 
1 & 10 & 0.5 & $1 \times 10^5$ 
\end{tabular} 
\end{center}
\caption{Parameters describing the artery used in the different test-cases, given in "cgs": the density $\rho$, the length $L$, the inlet radius $R_{in}$ and the inlet rigidity $K_{in}$.}
\label{table:Artery-Properties}
\end{table}
We study two geometrical configurations in which both the cross-sectional area at rest $A_0$ and the arterial wall rigidity $K$ vary. Both are idealized representations of variations of arteries' geometrical and mechanical properties encountered in arterial networks. The first configuration is a smooth stenosis and corresponds to a local reduction of the cross-sectional area at rest $A_0$. It is a classical arterial pathology caused by the formation of plaque that deposits on the arterial wall and slowly obliterates the vessel. The stenosis is represented in figure \ref{fig:Stenosis-Step-R0-K} and is defined by the following radius at rest $R_0$ and arterial wall rigidity $K$
\begin{equation}
\left\{
\begin{split}
& R_0\left( x \right) = &	\left\{
	\begin{split}
	&R_{in} & \: \: \: \text{ if } & x < x_s \text{ or } x > x_f \\
	&R_{in} \left( 1 - \frac{\Delta \mathcal{G}}{2} \left[ 1 + \cos \left( \pi +2 \pi \frac{x-x_s}{x_f -x_s} \right) \right] \right) & \: \: \: \text{ if } &  x_s \leq x \geq x_f \\
	\end{split}
	\right.\\
& K \left( x \right) = &\left\{
	\begin{split}
	&K_{in} & \: \: \: \text{ if } & x < x_s \text{ or } x > x_f \\
	&K_{in} \left( 1 + \frac{\Delta \mathcal{G}}{2} \left[ 1 + \cos \left( \pi +2 \pi \frac{x-x_s}{x_f -x_s} \right) \right] \right) & \: \: \: \text{ if } &  x_s \leq x \geq x_f .\\
	\end{split}
	\right.\\
\end{split}
\right.
\label{eq:Ex-Stenosis-Geom}
\end{equation}
We choose $x_s = \frac{3L}{10}$ and $x_f = \frac{7 L}{10}$. The second configuration we investigate is a decreasing step, or decreasing discontinuity. It is an idealized representation of a pointwize transition between a parent artery and a smaller daughter artery and is useful to evaluate the reflecting behavior of a numerical method. The decreasing step is represented in figure \ref{fig:Stenosis-Step-R0-K} and is defined by the following radius at rest $R_0$ and arterial wall rigidity $K$
\begin{equation}
\left\{
\begin{split}
& R_0\left( x \right) = &	\left\{
	\begin{split}
	&R_{in} & \: \: \: \text{ if } & x < x_m \\
	&R_{in} \left( 1 - \Delta \mathcal{G} \right) & \: \: \: \text{ if } & x \geq x_m\\
	\end{split}
	\right.\\
& K \left( x \right) = &\left\{
	\begin{split}
	&K_{in} & \: \: \: \text{ if } & x < x_m \\
	&K_{in} \left( 1 + \Delta \mathcal{G} \right) & \: \: \: \text{ if } & x \geq x_m . \\
	\end{split}
	\right.\\
\end{split}
\right.
\label{eq:Ex-Step-Geom}
\end{equation}
We choose $x_m = \frac{L}{2}$. In both configuration, the amplitude of the geometrical and mechanical variations depends on the wall deformation parameter $\Delta \mathcal{G}$. The values of $\Delta \mathcal{G}$ used in the following simulations are taken from table \ref{table:Sh-DG-Steady} and are chosen to test the limits of the well-balanced methods while staying in the subcritical flow regime. From a well-balanced point of view, each of these two configurations has a different behavior with respect to the cell size $\Delta x$. Indeed, the step configuration is a discontinuity of the cross-sectional area at rest $A_0$ and of the arterial wall rigidity $K$, and therefore the amplitude of the variation of the geometrical and mechanical properties of the artery, proportional to $\Delta \mathcal{G}$, is independent of $\Delta x$. On the contrary, the stenosis configuration is a smooth variation of $A_0$ and $K$, and therefore the local variation of the artery's geometrical and mechanical properties at each cell interface will decrease with the cell size $\Delta x$.\\

We now provide the values of the conservative variables at the inlet and outlet of the computational domain, based on methods detailed in section \ref{sec:BC}. We impose the flow rate $Q_{in}$ at the inlet of the computational domain, in $x=0$. In reality, to control the flow regime, we give the value of the inlet Shapiro number $S_{h,in}$ and compute the inlet flow rate $Q_{in}$ as a function of $S_{h,in}$
\begin{equation}
Q_{in} = S_{h,in} A_{in} c_{in}.
\label{bc:Qin-Shin}
\end{equation}
$A_{in}$ and $c_{in}$ are respectively the inlet cross-sectional area and Moens-Korteweg wave speed \eqref{eq:Moens-Korteweg-c} and are unknown. However, a dimensional analysis of system \eqref{eq:BF-Sys-Conservative} allows us to show that the inlet Shapiro number $S_{h,in}$ scales as the ratio of the perturbation of the wall's radius $\Delta R=R-R_0$ over the radius at rest $R_0$. With this scaling law, we can estimate a value of the inlet cross-sectional area $A_{in}$ consistent with the inlet Shapiro number $S_{h,in}$ and we compute $A_{in}$ as
\begin{equation}
A_{in} = A_0\left( x=0 \right) \left[ 1 + S_{h,in} \right]^2.
\label{bc:Ain-Shin}
\end{equation}
At the outlet of the computational domain, in $x=L$, we either impose the reflection coefficient $R_t=0$ or the cross-sectional area $A_{out}$, depending on the test case. Similarly to the inlet cross-sectional area $A_{in}$, we compute the outlet cross-sectional area as a function of $S_{h,in}$
\begin{equation}
A_{out} = A_0\left( x=L \right) \left[ 1 + S_{h,in} \right]^2.
\label{bc:Aout-Shin}
\end{equation}

The values of the inlet Shapiro number $S_{h,in}$ and the wall deformation parameter $\Delta \mathcal{G}$ used in the following simulations are presented in table \ref{table:Sh-DG-Steady}. They cover a wide range of physiological configurations, allowing us to assess the behavior of the three numerical schemes under consideration in the limit of the low-Shapiro number flow regime. We recall that in arteries the average Shapiro number is of the order of $S_h = 1 \times 10^{-2}$.
\begin{table}[H]
\begin{center}
\def\arraystretch{1.2}
\begin{tabular}{|c||c|c|c|c|} 
\hline
 $S_{h,in}$ & $0$ & $ 1 \times 10^{-3}$ & $ 1 \times 10^{-2}$ & $ 1 \times 10^{-1}$ \\ \hline
\end{tabular}
\quad
\begin{tabular}{|c||c|c|c|}
\hline 
 $\Delta \mathcal{G}$ & $1 \%$ & $10 \%$ & $30 \%$ \\ \hline
\end{tabular} 
\end{center}
\caption{Values of the inlet Shapiro number $S_{h,in}$ and the wall deformation parameter $ \Delta \mathcal{G} $ used in the single artery test-cases. These values are chosen to test the well-balanced methods in the limits of the low-Shapiro number flow regime.}
\label{table:Sh-DG-Steady}
\end{table}

\subsection{Steady solutions}
\label{subsec:Steady}

We evaluate the well-balanced properties of HR, HR-LS and HR-S by computing steady solutions of system \eqref{eq:BF-Sys} in the smooth stenosis \eqref{eq:Ex-Stenosis-Geom} and the decreasing step \eqref{eq:Ex-Step-Geom}. Steady flow configurations in arterial geometries similar to the stenosis \eqref{eq:Ex-Stenosis-Geom} have been studied by M\"uller \cite{Muller2013}, where only variations of the wall rigidity $K$ are taken into account. In \cite{Murillo2015}, the authors computed steady solutions in tapered tubes. In the context of the shallow water equations, steady flow solutions over a bump (analogue of the stenosis) or a step have been studied by many authors \cite{Castro2007,Noelle2007,Castro2013,Delestre2013}.\\

The steady numerical solutions are obtained for $t=200$ $s$. The time step $\Delta t$ is constant and chosen such that the CFL condition \eqref{eq:CFL-kin} is always satisfied. We impose the flow rate $Q_{in}$ \eqref{bc:Qin-Shin} at the inlet and the cross-sectional area $A_{out}$ \eqref{bc:Aout-Shin} at the outlet. We therefore select a specific steady state characterized by its associated flow rate $Q_{st}$ and energy discharge $E_{st}$. These values can be computed analytically and provide exact solutions to compare with our numerical results
\begin{equation}
\left\{
\begin{split}
&Q_{st} = Q_{in} \\
&E_{st} = \frac{1}{2} \frac{Q_st^2}{A_{out}^2} + \frac{K\left( x=L \right)}{\rho}\left( \sqrt{A_{out}} - \sqrt{A_0\left( x=L \right)} \right). \\
\end{split}
\right.
\label{eq:Ex-Steady-Solution}
\end{equation}

In both configurations \eqref{eq:Ex-Stenosis-Geom} and \eqref{eq:Ex-Step-Geom}, we perform a series of 12 numerical computations for all combinations of the inlet Shapiro number $S_{h,in}$ and the wall deformation parameter $\Delta \mathcal{G}$ taken from table \ref{table:Sh-DG-Steady}. Table \ref{table:Steady-Err-Sh-DG} shows $L^1$ relative errors between the analytic solutions and the results obtained with HR, HR-LS and HR-S for a fixed number of cells $N=50$.
\begin{table}[!h]
\begin{footnotesize}
\begin{center}
\def\arraystretch{1.2}
{\setlength{\tabcolsep}{0.5em}
\begin{tabular}{c|c||c|c|c||c|c|c}

\multicolumn{2}{c||}{} & \multicolumn{3}{c||}{Stenosis} & \multicolumn{3}{c}{Step} \\  \hline \hline

\multicolumn{2}{c||}{$S_{h,in}$} & \multicolumn{6}{c}{0} \\  \cline{3-8} 
 
\multicolumn{2}{c||}{$\Delta	\mathcal{G}$} & $1\%$ & $10\%$ & $30\%$ & $1\%$ & $10\%$ & $30\%$  \\ 
 
\hline

 & HR &
 	$0$ & $0$ & $0$ & 
 	$0$& $0$& $0$
  \\
   
$L^1\left[Q\right]$ & HR-LS & 
	$0$ & $0$ & $0$ & 
	$0$ & $0$ & $0$ 
\\ 

 & HR-S &
 	$0$ & $0$ & $0$ & 
	$0$ & $0$ & $0$ 
  \\ 

\hline

 & HR &
	$0$ & $0$ & $0$ & 
	$0$ & $0$ & $0$ 
 \\
 
$L^1\left[E\right]$ & HR-LS &
	$0$ & $0$ & $0$ & 
	$0$ & $0$ & $0$ 
 \\

 & HR-S &
 	$0$ & $0$ & $0$ & 
	$0$ & $0$ & $0$ 
 \\  \hline \hline
  
\multicolumn{2}{c||}{$S_{h,in}$} & \multicolumn{6}{c}{$1 \times 10^{-3}$} \\  \cline{3-8} 
 
\multicolumn{2}{c||}{$\Delta	\mathcal{G}$} & $1\%$ & $10\%$ & $30\%$ & $1\%$ & $10\%$ & $30\%$  \\ 
 
\hline

 & HR & 
 $4.0 \times 10^{-4}$& $4.2 \times 10^{-3}$& $1.4 \times 10^{-2}$& 
 $2.2 \times 10^{-4}$& $2.3 \times 10^{-3}$& $7.4 \times 10^{-3}$ 
  \\ 
  
$L^1\left[Q\right]$ & HR-LS & 
$3.6 \times 10^{-7}$& $4.1 \times 10^{-6}$& $1.9 \times 10^{-5}$ &
$1.8 \times 10^{-7}$& $2.1 \times 10^{-6}$& $9.4 \times 10^{-6}$
 \\ 
 
 & HR-S &  
 $5.4 \times 10^{-13}$& $5.3 \times 10^{-13}$& $4.2 \times 10^{-14}$&
 $2.9 \times 10^{-13}$& $3.1 \times 10^{-13}$& $5.8 \times 10^{-13}$
 \\ 
 
\hline

 & HR & 
 $3.0 \times 10^{-4}$&  $5.1 \times 10^{-3}$&  $4.2 \times 10^{-2}$& 
 $2.1 \times 10^{-4}$&  $9.4 \times 10^{-3}$&  $1.3 \times 10^{-1}$
 \\
 
$L^1\left[E\right]$ & HR-LS & 
$2.1 \times 10^{-7}$& 	$2.6 \times 10^{-6}$& $1.5 \times 10^{-5}$&
$1.1 \times 10^{-7}$& 	$1.4 \times 10^{-6}$& $1.0 \times 10^{-5}$ 
 \\
 
 & HR-S &  
 $4.6 \times 10^{-13}$& $4.9 \times 10^{-13}$& $6.1 \times 10^{-13}$&
 $6.7 \times 10^{-13}$& $6.5 \times 10^{-13}$& $1.4 \times 10^{-12}$ \\ \hline \hline
  
\multicolumn{2}{c||}{$S_{h,in}$} & \multicolumn{6}{c}{$1 \times 10^{-2}$} \\  \cline{3-8} 
 
\multicolumn{2}{c||}{$\Delta	\mathcal{G}$} & $1\%$ & $10\%$ & $30\%$ & $1\%$ & $10\%$ & $30\%$  \\ 
 
\hline

 & HR & 
 $4.0 \times 10^{-4}$& $4.2 \times 10^{-3}$& $1.4 \times 10^{-2}$& 
 $2.3 \times 10^{-4}$& $2.3 \times 10^{-3}$& $7.4 \times 10^{-3}$ 
  \\ 
  
$L^1\left[Q\right]$ & HR-LS & 
$3.6 \times 10^{-6}$& $4.1 \times 10^{-5}$& $1.9 \times 10^{-4}$ &
$1.8 \times 10^{-6}$& $2.1 \times 10^{-5}$& $9.4 \times 10^{-5}$
 \\ 
 
 & HR-S &  
 $2.6 \times 10^{-13}$& $2.7 \times 10^{-13}$& $9.6 \times 10^{-14}$&
 $2.4 \times 10^{-13}$& $9.5 \times 10^{-14}$& $1.8 \times 10^{-13}$
 \\ 
 
\hline

 & HR & 
 $3.0 \times 10^{-4}$&  $5.1 \times 10^{-3}$&  $4.2 \times 10^{-2}$& 
 $2.1 \times 10^{-4}$&  $9.4 \times 10^{-3}$&  $1.2 \times 10^{-1}$
 \\
 
$L^1\left[E\right]$ & HR-LS & 
$2.1 \times 10^{-6}$& 	$2.6 \times 10^{-5}$& $1.5 \times 10^{-4}$&
$1.1 \times 10^{-6}$& 	$1.4 \times 10^{-5}$& $8.1 \times 10^{-5}$ 
 \\
 
 & HR-S &  
 $2.7 \times 10^{-13}$& $2.7 \times 10^{-13}$& $3.3 \times 10^{-13}$&
 $2.7 \times 10^{-13}$& $3.4 \times 10^{-13}$& $5.9 \times 10^{-13}$ \\ \hline \hline
   
\multicolumn{2}{c||}{$S_{h,in}$} & \multicolumn{6}{c}{$1 \times 10^{-1}$} \\  \cline{3-8} 
 
\multicolumn{2}{c||}{$\Delta	\mathcal{G}$} & $1\%$ & $10\%$ & $30\%$ & $1\%$ & $10\%$ & $30\%$  \\ 
 
\hline

 & HR & 
 $4.0 \times 10^{-4}$& $4.2 \times 10^{-3}$& $1.4 \times 10^{-2}$& 
 $2.3 \times 10^{-4}$& $2.3 \times 10^{-3}$& $7.5 \times 10^{-3}$ 
  \\ 
  
$L^1\left[Q\right]$ & HR-LS & 
$3.6 \times 10^{-5}$& $4.1 \times 10^{-4}$& $1.8 \times 10^{-3}$ &
$1.8 \times 10^{-5}$& $2.1 \times 10^{-4}$& $9.0 \times 10^{-4}$
 \\ 
 
 & HR-S &  
 $2.6 \times 10^{-13}$& $3.4 \times 10^{-13}$& $2.0 \times 10^{-13}$&
 $2.8 \times 10^{-13}$& $2.4 \times 10^{-13}$& $1.4 \times 10^{-13}$
 \\ 
 
\hline

 & HR & 
 $3.2 \times 10^{-4}$&  $5.4 \times 10^{-3}$&  $4.4 \times 10^{-2}$& 
 $2.2 \times 10^{-4}$&  $9.9 \times 10^{-3}$&  $1.2 \times 10^{-1}$
 \\
 
$L^1\left[E\right]$ & HR-LS & 
$2.2 \times 10^{-5}$& 	$2.8 \times 10^{-4}$& $1.8 \times 10^{-3}$&
$1.2 \times 10^{-5}$& 	$2.0 \times 10^{-4}$& $2.2 \times 10^{-3}$ 
 \\
 
 & HR-S &  
 $2.3 \times 10^{-13}$& $2.4 \times 10^{-13}$& $2.9 \times 10^{-13}$&
 $2.3 \times 10^{-13}$& $2.9 \times 10^{-13}$& $3.4 \times 10^{-13}$
\end{tabular} 
}
\end{center}
\end{footnotesize}
\caption{Steady solutions: Relative errors $L^1\left[Q\right]$ and $L^1\left[E\right]$ computed in the stenosis \eqref{eq:Ex-Stenosis-Geom} and the step \eqref{eq:Ex-Step-Geom} for $N=50$ cells for all combinations of values of the inlet Shapiro number $S_{h,in}$ and the wall deformation parameter $\Delta \mathcal{G}$ taken for table \ref{table:Sh-DG-Steady}. Only HR-S is exactly well-balanced, but HR-LS is more accurate than HR.}
\label{table:Steady-Err-Sh-DG}
\end{table}
In both the stenosis \eqref{eq:Ex-Stenosis-Geom} and the step \eqref{eq:Ex-Step-Geom} configurations, the results are similar and indicate that, as expected, each numerical method is exactly well-balanced for the steady states at rest ($S_{h,in}=0$). Only HR-S is exactly well-balanced for all considered subcritical steady states. For the low-Shapiro number steady states ($S_{h,in}=10^{-3},10^{-2},10^{-1}$), HR-LS is more accurate than HR. However, the accuracy of HR-LS diminishes when the values of $S_{h,in}$ and $\Delta \mathcal{G}$ increase, and for $S_{h,in}=1 \times 10^{-1}$ and $\Delta \mathcal{G} = 30 \%$, in the limit of the low-Shapiro number flow regime, HR-LS is only one order of magnitude more accurate than HR. Interestingly, the errors obtained with HR are independent of the inlet Shapiro number $S_{h,in}$, but increase significantly with the wall deformation parameter $\Delta \mathcal{G}$. \\
 
To test the consistency and the order of convergence of the different methods, we perform a convergence study for the average low-Shapiro steady configuration $S_{h,in} = 1 \times 10^{-2}$ and $\Delta \mathcal{G}=10 \%$ in both the stenosis and the step configurations. $L^1$ relative errors with analytic solutions are presented in table \ref{table:Steady-CV-Sh-DG} for the following number of cells $N \in \left\{ 50,100,200,400\right\}$. 

\begin{table}[!h]
\begin{footnotesize}
\begin{center}
\def\arraystretch{1.2}
{\setlength{\tabcolsep}{0.5em}
\begin{tabular}{c || c|c|c|c|| c|c|c|c }

	& \multicolumn{4}{c||}{Stenosis} & \multicolumn{4}{c}{Step} \\  \hline \hline

	& \multicolumn{8}{c}{HR}  \\  \cline{2-9} 
 $N$ & $L^1\left[Q\right]$ & Order & $L^1\left[E\right]$ & Order & $L^1\left[Q\right]$ & Order & $L^1\left[E\right]$ & Order   \\
\hline
$50$& 	$4.22 \times 10^{-3}$& 		$-$ 		& $5.09 \times 10^{-3}$	& 		$-$ &
		$2.34 \times 10^{-3}$& 		$-$ 		& $9.41 \times 10^{-3}$	& 		$-$
	\\ 
$100$& 	$2.11 \times 10^{-3}$& 		$-1.01$ & $2.56 \times 10^{-3}$	& 		$-1.01$ &
		$1.17 \times 10^{-3}$& 		$-1.01$ & $8.64 \times 10^{-3}$	& 		$-0.12$
	 \\
$200$& 	$1.05 \times 10^{-3}$& 		$-1.01$ & $1.28 \times 10^{-3}$	& 		$-1.01$ &
		$5.86 \times 10^{-4}$& 		$-1.01$ & $8.26 \times 10^{-3}$	& 		$-0.07$
	 \\
$400$& 	$5.26 \times 10^{-4}$& 		$-1.00$ & $6.38 \times 10^{-4}$	& 		$-1.00$ &
		$2.93 \times 10^{-4}$& 		$-1.00$ & $8.07 \times 10^{-3}$	& 		$-0.03$
	 \\
\hline \hline
 & \multicolumn{8}{c}{HR-LS}  \\  \cline{2-9} 
 $N$ & $L^1\left[Q\right]$ & Order & $L^1\left[E\right]$ & Order & $L^1\left[Q\right]$ & Order & $L^1\left[E\right]$ & Order   \\
\hline
$50$& 	$4.14 \times 10^{-5}$& 		$-$ 		& $2.61 \times 10^{-5}$	& 		$-$ &
		$2.08 \times 10^{-5}$& 		$-$ 		& $1.39 \times 10^{-5}$	& 		$-$
	\\ 
$100$& 	$2.07 \times 10^{-5}$& 		$-1.01$ & $1.31 \times 10^{-5}$	& 		$-1.01$ &
		$1.04 \times 10^{-5}$& 		$-1.01$ & $7.24 \times 10^{-6}$	& 		$-0.96$
	 \\
$200$& 	$1.04 \times 10^{-5}$& 		$-1.01$ & $6.58 \times 10^{-6}$	& 		$-1.00$ &
		$5.19 \times 10^{-6}$& 		$-1.01$ & $3.91 \times 10^{-6}$	& 		$-0.90$
	 \\
$400$& 	$5.19 \times 10^{-6}$& 		$-1.00$ & $3.30 \times 10^{-6}$	& 		$-1.00$ &
		$2.59 \times 10^{-6}$& 		$-1.00$ & $2.24 \times 10^{-6}$	& 		$-0.80$
	 \\
\hline \hline
& \multicolumn{8}{c}{HR-S}  \\  \cline{2-9} 
 $N$ & $L^1\left[Q\right]$ & Order & $L^1\left[E\right]$ & Order & $L^1\left[Q\right]$ & Order & $L^1\left[E\right]$ & Order   \\
\hline
$50$& 	$2.68 \times 10^{-13}$& 		$-$ 		& $2.73 \times 10^{-13}$	& 		$-$ &
		$9.53 \times 10^{-14}$& 		$-$ 		& $3.43 \times 10^{-13}$	& 		$-$
	\\ 
$100$& 	$1.40 \times 10^{-15}$& 		$-$ 		& $3.39	 \times 10^{-13}$& 		$-$ &
		$9.20 \times 10^{-14}$& 		$-$ 		& $3.97 \times 10^{-13}$	& 		$-$
	 \\
$200$& 	$1.94 \times 10^{-12}$& 		$-$ 		& $7.30 \times 10^{-13}$	& 		$-$ &
		$2.26 \times 10^{-12}$& 		$-$ 		& $8.44 \times 10^{-13}$	& 		$-$
	 \\
$400$& 	$8.83 \times 10^{-12}$& 		$-$ 		& $1.45 \times 10^{-12}$	& 		$-$ &
		$1.01 \times 10^{-11}$& 		$-$ 		& $1.63 \times 10^{-12}$	& 		$-$
\end{tabular} 
}
\end{center}
\end{footnotesize}
\caption{Steady solutions: Relative errors $L^1\left[Q\right]$ and $L^1\left[E\right]$ computed in the stenosis \eqref{eq:Ex-Stenosis-Geom} and the step \eqref{eq:Ex-Step-Geom} for $S_{h,in}=1 \times 10^{-2}$ and $\Delta \mathcal{G} = 10 \%$ obtained for $N \in \left\{ 50,100,200,400\right\}$. HR and HR-LS converge with order 1 whereas HR-S is exactly well-balanced up to machine precision.}
\label{table:Steady-CV-Sh-DG}
\end{table}

\begin{figure}[!h]
\makebox[1.\textwidth][c]{
\begin{minipage}[t]{0.5\textwidth}
  \centering
  \includegraphics[scale=0.40,angle=0]{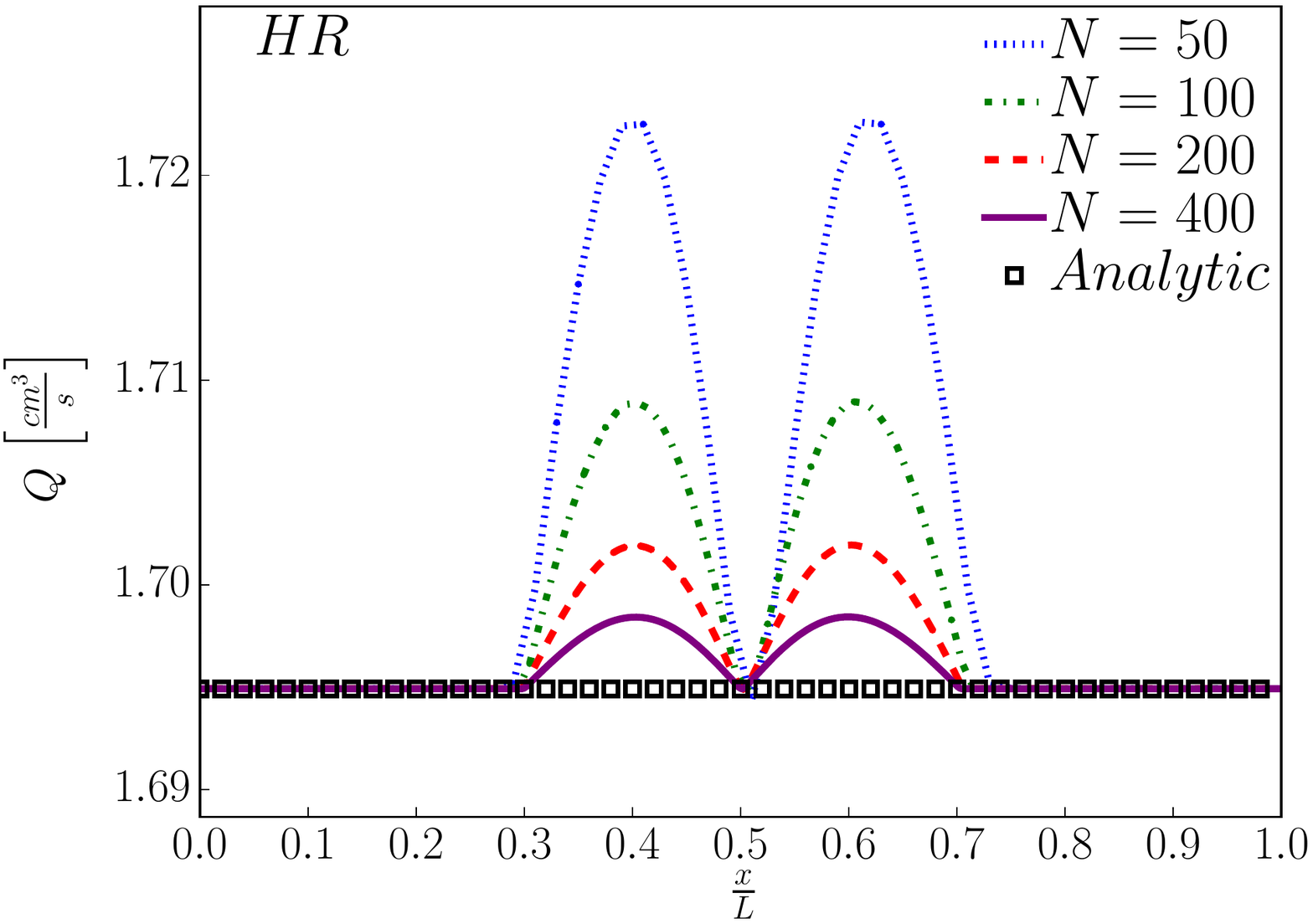}\\
\end{minipage} 
\begin{minipage}[t]{0.5\textwidth}
  \centering
  \includegraphics[scale=0.40,angle=0]{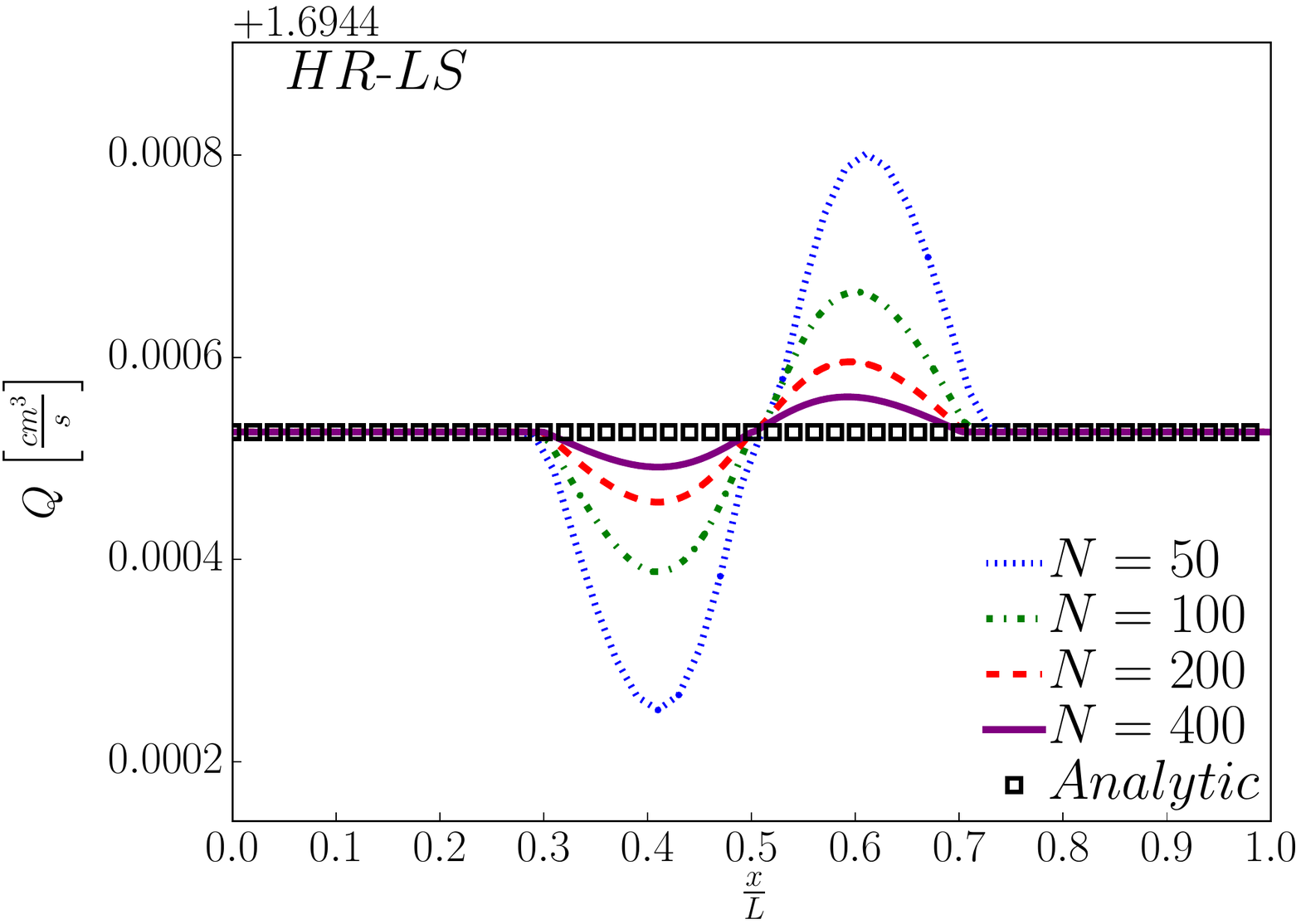}\\
\end{minipage} 
}
\makebox[1.\textwidth][c]{
\begin{minipage}[t]{0.5\textwidth}
  \centering
  \includegraphics[scale=0.40,angle=0]{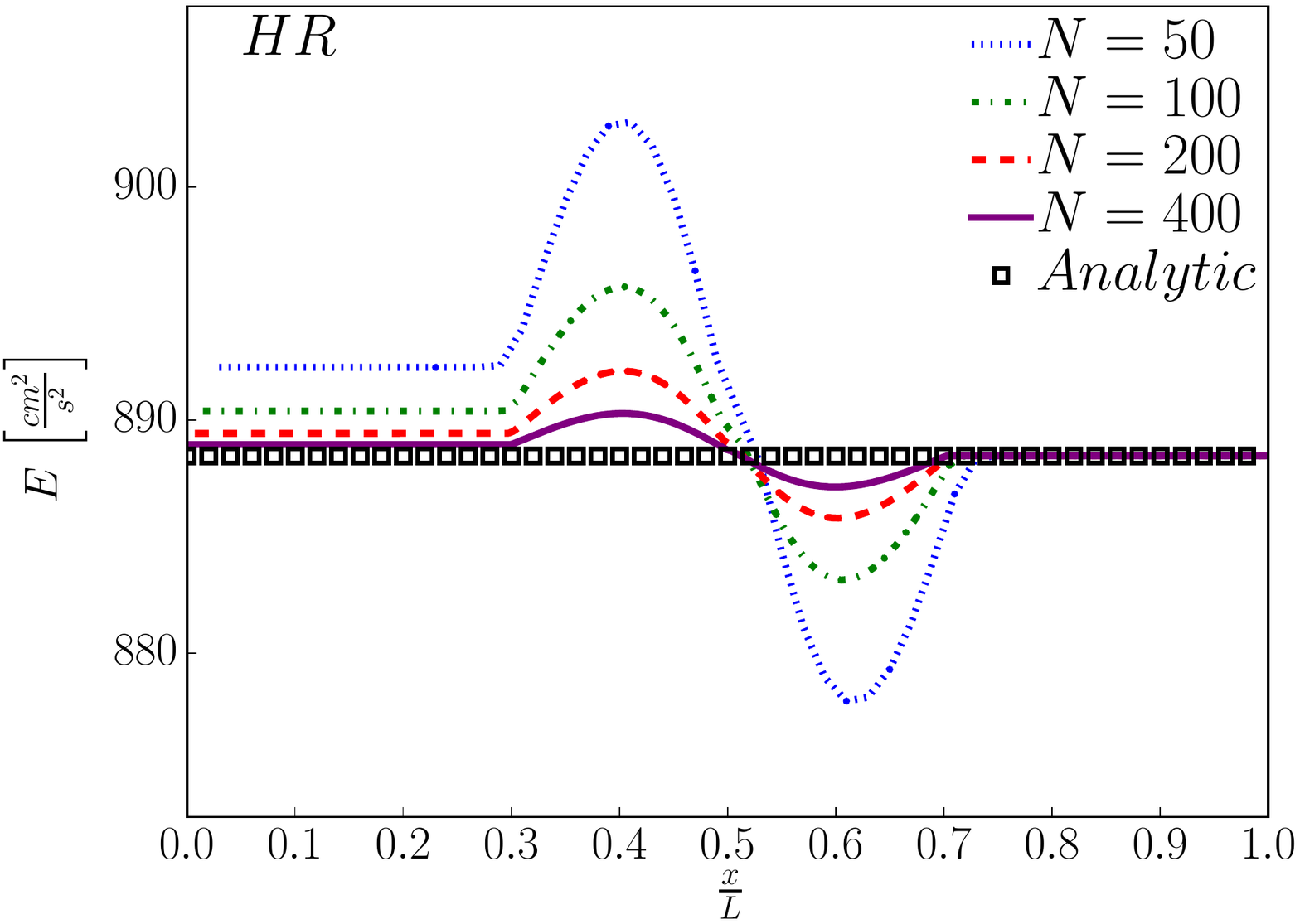}\\
\end{minipage} 
\begin{minipage}[t]{0.5\textwidth}
  \centering
  \includegraphics[scale=0.40,angle=0]{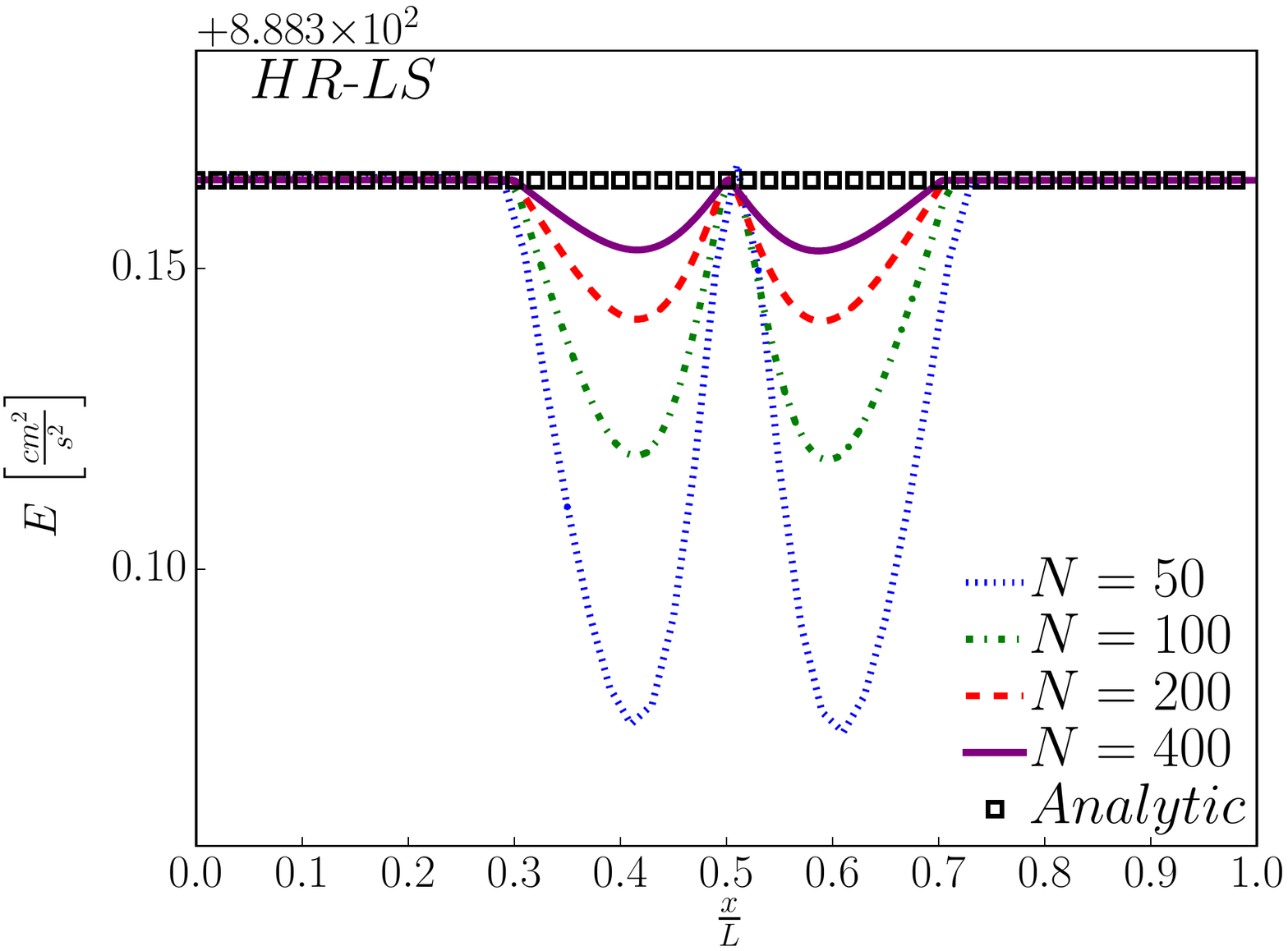}\\
\end{minipage} 
}
\caption{Steady solutions: Spatial evolution of the flow rate $Q$ (top) and the energy discharge $E$ (bottom) in the stenosis configuration \eqref{eq:Ex-Stenosis-Geom}, at $t=200$ s for $S_{h,in}=1 \times 10^{-2}$ and $\Delta \mathcal{G} = 10 \%$ obtained with different numbers of cells $N = \left\{ 50 \mathrm{ (blue)},100 \mathrm{ (green)},200 \mathrm{ (red)},400 \mathrm{ (purple)} \right\}$ and compared to the analytic solution \eqref{eq:Ex-Steady-Solution} (black). \underline{\textit{Left}}: HR; \underline{\textit{Right}}: HR-LS. We observe that for both HR and HR-LS, the errors with the analytic solution decrease when the number of cells $N$ increases, indicating the convergence of the method.  }
\label{fig:Steady-Stenosis-Fr-10m2-dR-10}
\end{figure}

\begin{figure}[!h]
\makebox[1.\textwidth][c]{
\begin{minipage}[t]{0.5\textwidth}
  \centering
  \includegraphics[scale=0.40,angle=0]{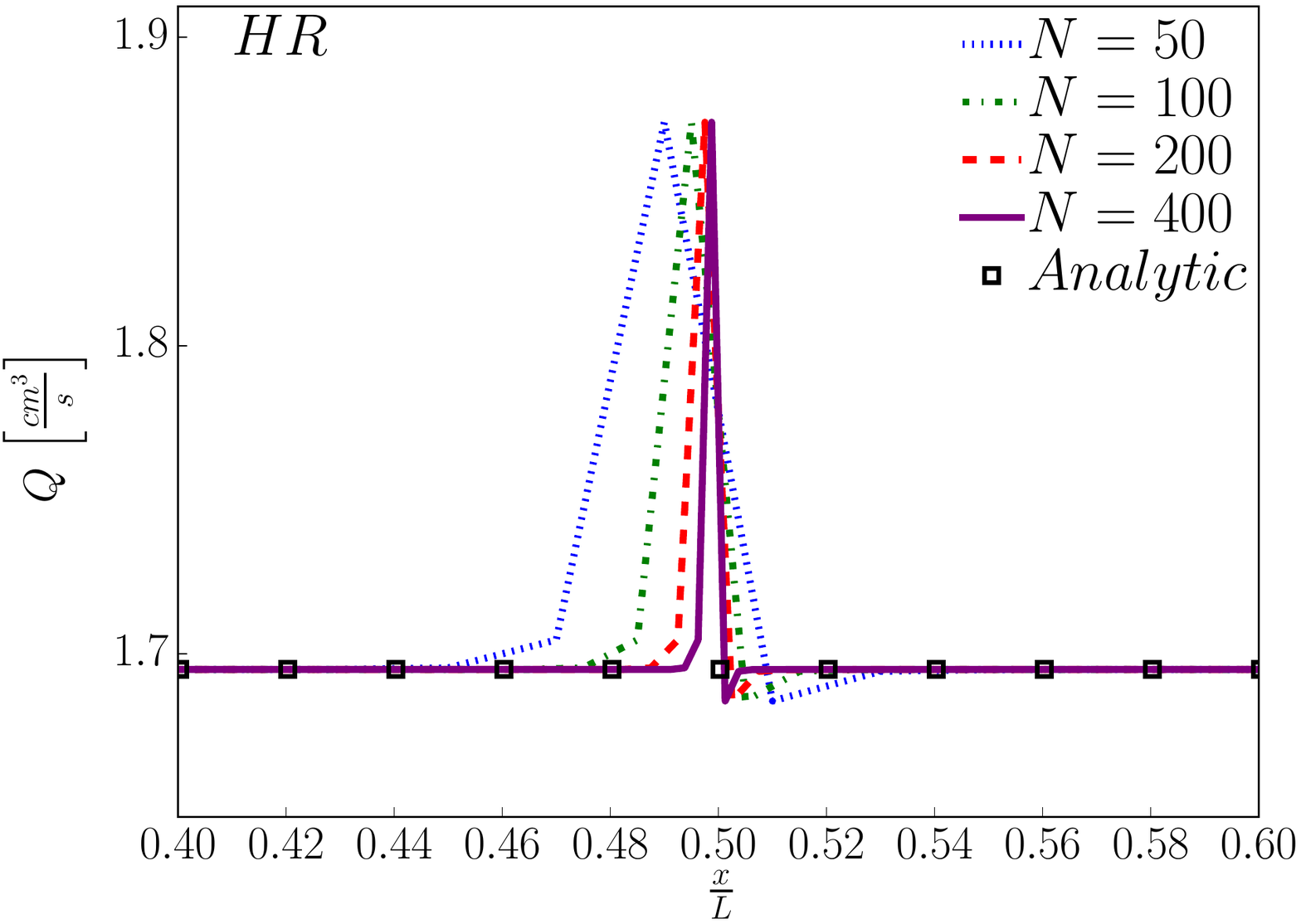}\\
\end{minipage} 
\begin{minipage}[t]{0.5\textwidth}
  \centering
  \includegraphics[scale=0.40,angle=0]{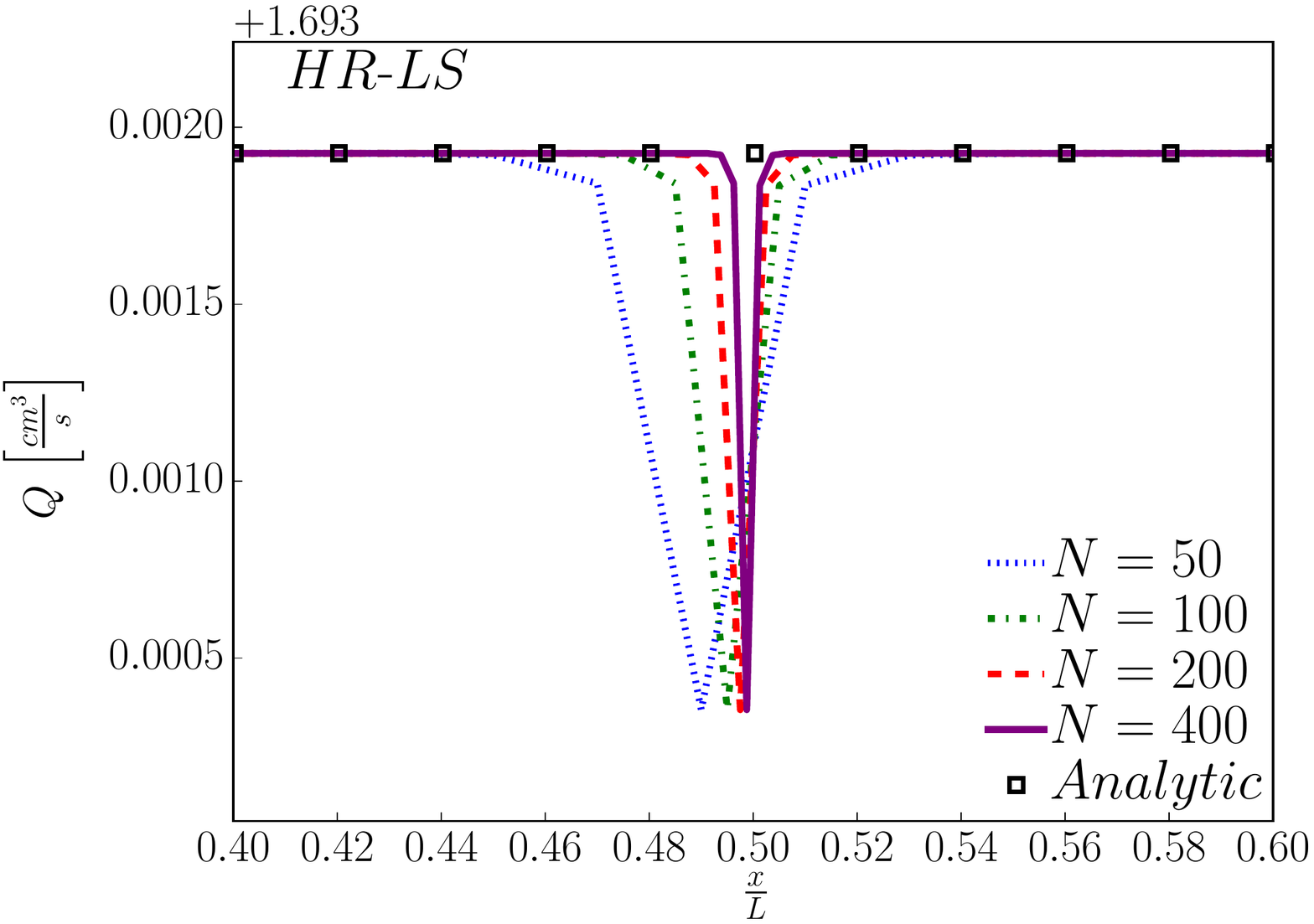}\\
\end{minipage} 
}
\makebox[1.\textwidth][c]{
\begin{minipage}[t]{0.5\textwidth}
  \centering
  \includegraphics[scale=0.40,angle=0]{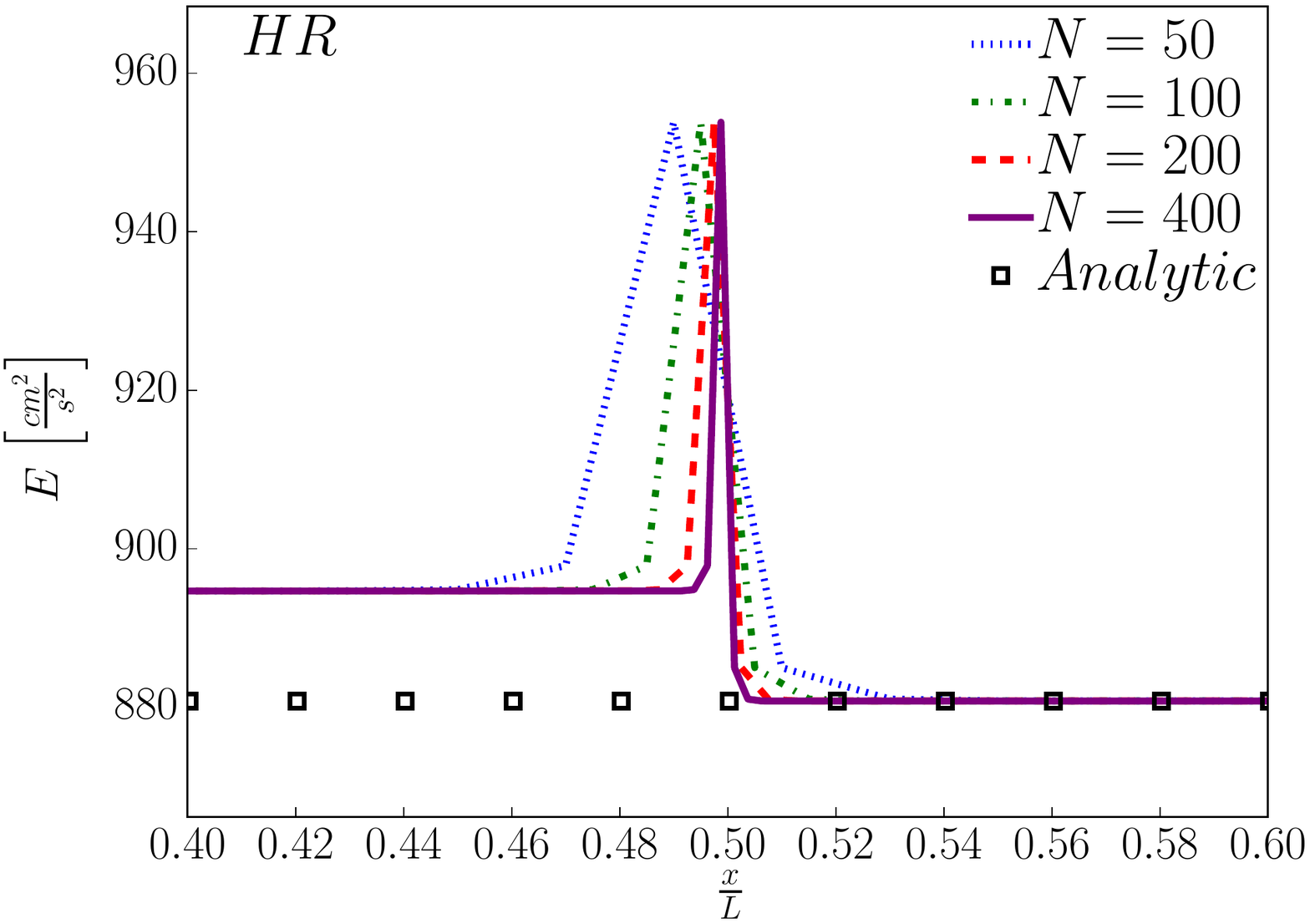}\\
\end{minipage} 
\begin{minipage}[t]{0.5\textwidth}
  \centering
  \includegraphics[scale=0.40,angle=0]{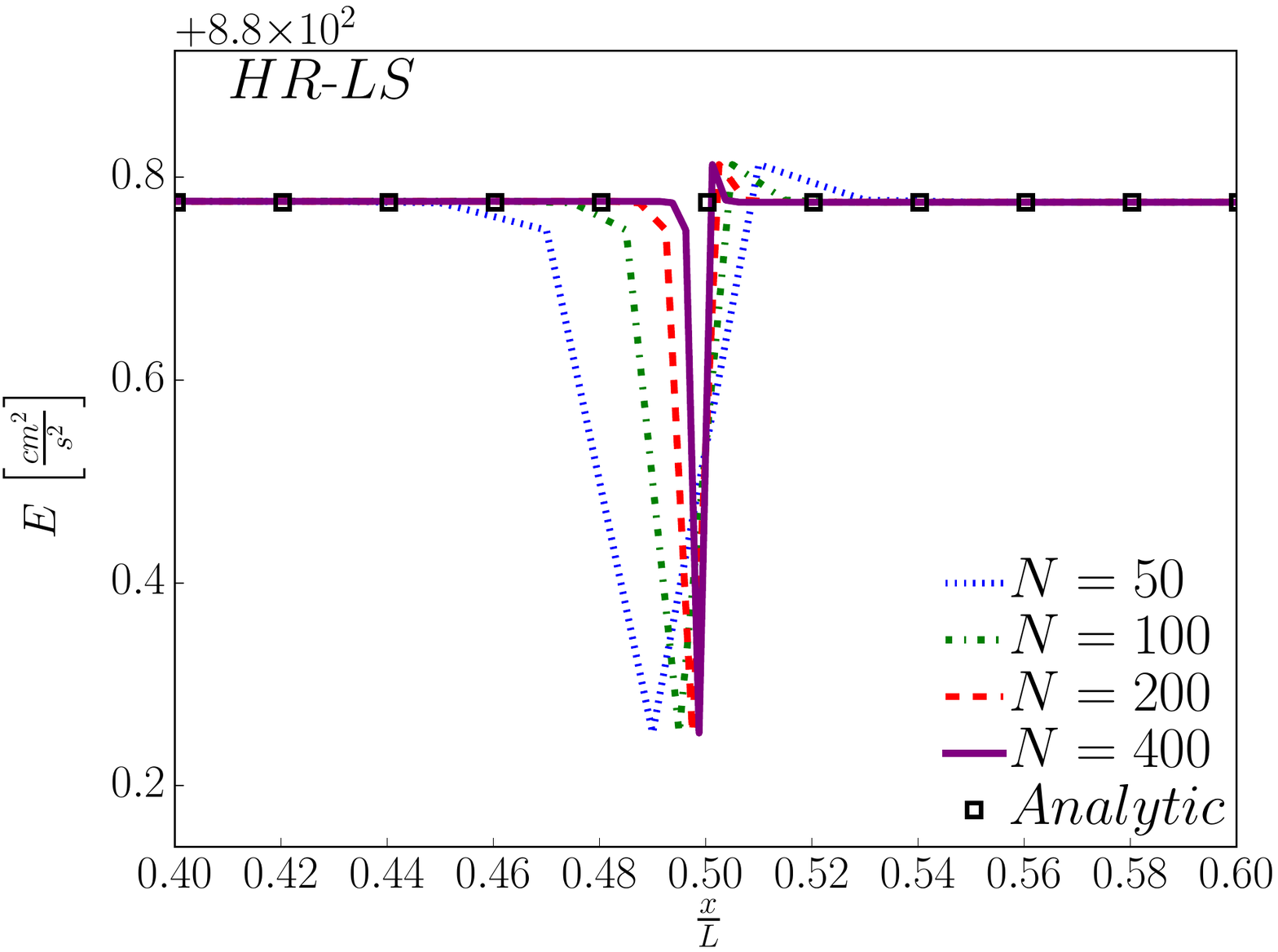}\\
\end{minipage} 
}
\caption{Steady solutions: Spatial evolution (zoom for $0.4 \leq \frac{x}{L} \leq 0.6$) of the flow rate $Q$ (top) and the energy discharge $E$ (bottom) in the step configuration \eqref{eq:Ex-Step-Geom}, at $t=200$ s for $S_{h,in}=1 \times 10^{-2}$ and $\Delta \mathcal{G} = 10 \%$ obtained with different numbers of cells $N = \left\{ 50 \mathrm{ (blue)},100 \mathrm{ (green)},200 \mathrm{ (red)},400 \mathrm{ (purple)} \right\}$ and compared to the analytic solution \eqref{eq:Ex-Steady-Solution} (black). \underline{\textit{Left}}: HR; \underline{\textit{Right}}: HR-LS. We observe that for both HR and HR-LS, the maximal amplitude of the errors with the analytic solution remains unchanged when the number of cells $N$ increases. However, the region of error is more localized when the number of cells increases, explaining why the error decreases.  }
\label{fig:Steady-Step-Fr-10m2-dR-10}
\end{figure}

In the stenosis configuration \eqref{eq:Ex-Stenosis-Geom}, both HR and HR-LS converge with order 1, whereas in the step configuration \eqref{eq:Ex-Step-Geom}, they do not achieve order 1 convergence. Indeed, in the stenosis configuration, the variations of the artery's geometrical and mechanical properties at each cell interface decrease when the number of cells $N$ increases, enabling the convergence of both methods. On the contrary, the geometrical and mechanical variations remain unchanged in the step configuration when the number of cells $N$ increases. These observations are illustrated by figures \ref{fig:Steady-Stenosis-Fr-10m2-dR-10} and \ref{fig:Steady-Step-Fr-10m2-dR-10}, where we respectively plot the spatial evolution of the flow rate $Q$ and the energy discharge $E$ with the number of cells in the stenosis and step configurations.

In both configurations, the values of the errors obtained in table \ref{table:Steady-CV-Sh-DG} with HR-S are of the order of machine precision, indicating that HR-S is exactly well-balanced for the considered low-Shapiro steady state. However, the errors increase slightly with the number of cells.  Similar behaviors are observed in convergence studies presented in \cite{Castro2013} for an exactly well-balanced method. In our case, this phenomenon is due to a small error between the computed boundary conditions and those required to obtain the desired steady state, and is not caused by HR-S.\\

The results indicate that among the three well-balanced methods considered, HR is the least accurate when computing low-Shapiro number steady solutions in an artery presenting smooth and discontinuous variations of its cross-sectional area at rest $A_0$ and of its arterial wall rigidity $K$. On the contrary, HR-S is the only exactly well-balanced method for the considered low-Shapiro number steady states. Finally, even though HR-LS is not exactly well-balanced for the considered low-Shapiro number steady states, it allows to compute with satisfying accuracy steady solutions for smooth and discontinuous variations of the artery's geometrical and mechanical properties. These results show that the system \eqref{eq:WB-HRQ-sys-SW} (used by HR-LS) is a better approximation than system \eqref{eq:WB-HRU-sys-SW} (used by HR) of the steady state system \eqref{eq:BF-Steady-State-Equation-SW} (used by HR-S) in low-Shapiro flow configurations.

\subsection{Single wave propagation}

The wave-capturing properties of HR, HR-LS and HR-S are now evaluated. We simulate the propagation of a single wave in the smooth stenosis \eqref{eq:Ex-Stenosis-Geom} and the decreasing step \eqref{eq:Ex-Step-Geom}. The step configuration was studied in \cite{Delestre2012,Delestre2016,Wang2016} for an artery with only variations of its cross-sectional area at rest $A_0$. \\

\begin{figure}[!h]
\makebox[1.\textwidth][c]{
\begin{minipage}[t]{0.5\textwidth}
  \centering
  \includegraphics[scale=0.40,angle=0]{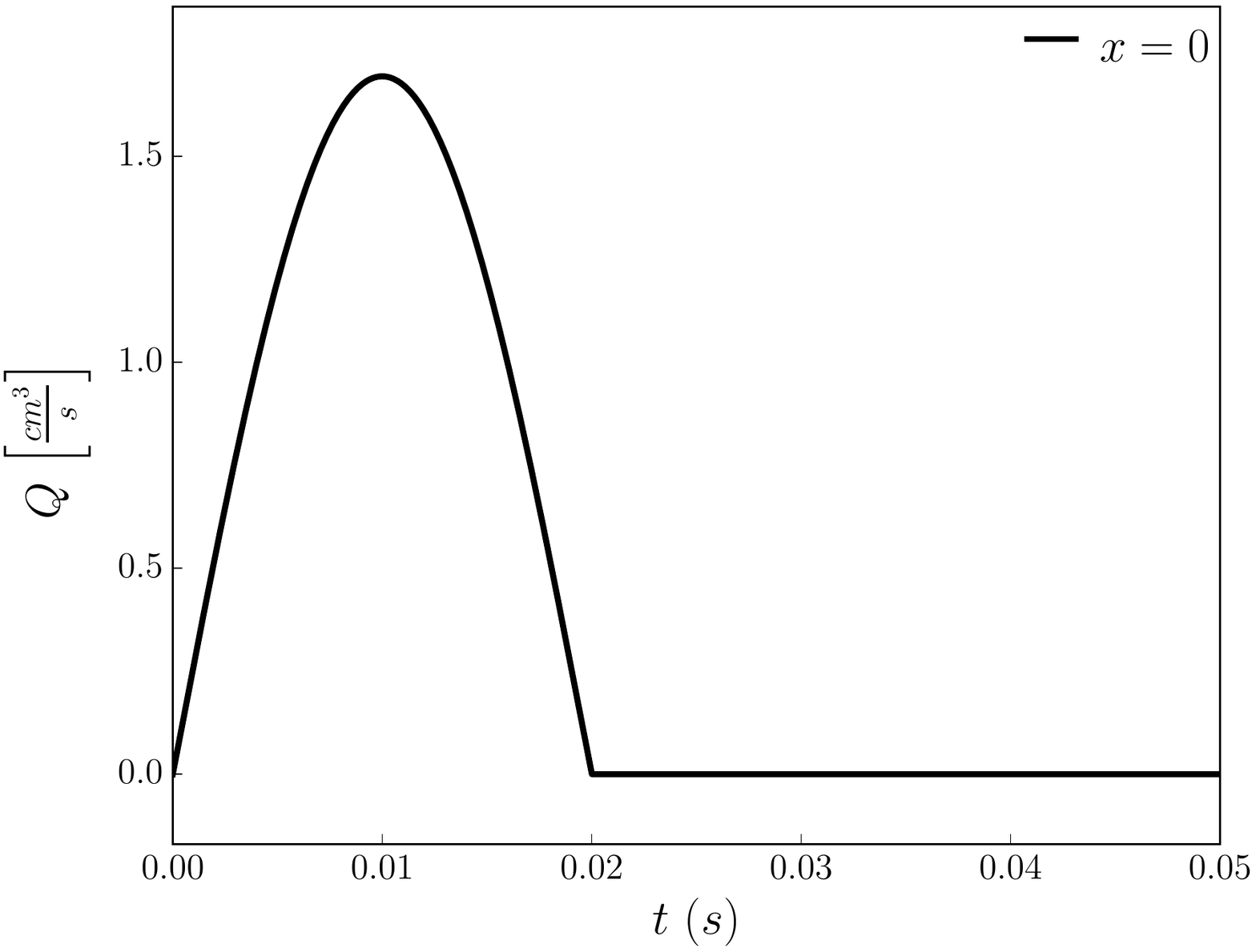}\\
\end{minipage} 
}
\caption{Wave propagation: Time evolution of the inlet flow rate $Q_{in}$ for $S_{h,in} = 1 \times 10^{-2}$ and $T_{pulse} = 0.04 $.}
\label{fig:Inlet-Flow-Rate-Wave}
\end{figure}

The results are obtained for $t=0.045$ s. The time step $\Delta t$ is constant and chosen such that the CFL condition \eqref{eq:CFL-kin} is always satisfied. We impose a single pulse of flow at the inlet of the computational domain and the unsteady inlet flow rate $Q_{in}\left( t \right)$ is defined as
\begin{equation}
\left.
\begin{split}
&Q_{in}\left( t \right) = 
\left\{
\begin{split}
& Q_{pulse} \sin \left(  2 \pi \frac{t}{T_{pulse}} \right) & \: \: \: &\text{ if } t \leq \frac{T_{pulse}}{2}\\
& 0 & \: \: \: &\text{ else } & .\\
\end{split}
\right.
\end{split}
\right.
\label{bc:Inlet-Flow-Pulse}
\end{equation}
We choose $T_{pulse} = 0.04 $ s to artificially reduce the wave length of the pulse for visualization purposes and the value of $Q_{pulse}$ is a function of the inlet Shapiro number $S_{h,in}$ and is defined as in equation \eqref{bc:Qin-Shin}. Figure \ref{fig:Inlet-Flow-Rate-Wave} represents the function $Q_{in}$ for $S_{h,in} = 1 \times 10^{-2}$. At the outlet of the computational domain, we set the reflection coefficient $R_t=0$ to remove any terminal reflection. 

\subsubsection{The step configuration}
\label{subsec:Wave-Step}

\begin{figure}[!h]
\makebox[1.\textwidth][c]{
\begin{minipage}[t]{.33\textwidth}
  \centering
  \includegraphics[scale=0.25,angle=0]{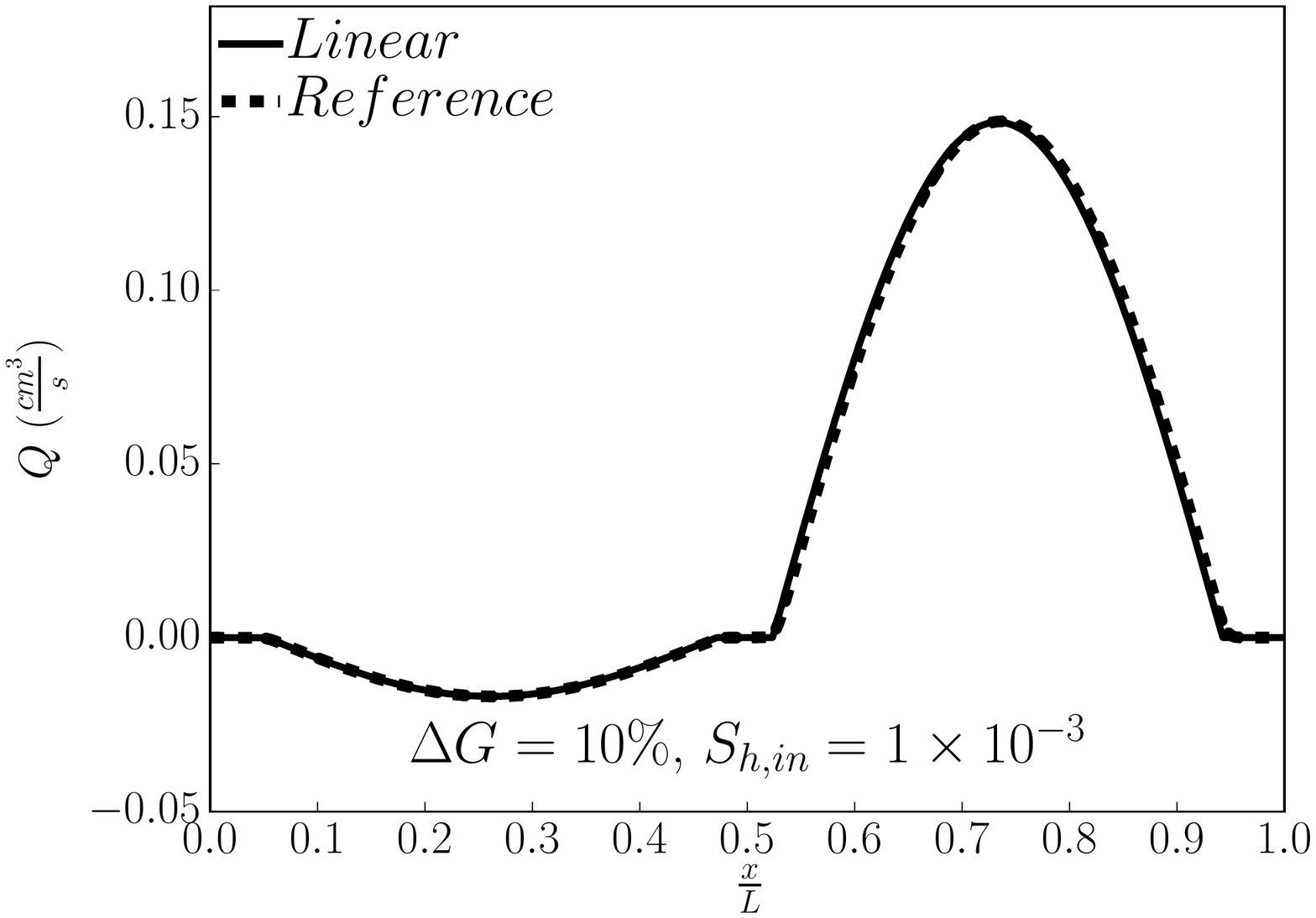}\\
\end{minipage}%
\begin{minipage}[t]{.33\textwidth}
  \centering
   \includegraphics[scale=0.25,angle=0]{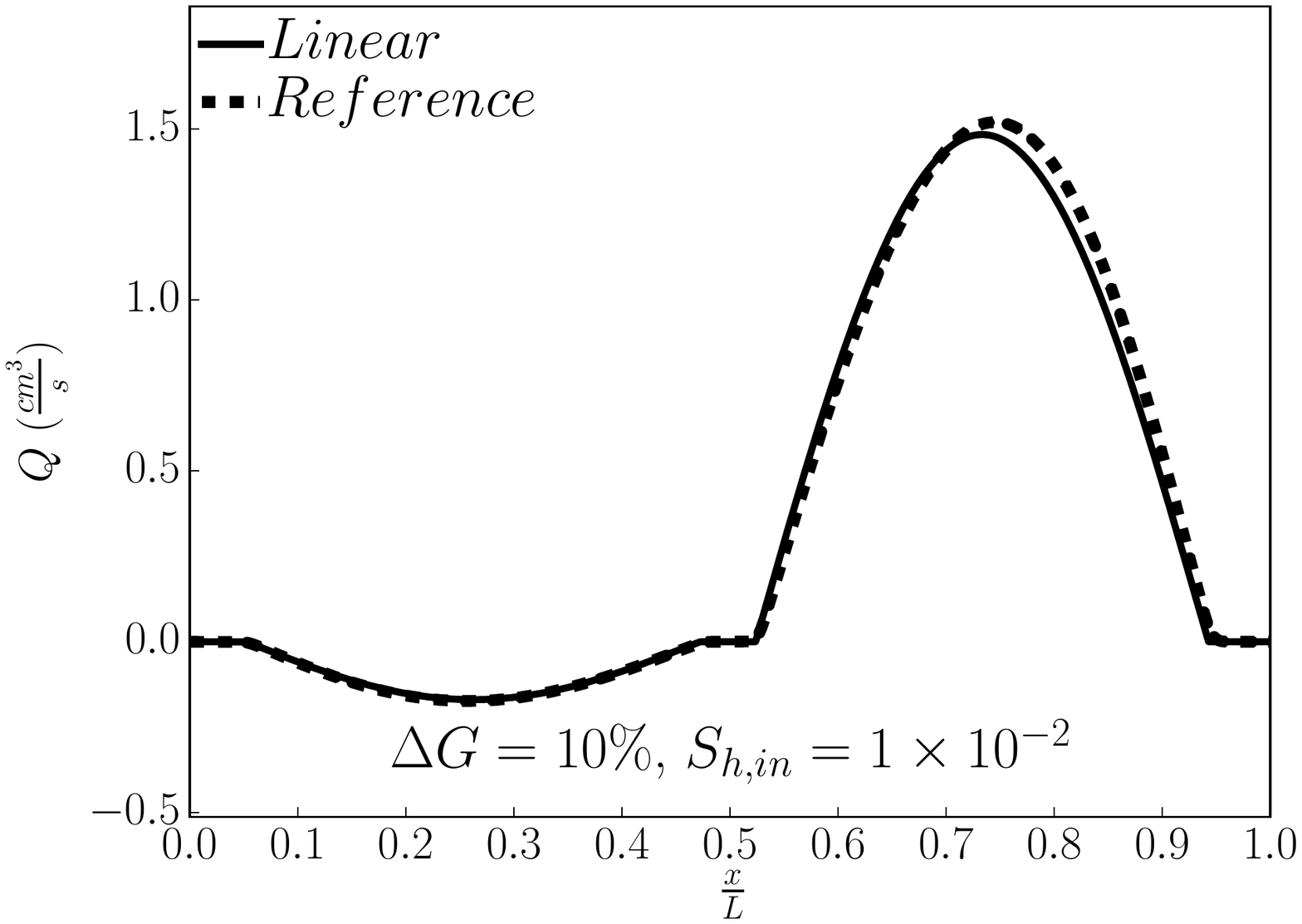}\\
\end{minipage}
\begin{minipage}[t]{.33\textwidth}
  \centering
   \includegraphics[scale=0.25,angle=0]{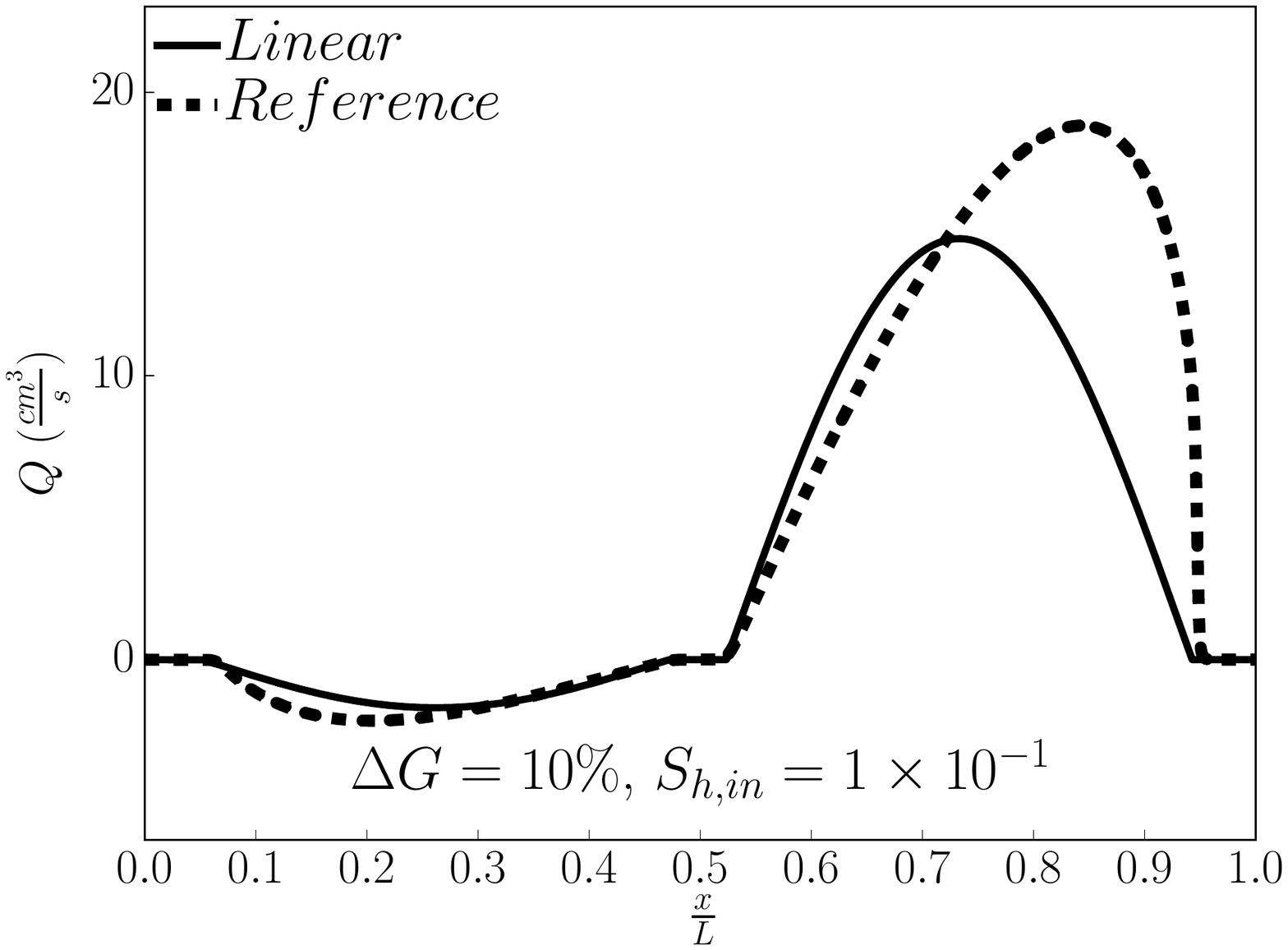}\\
\end{minipage}
}
\caption{Wave propagation: Comparison between the linear solution (full black line) and the reference solution (dashed black line) for the step configuration \eqref{eq:Ex-Step-Geom}, obtained using HR-S for $N=25600$, for the flow rate $Q$ at $t=0.045$ s for $\Delta \mathcal{G}=10 \%$. \underline{\textit{Left}}: $S_h = 1 \times 10^{-3}$; \underline{\textit{Center}}: $S_h = 1 \times 10^{-2}$; \underline{\textit{Right}}: $S_h = 1 \times 10^{-1}$.}
\label{fig:Step-Q-Analytic}
\end{figure}

We focus on the decreasing step configuration \eqref{eq:Ex-Step-Geom}. Given the inlet condition \eqref{bc:Inlet-Flow-Pulse}, the pulse wave propagates in the artery starting from the left-hand side of the domain until it reaches the step. The change of impedance of the vessel creates reflected and transmitted waves that need to be captured by the numerical scheme. A linear analytic solution was proposed in \cite{Raines1974} and validated in \cite{Delestre2012,Delestre2016,Wang2015}, and gives the expression of the reflection coefficient $R_t$ and the transmission coefficient $T_t$, based on the conservation properties \eqref{eq:bc-conj-Q-P-Low-Fr}
\begin{equation}
\left\{
\begin{split}
& R_t = \frac{Y_L - Y_R}{Y_L + Y_R}\\
& T_t = 1 + R_t,\\
\end{split}
\right.
\label{eq:Rt-Tt}
\end{equation}
where $Y = A / \left(\rho c\right)$ is the vessel admittance. Subscripts $L$ and $R$ respectively refer to the values at the left and right of the step. As the coefficients $R_t$ and $T_t$ do not depend on the frequency of the incoming wave, we can analytically predict the position, shape and amplitude of the linear reflected and transmitted waves. However, as the inlet Shapiro number $S_{h,in}$ is non-zero, the flow is nonlinear and the linear analytic solution \eqref{eq:Rt-Tt} is only valid in the asymptotic limit $S_{h,in} \rightarrow 0$. To evaluate the quality of the results obtained with HR, HR-LS and HR-S, we compute reference solutions, obtained with HR-S for $N=25600$ and values of $S_{h,in}$ and $\Delta \mathcal{G}$ taken from table \ref{table:Sh-DG-Steady}. To assess the validity of these reference solutions, we compare them to the linear analytic solutions \eqref{eq:Rt-Tt} in figure \ref{fig:Step-Q-Analytic}. We observe that for low values of the inlet Shapiro number $S_{h,in}$ (figure \ref{fig:Step-Q-Analytic} left), for which the linear approximation is valid, the analytic and reference solutions match. As expected, for higher values of $S_{h,in}$, the flow is no longer linear and the propagation speed as well as the amplitude of the reflected and transmitted waves change (figure \ref{fig:Step-Q-Analytic} center and right).\\

We present results only for the flow rate $Q$ to reduce the number of variables and simplify the analysis of the results. Similar conclusions to those presented hereafter would have been drawn if we had considered the pressure $P$ or the wall perturbation $R-R_0$.\\

We perform a series of 9 numerical computations with different combinations of the non-zero inlet Shapiro number $S_{h,in}$ and the wall deformation parameter $\Delta \mathcal{G}$ taken from table \ref{table:Sh-DG-Steady}. Table \ref{table:Step-Err-Sh-DG} shows $L^1\left[Q\right]$ relative errors between the reference solutions and the results obtained with HR, HR-LS and HR-S for a fixed number of cells $N=1600$. We choose a high value of $N$ to reduce the numerical dissipation and highlight the effects of the well-balanced methods. 
\begin{table}[!h]
\begin{footnotesize}
\begin{center}
\def\arraystretch{1.2}
{\setlength{\tabcolsep}{0.5em}
\begin{tabular}{c | c|| c|c|c }
\multicolumn{2}{c||}{$S_{h,in}$} & \multicolumn{3}{c}{$1 \times 10^{-3}$} \\
 \cline{3-5} 
\multicolumn{2}{c||}{$\Delta	\mathcal{G}$} & $1\%$ & $10\%$ & $30\%$  \\ 
 
\hline

 & HR &
 $2.3 \times 10^{-2}$ &  $5.5 \times 10^{-2}$ & $5.5 \times 10^{-1}$ 
  \\
   
$L^1\left[Q\right]$ & HR-LS & 
$2.3 \times 10^{-2}$ & $2.8 \times 10^{-2}$ & $6.6 \times 10^{-2}$  
\\ 

 & HR-S &
 $2.3 \times 10^{-2}$ & $2.8 \times 10^{-2}$ & $6.6 \times 10^{-2}$   
  \\ 
  
 \hline \hline 
 \multicolumn{2}{c||}{$S_{h,in}$} & \multicolumn{3}{c}{$1 \times 10^{-2}$}  \\  \cline{3-5} 
 
\multicolumn{2}{c||}{$\Delta	\mathcal{G}$} & $1\%$ & $10\%$ & $30\%$   \\ 
 
\cline{1-5} 

 & HR & 
 $2.3 \times 10^{-2}$ &  $5.5 \times 10^{-2}$ & $5.5 \times 10^{-1}$ 
  \\ 
  
$L^1\left[Q\right]$ & HR-LS & 
$2.3 \times 10^{-2}$ & $2.8 \times 10^{-2}$ & $6.6 \times 10^{-2}$  
 \\ 
 
 & HR-S &  
 $2.3 \times 10^{-2}$ & $2.8 \times 10^{-2}$ & $6.6 \times 10^{-2}$  
 \\
  \hline \hline 
 \multicolumn{2}{c||}{$S_{h,in}$} & \multicolumn{3}{c}{$1 \times 10^{-1}$}  \\  \cline{3-5} 
 
\multicolumn{2}{c||}{$\Delta	\mathcal{G}$} & $1\%$ & $10\%$ & $30\%$   \\ 
 
\cline{1-5} 

 & HR & 
 $2.9 \times 10^{-2}$& $6.1 \times 10^{-2}$& $5.1 \times 10^{-1}$
  \\ 
  
$L^1\left[Q\right]$ & HR-LS & 
$2.9 \times 10^{-2}$& $3.5 \times 10^{-2}$& $7.6 \times 10^{-2}$ 
 \\ 
 
 & HR-S &  
 $2.9 \times 10^{-2}$& $3.5 \times 10^{-2}$& $7.5 \times 10^{-2}$
 
\end{tabular} 
}
\end{center}
\end{footnotesize}
\caption{Wave propagation: Relative error $L^1\left[Q\right]$ computed in the step \eqref{eq:Ex-Step-Geom} for values of $S_{h,in}$ and $\Delta \mathcal{G}$ taken from table \ref{table:Sh-DG-Steady} obtained for $N =1600$. HR, HR-LS and HR-S present similar results except for $\Delta \mathcal{G} = 30 \%$.}
\label{table:Step-Err-Sh-DG}
\end{table}
The results obtained with HR, HR-LS and HR-S are almost identical and indicate that each method is able to correctly compute the expected reflected and transmitted waves. For each method, the error $L^1\left[Q\right]$ is independent of the inlet Shapiro number $S_{h,in}$ but increases with the wall deformation parameter $\Delta \mathcal{G}$. However, the error obtained with HR increases faster with $\Delta \mathcal{G}$ than with the other methods. In particular, for $\Delta \mathcal{G} = 30 \%$, the value of $L^1\left[Q\right]$ obtained with HR is one order of magnitude higher than the one obtained with HR-LS or HR-S.\\

\begin{figure}[!h]
\makebox[1.\textwidth][c]{
\begin{minipage}[t]{.5\textwidth}
  \centering
  \includegraphics[scale=0.40,angle=0]{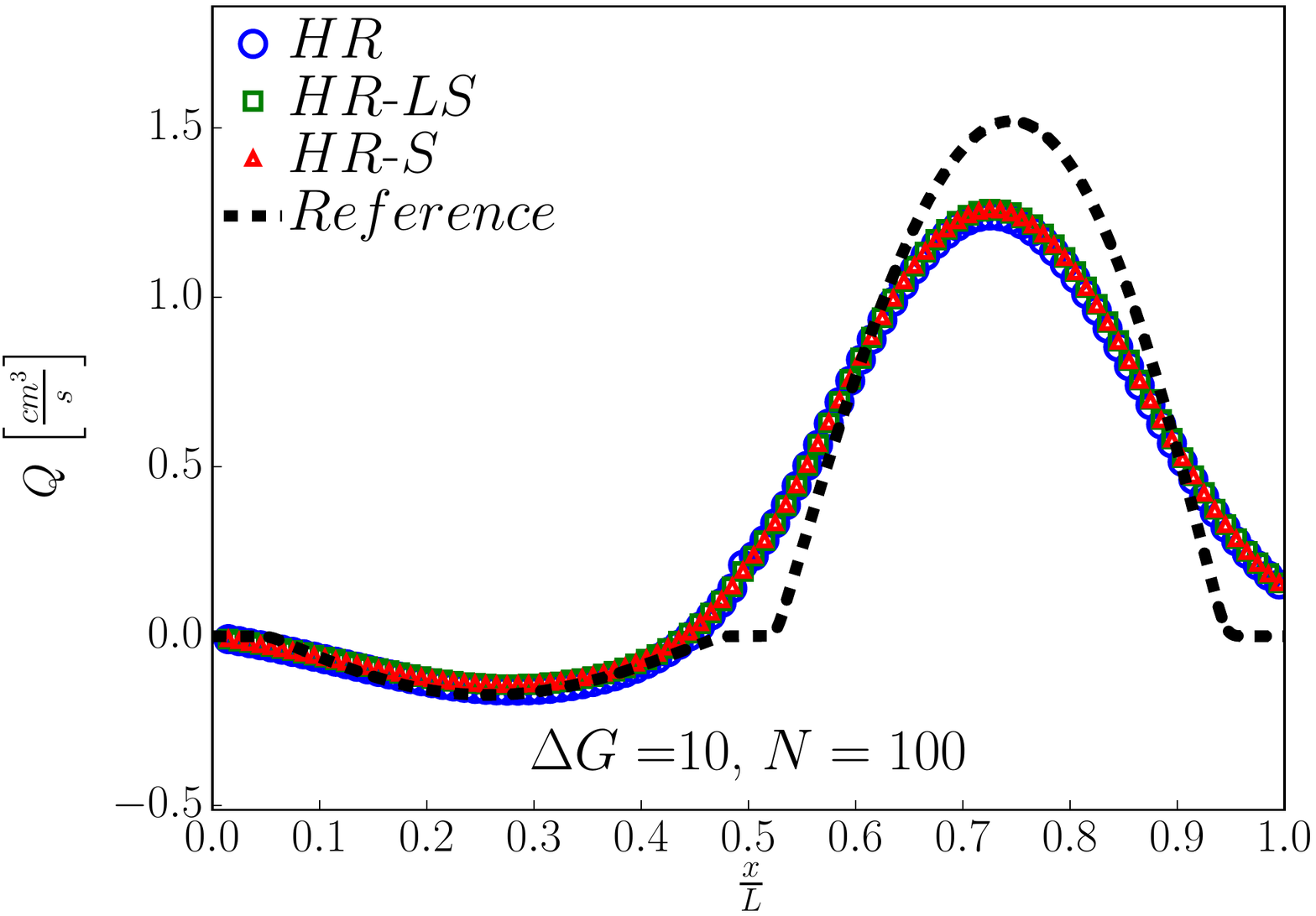}\\
\end{minipage}%
\begin{minipage}[t]{.5\textwidth}
  \centering
   \includegraphics[scale=0.40,angle=0]{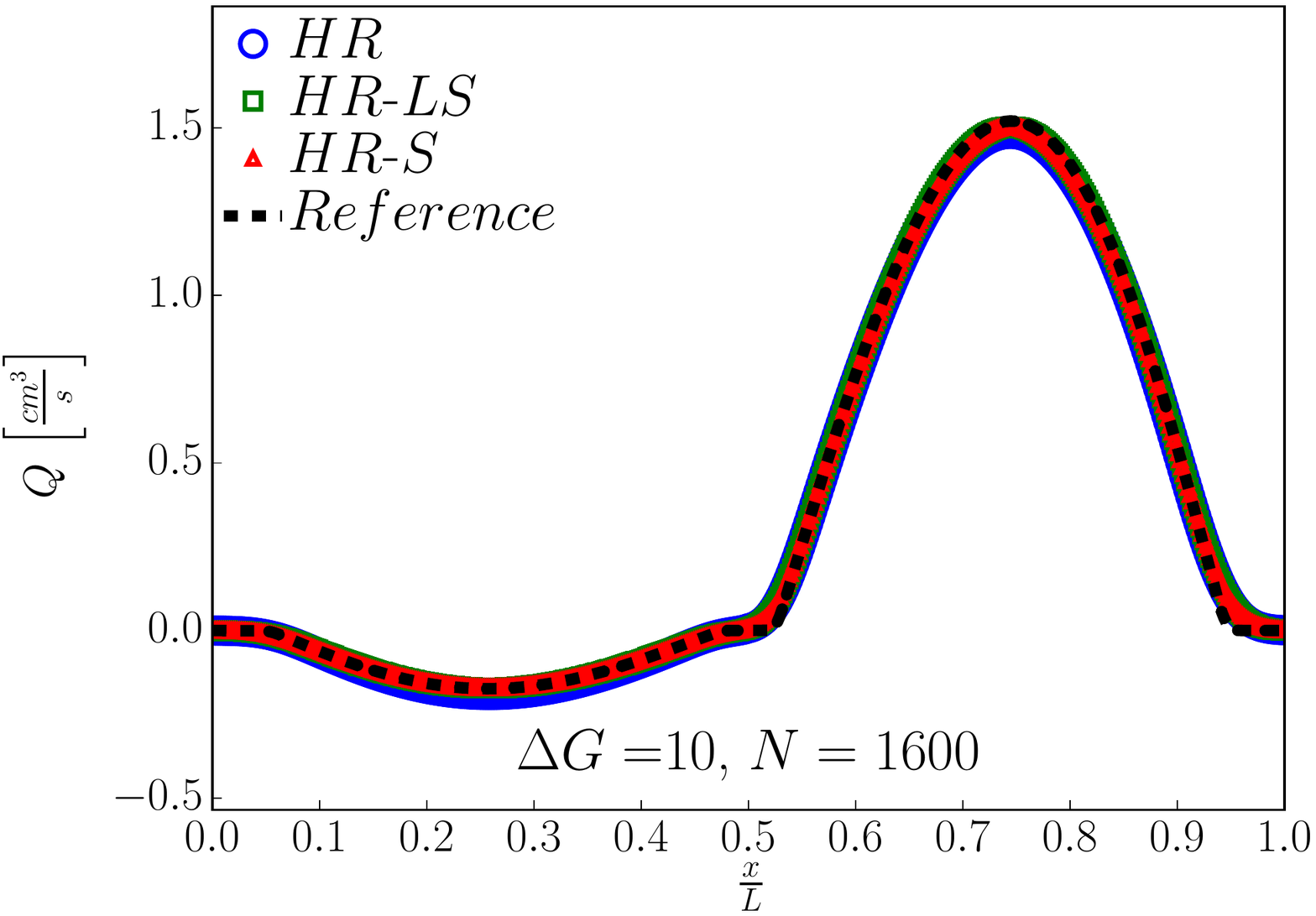}\\
\end{minipage}
}
\caption{Wave propagation: Flow rate $Q\left( x \right)$ for the step \eqref{eq:Ex-Step-Geom} at $t=0.045$ s for the reference solution (black dashed line), HR (blue circle), HR-LS (green square) and HR-S (red triangle) for $S_h = 1 \times 10^{-2}$ and $\Delta \mathcal{G} = 10 \%$; \underline{\textit{Left}}: $N=100$; \underline{\textit{Right}}: $N=1600$. For $N=100$ and $N=1600$, all solutions are comparable, and for $N=1600$, HR, HR-LS and HR-S converge towards the reference solution.}
\label{fig:Step-Fr-10m2-dR-10}
\end{figure}

\begin{figure}[!h]
\makebox[1.\textwidth][c]{
\begin{minipage}[t]{.5\textwidth}
  \centering
  \includegraphics[scale=0.40,angle=0]{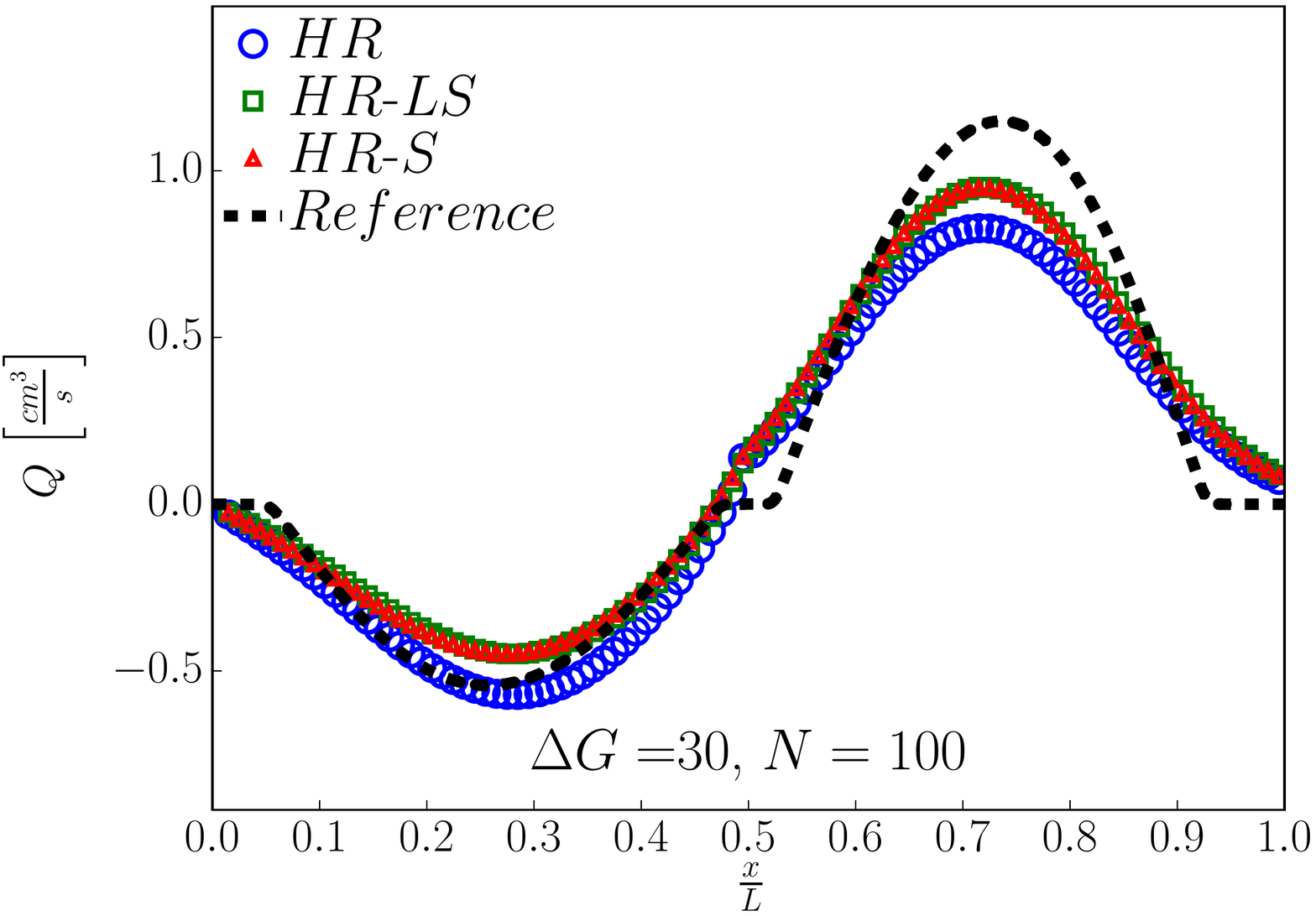}\\
\end{minipage}%
\begin{minipage}[t]{.5\textwidth}
  \centering
   \includegraphics[scale=0.40,angle=0]{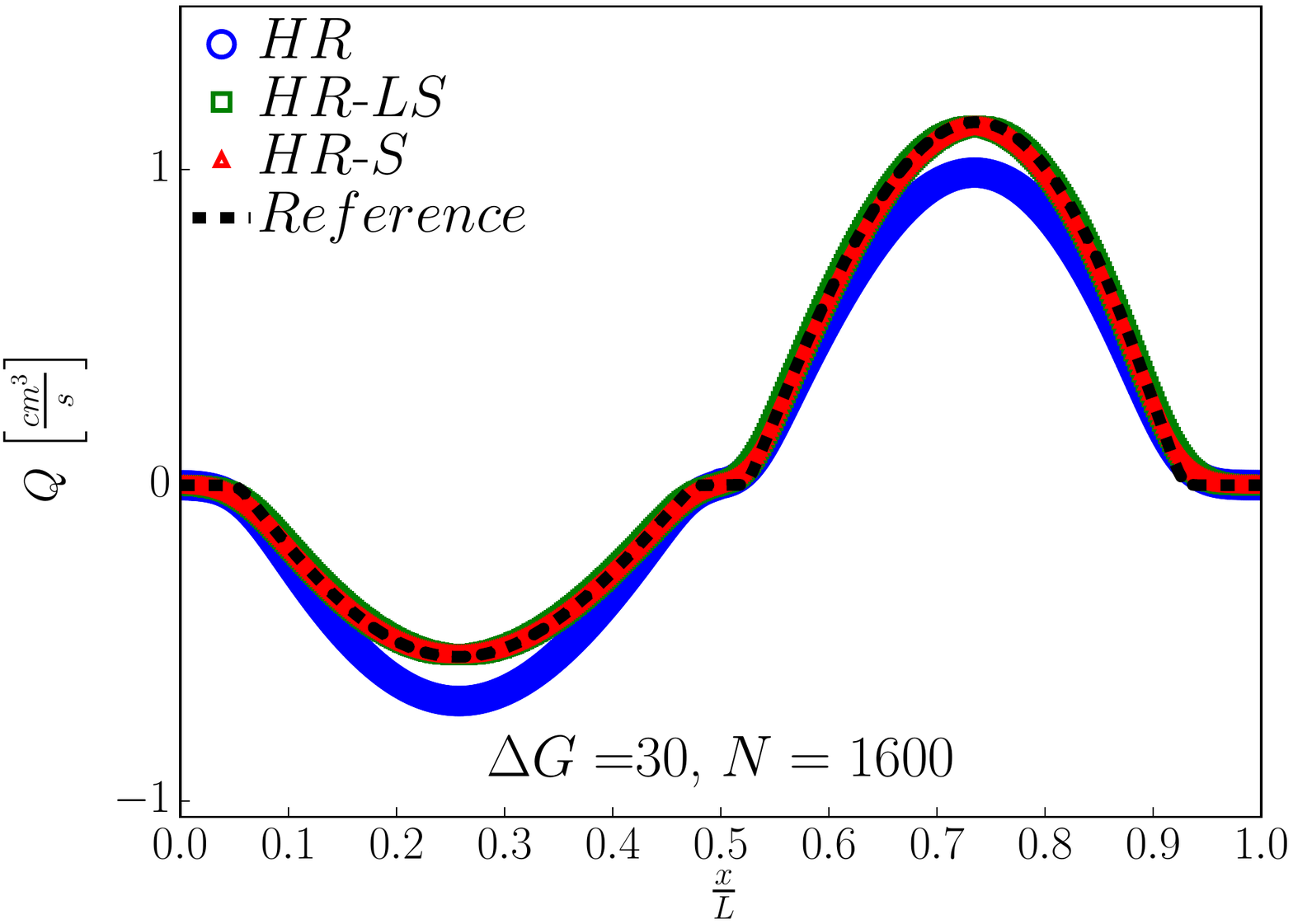}\\
\end{minipage}
}
\caption{Wave propagation: Flow rate $Q\left( x \right)$ for the step \eqref{eq:Ex-Step-Geom} at $t=0.045$ s for the reference solution (black dashed line), HR (blue circle), HR-LS (green square) and HR-S (red triangle) for $S_h = 1 \times 10^{-2}$ and $\Delta \mathcal{G} = 30 \%$; \underline{\textit{Left}}: $N=100$; \underline{\textit{Right}}: $N=1600$. HR-LS and HR-S converge towards the reference solution while HR does not.}
\label{fig:Step-Fr-10m2-dR-30}
\end{figure}

\begin{figure}[!h]
\makebox[1.\textwidth][c]{
\begin{minipage}[t]{.5\textwidth}
  \centering
  \includegraphics[scale=0.40,angle=0]{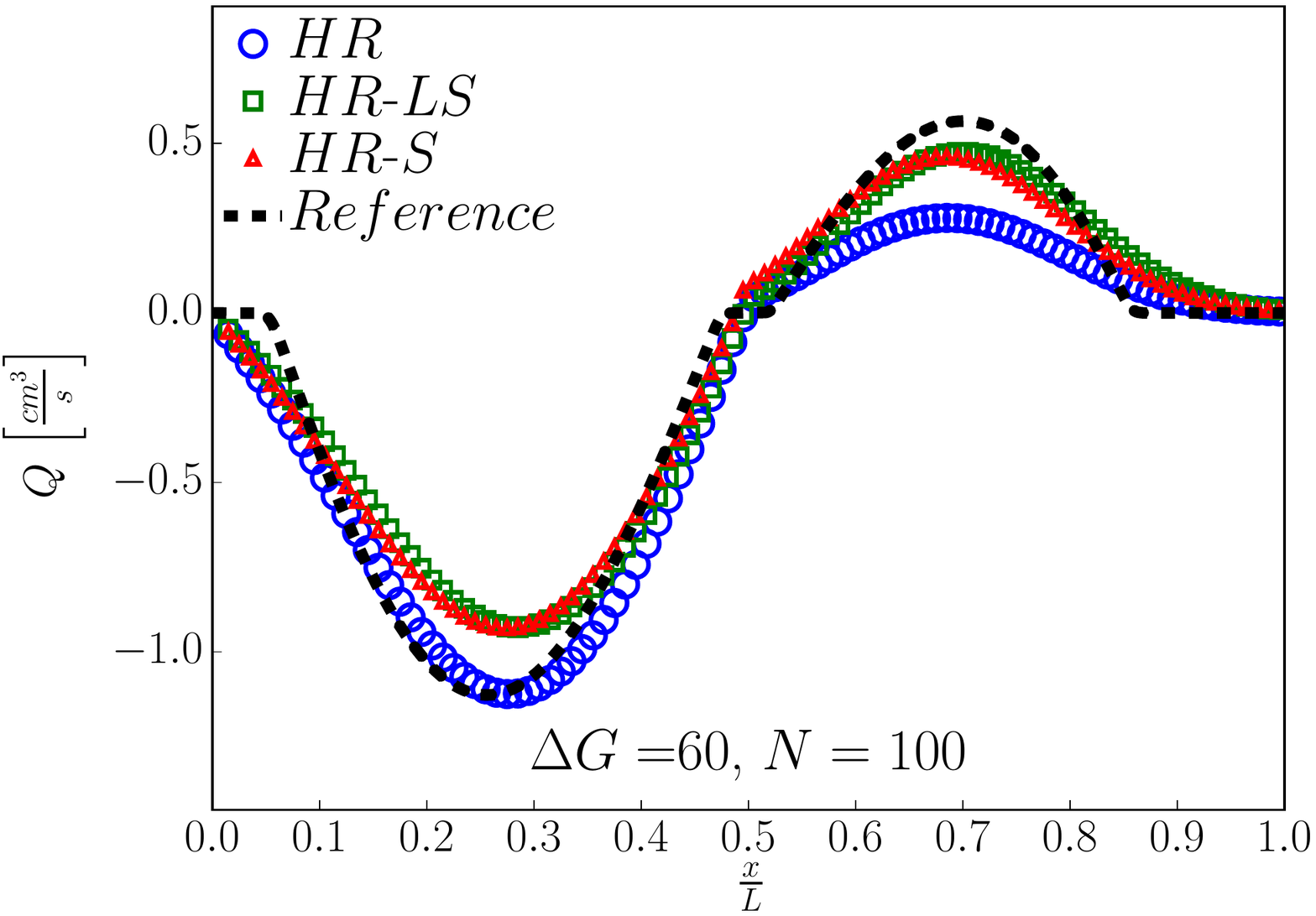}\\
\end{minipage}%
\begin{minipage}[t]{.5\textwidth}
  \centering
   \includegraphics[scale=0.40,angle=0]{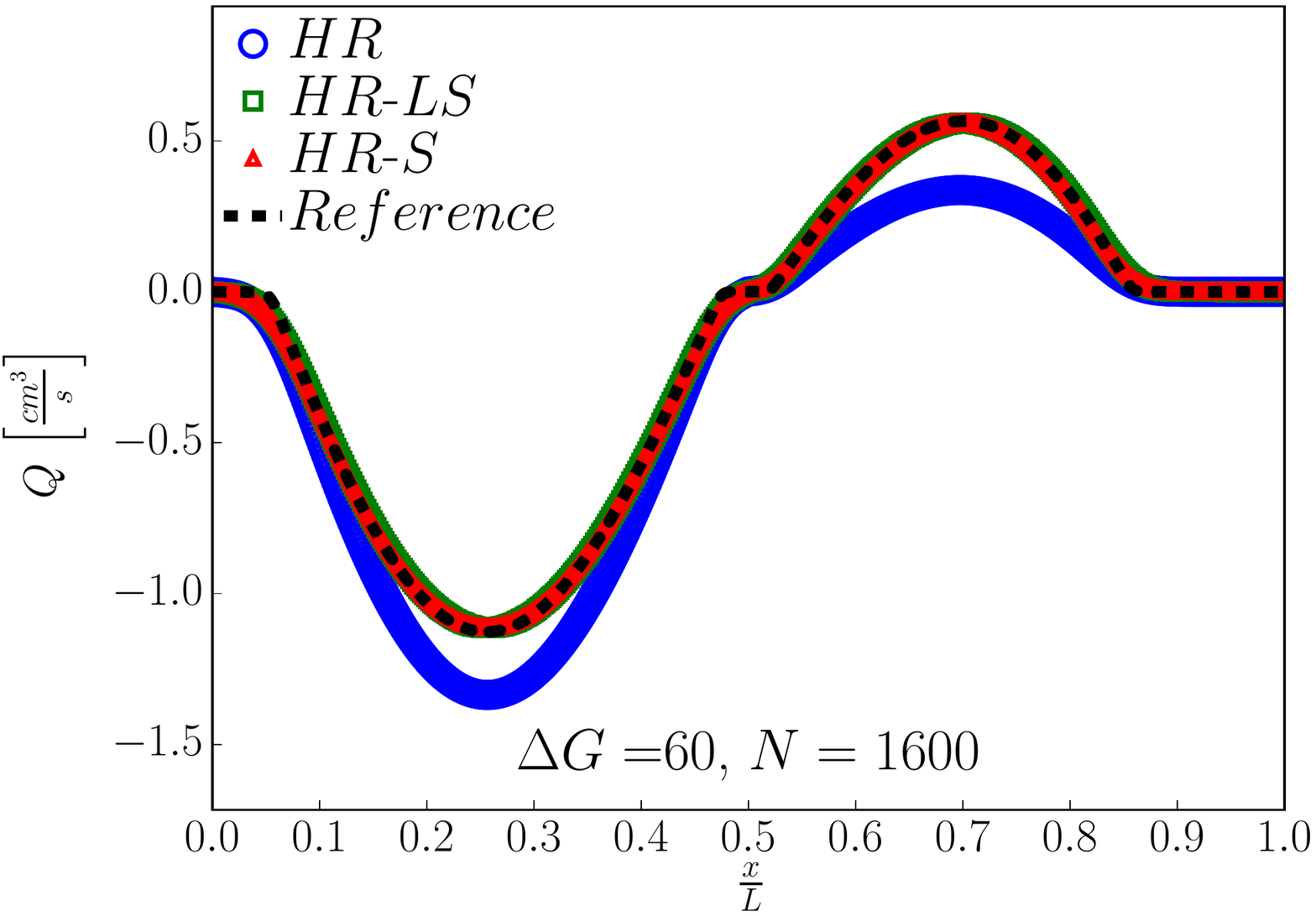}\\
\end{minipage}
}
\caption{Wave propagation: Flow rate $Q\left( x \right)$ for the step \eqref{eq:Ex-Step-Geom} at $t=0.045$ s for the reference solution (black dashed line), HR (blue circle), HR-LS (green square) and HR-S (red triangle) for $S_h = 1 \times 10^{-2}$ and $\Delta \mathcal{G} = 60 \%$; \underline{\textit{Left}}: $N=100$; \underline{\textit{Right}}: $N=1600$. HR-LS and HR-S converge towards the reference solution while HR does not.}
\label{fig:Step-Fr-10m2-dR-60}
\end{figure}

 This last point is corroborated by figures \ref{fig:Step-Fr-10m2-dR-10}, \ref{fig:Step-Fr-10m2-dR-30} and \ref{fig:Step-Fr-10m2-dR-60}, where we represent the spatial evolution of the flow rate $Q$ at $t=0.045$ s, obtained using $N=100$ (left) and $N=1600$ (right) for $S_{h,in}=1\times 10^{-2}$ and $\Delta \mathcal{G}= \left\{ 10 \%,30 \%,60 \%\right\}$ respectively. In each figure, we compare the results obtained using HR, HR-LS and HR-S to the corresponding reference solution and observe if increasing the number of cells allows the numerical solution to converge towards the reference solution. In figure \ref{fig:Step-Fr-10m2-dR-10}, the results obtained for $\Delta \mathcal{G}=10 \%$ with HR, HR-LS and HR-S are similar and indicate that each numerical solution converges towards the reference solution. On the contrary, in figure \ref{fig:Step-Fr-10m2-dR-30} for $\Delta \mathcal{G} = 30 \%$ and in figure \ref{fig:Step-Fr-10m2-dR-60} for $\Delta \mathcal{G} = 60 \%$, only the solutions obtained with HR-LS and HR-S converge towards the reference solution. HR is unable to compute the expected amplitude of the reflected and transmitted waves and overestimates the amplitude of the reflected wave and underestimates the amplitude of the transmitted wave. \\
  
The results indicate that HR-LS and HR-S are able to compute wave reflections and transmissions in an artery presenting an arbitrary large discontinuous variation of its cross-sectional area at rest $A_0$ and arterial wall rigidity $K$. On the contrary, HR is unable to compute the correct amplitude of the reflected and transmitted waves when the discontinuous variation of the artery's geometrical and mechanical properties is too large, independently of the number of cells $N$. Moreover, these results show that the system \eqref{eq:WB-HRQ-sys-SW} (used by HR-LS) has the appropriate conservation properties to compute wave reflections for arbitrary large discontinuous geometrical and mechanical variations in low-Shapiro number flow regimes. On the contrary, HR, using the system \eqref{eq:WB-HRU-sys-SW}, is only able to compute wave reflections for small discontinuous variations of the artery's properties ($\Delta \mathcal{G}=10 \%$, see figure \ref{fig:Step-Fr-10m2-dR-10}). This last point can be problematic as large variations of the artery's geometrical and mechanical properties can be encountered when modeling arterial pathologies such as stenoses.

\subsubsection{The stenosis configuration}
\label{subsec:Wave-Stenosis}

In this subsection we focus on the stenosis configuration \eqref{eq:Ex-Stenosis-Geom}. To evaluate the quality of the results obtained with HR, HR-LS and HR-S, we compute reference solutions, obtained with HR-S for $N=25600$ and values of $S_{h,in}$ and $\Delta \mathcal{G}$ taken from table \ref{table:Sh-DG-Steady}. As the variation of geometrical and mechanical properties of the artery is smooth, the observed flow rate is constituted of a continuum of reflected and transmitted waves that are created at each cell interface, where the artery's geometrical and mechanical properties are discontinuous.\\

\begin{figure}[!h]
\makebox[1.\textwidth][c]{
\begin{minipage}[t]{.5\textwidth}
  \centering
  \includegraphics[scale=0.40,angle=0]{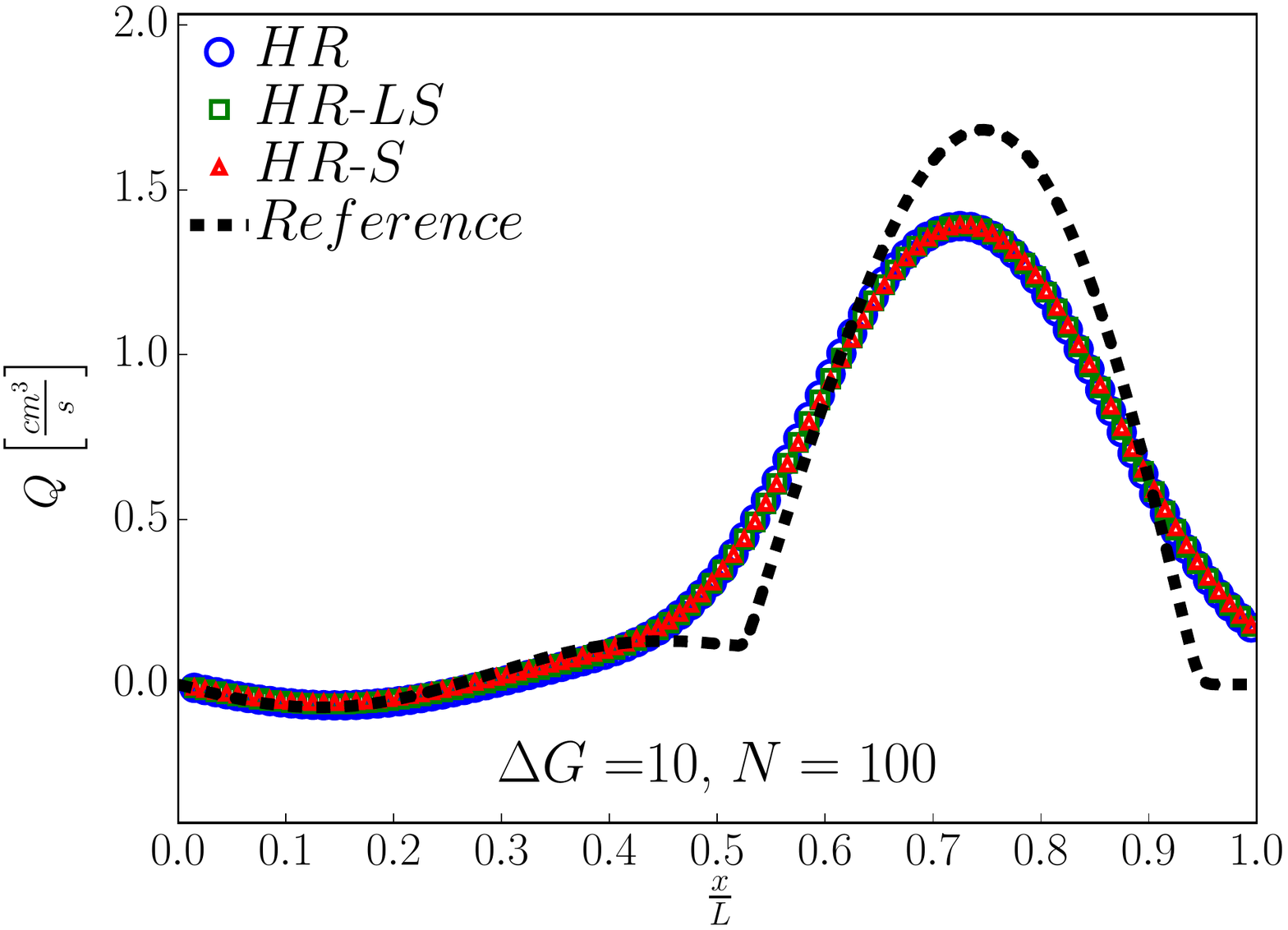}\\
\end{minipage}%
\begin{minipage}[t]{.5\textwidth}
  \centering
   \includegraphics[scale=0.40,angle=0]{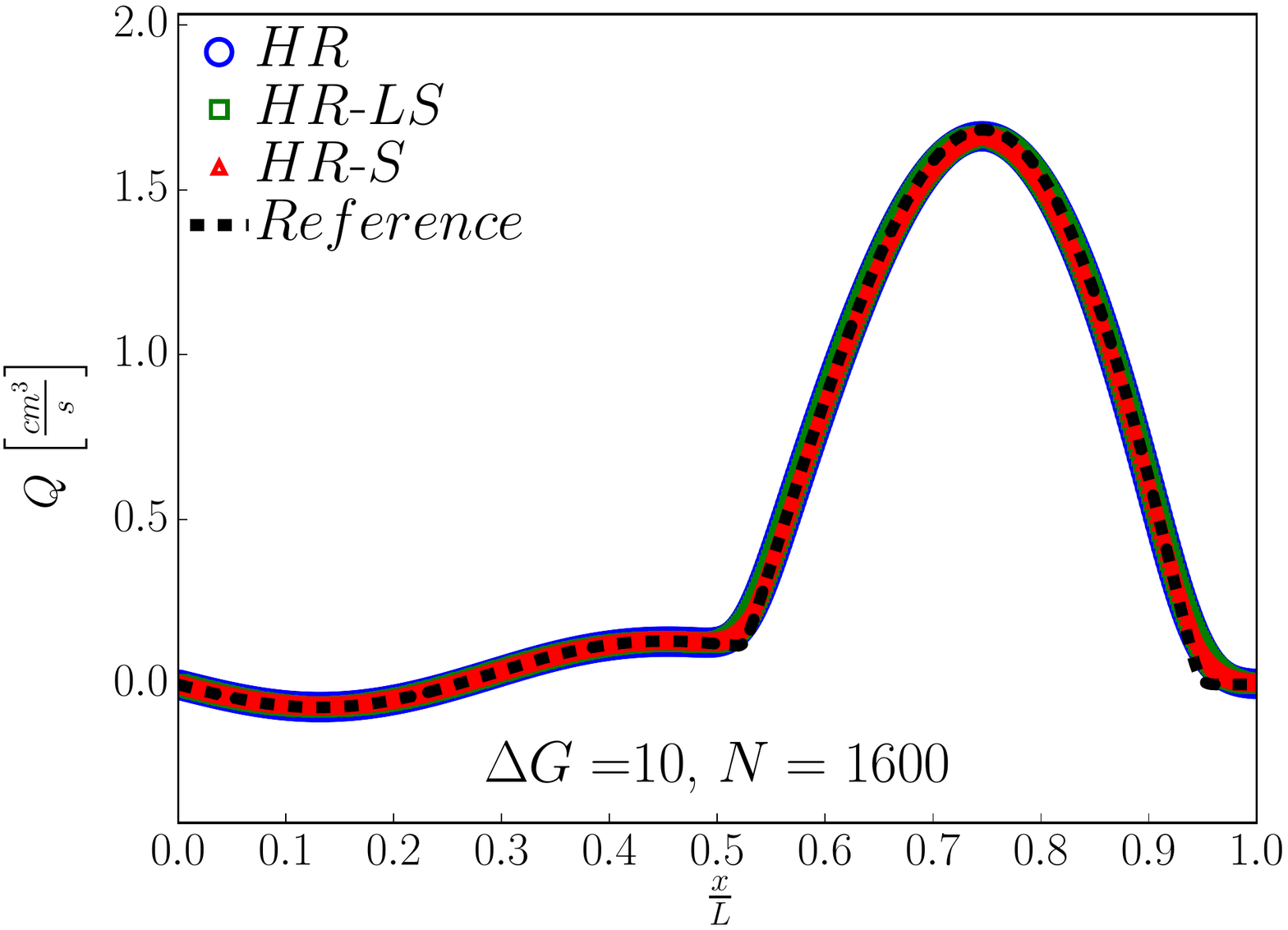}\\
\end{minipage}
}
\caption{Wave propagation: Flow rate $Q\left( x \right)$ for the stenosis \eqref{eq:Ex-Stenosis-Geom} at $t=0.045$ s for the reference solution (black dashed line), HR (blue circle), HR-LS (green square) and HR-S (red triangle) for $S_h = 1 \times 10^{-2}$ and $\Delta \mathcal{G} = 10 \%$; \underline{\textit{Left}}: $N=100$; \underline{\textit{Right}}: $N=1600$. For $N=100$ and $N=1600$, all solutions are comparable, and for $N=1600$, HR, HR-LS and HR-S converge towards the reference solution.}
\label{fig:Stenosis-Fr-10m2-dR-10}
\end{figure}

\begin{figure}[!h]
\makebox[1.\textwidth][c]{
\begin{minipage}[t]{.5\textwidth}
  \centering
  \includegraphics[scale=0.40,angle=0]{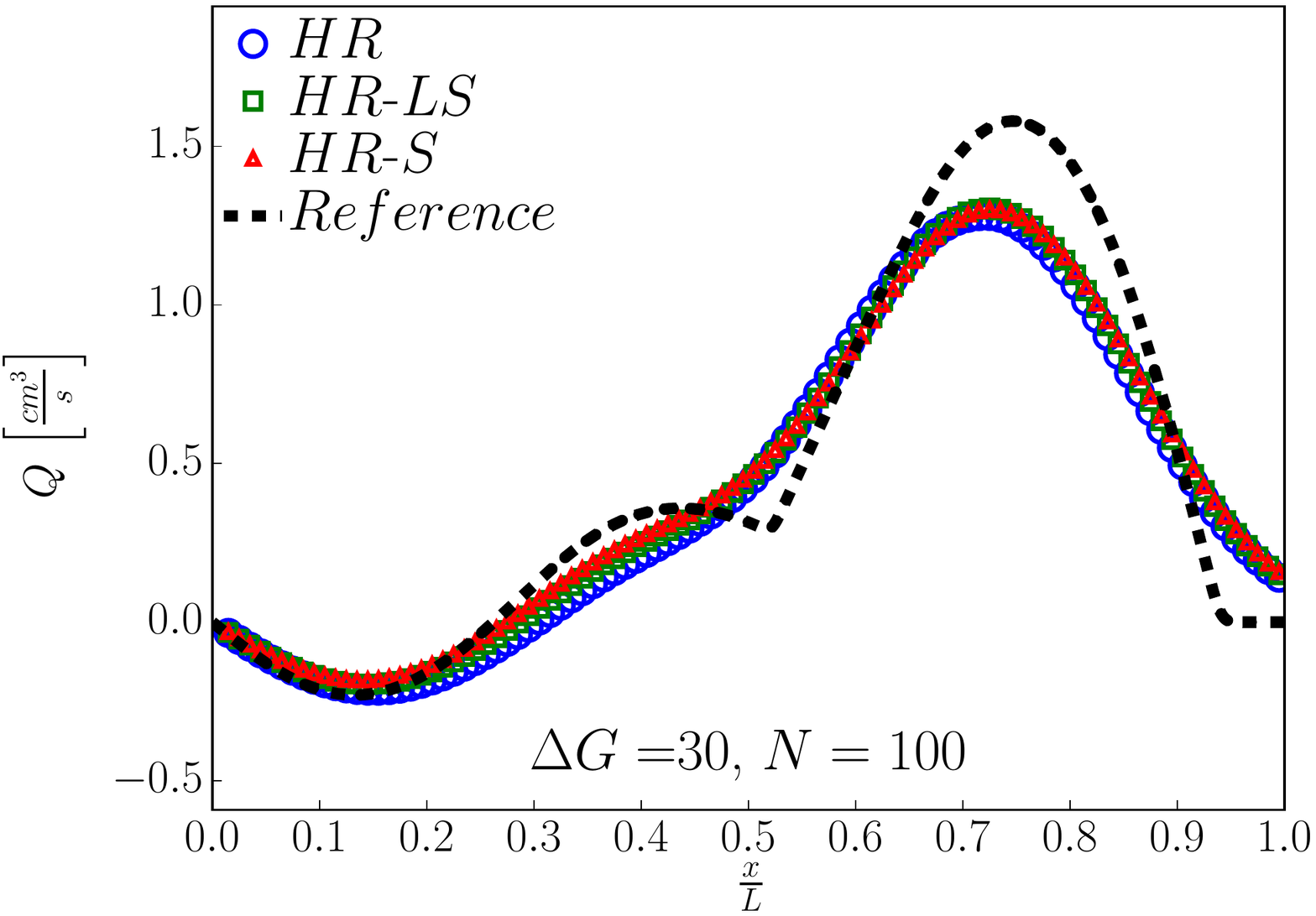}\\
\end{minipage}%
\begin{minipage}[t]{.5\textwidth}
  \centering
   \includegraphics[scale=0.40,angle=0]{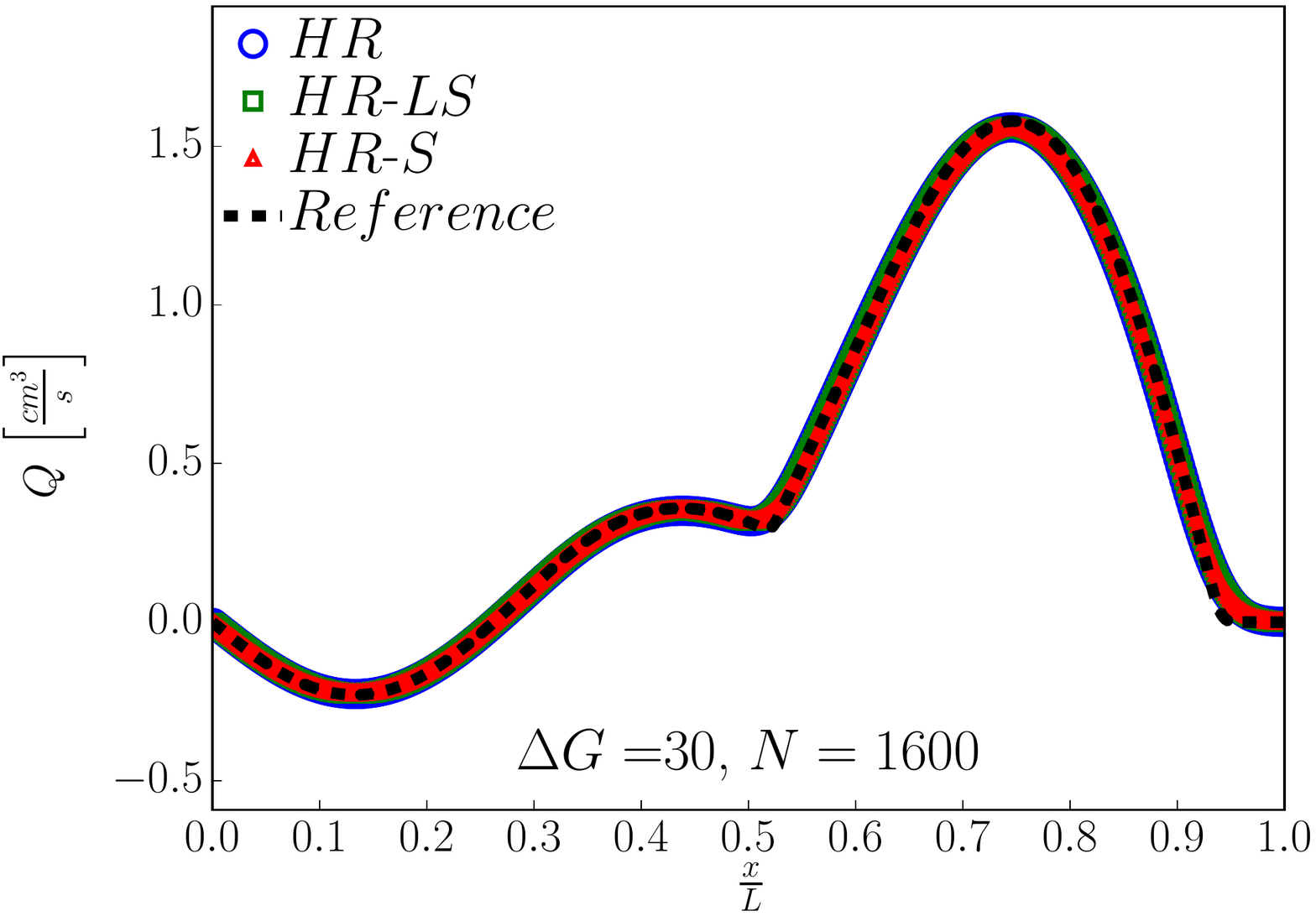}\\
\end{minipage}
}
\caption{Wave propagation: Flow rate $Q\left( x \right)$ for the stenosis \eqref{eq:Ex-Stenosis-Geom} at $t=0.045$ s for the reference solution (black dashed line), HR (blue circle), HR-LS (green square) and HR-S (red triangle) for $S_h = 1 \times 10^{-2}$ and $\Delta \mathcal{G} = 30 \%$; \underline{\textit{Left}}: $N=100$; \underline{\textit{Right}}: $N=1600$. For $N=100$ and $N=1600$, all solutions are comparable, and for $N=1600$, HR, HR-LS and HR-S converge towards the reference solution.}
\label{fig:Stenosis-Fr-10m2-dR-30}
\end{figure}

\begin{figure}[!h]
\makebox[1.\textwidth][c]{
\begin{minipage}[t]{.5\textwidth}
  \centering
  \includegraphics[scale=0.40,angle=0]{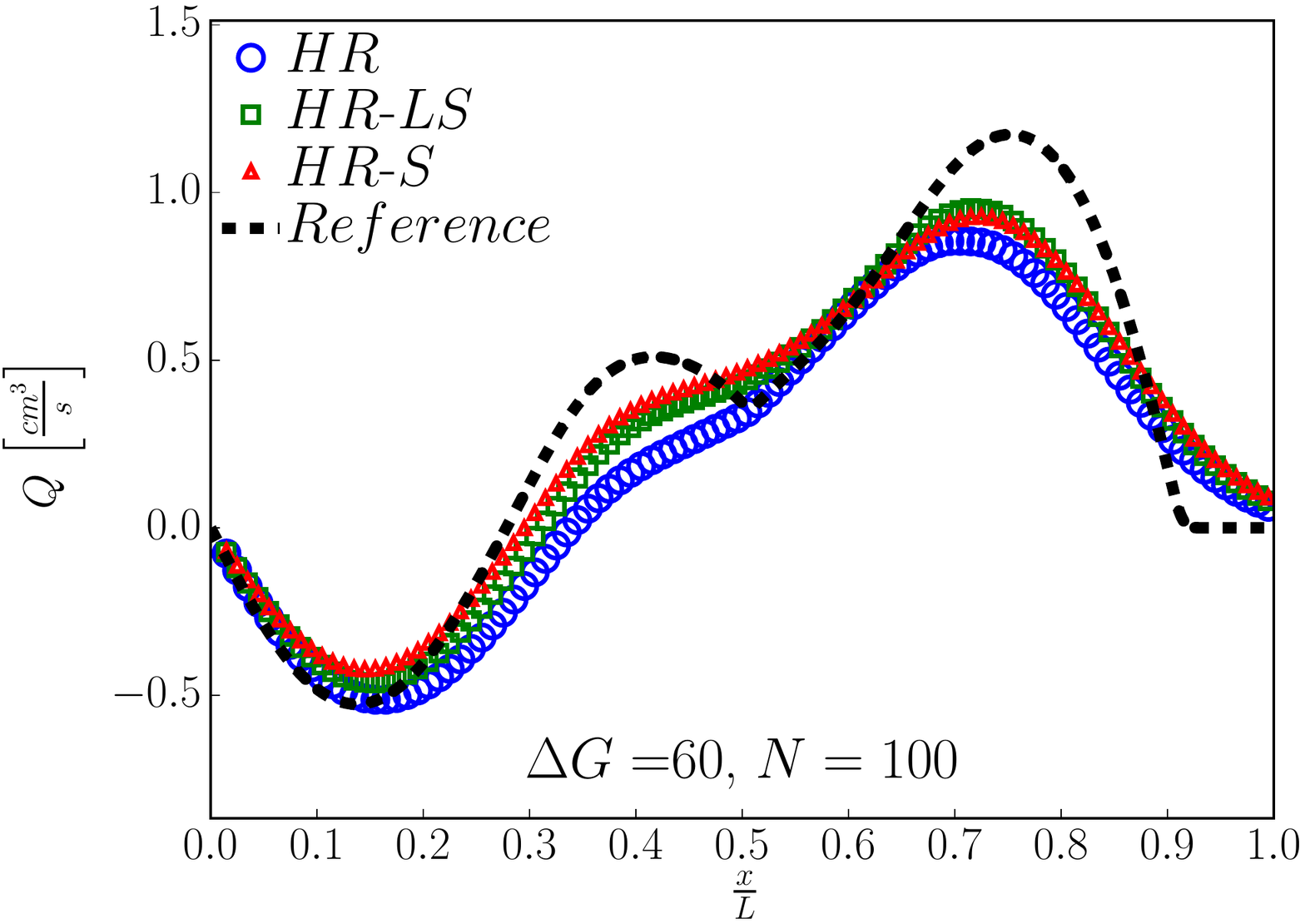}\\
\end{minipage}%
\begin{minipage}[t]{.5\textwidth}
  \centering
   \includegraphics[scale=0.40,angle=0]{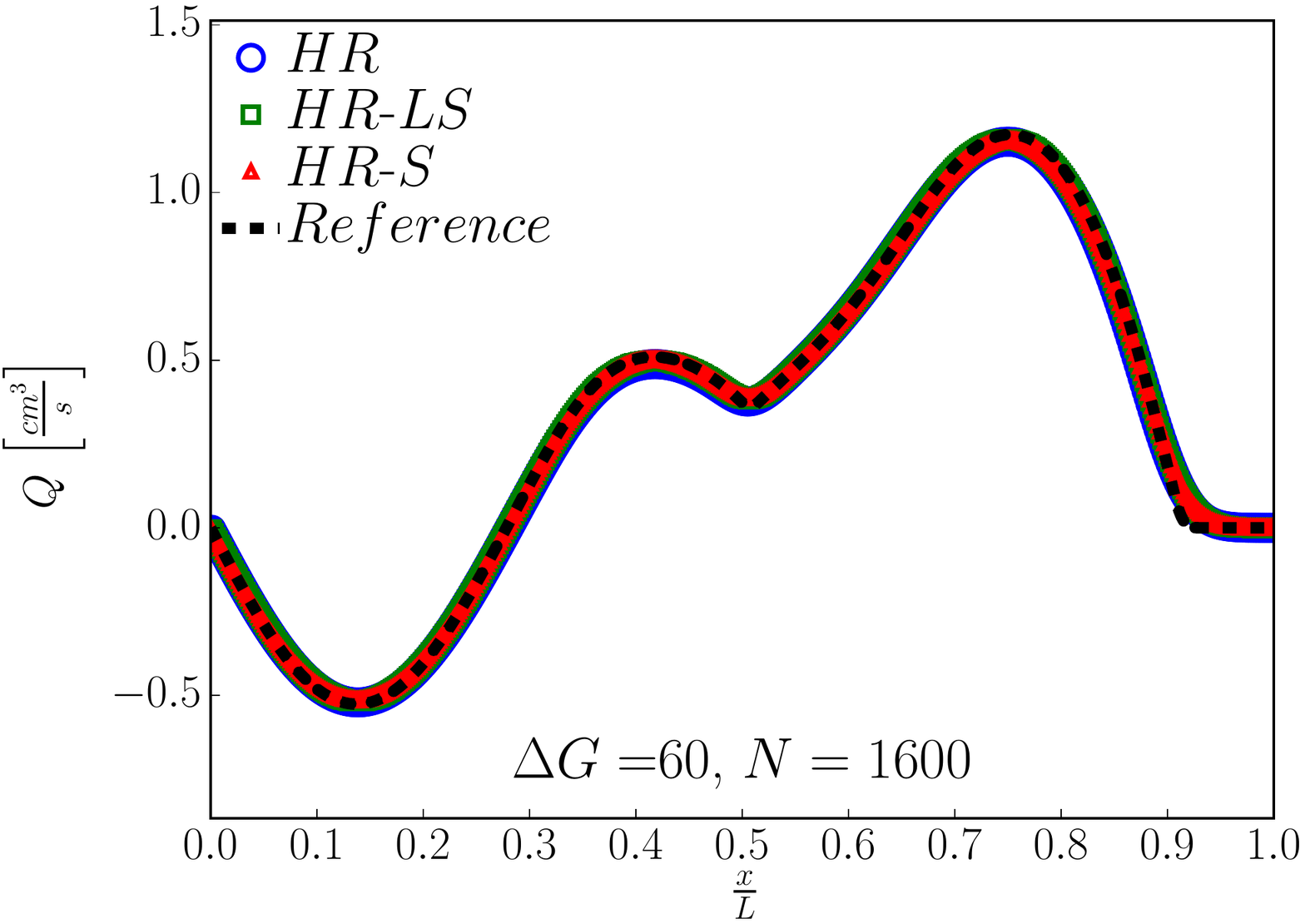}\\
\end{minipage}
}
\caption{Wave propagation: Flow rate $Q\left( x \right)$ for the stenosis \eqref{eq:Ex-Stenosis-Geom} at $t=0.045$ s for the reference solution (black dashed line), HR (blue circle), HR-LS (green square) and HR-S (red triangle) for $S_h = 1 \times 10^{-2}$ and $\Delta \mathcal{G} = 60 \%$; \underline{\textit{Left}}: $N=100$; \underline{\textit{Right}}: For $N=100$, HR is less accurate than HR-LS and HR-S, and for $N=1600$, HR, HR-LS and HR-S converge towards the reference solution.}
\label{fig:Stenosis-Fr-10m2-dR-60}
\end{figure}

Similar results to those of subsection \ref{subsec:Wave-Step} are obtained, and therefore we do not completely repeat the previous analysis. In figures \ref{fig:Stenosis-Fr-10m2-dR-10}, \ref{fig:Stenosis-Fr-10m2-dR-30} and \ref{fig:Stenosis-Fr-10m2-dR-60}, we present the spatial evolution of the flow rate $Q$ at $t=0.045$ s, obtained using $N=100$ (left) and $N=1600$ (right) for $S_{h,in}=1\times 10^{-2}$ and $\Delta \mathcal{G}= \left\{ 10 \%,30 \%,60 \%\right\}$ respectively. In each figure, we compare the results obtained using HR, HR-LS and HR-S to the corresponding reference solution and observe if increasing the number of cells allows the numerical solution to converge towards the reference solution. Contrary to the step configuration studied in subsection \ref{subsec:Wave-Step}, the results obtained with HR, HR-LS and HR-S are similar and indicate that each numerical solution converges towards the reference solution. However, for $\Delta \mathcal{G}=\left\{30\%,60\%\right\}$ and $N=100$, HR is less accurate than HR-LS and HR-S.\\

These results are coherent with those of subsection \ref{subsec:Wave-Step}. Indeed, when studying the step configuration, we showed that contrary to HR-LS and HR-S, HR overestimates the amplitude of the reflected wave and underestimates the amplitude of the transmitted wave when a large discontinuous variation of the artery's geometrical and mechanical properties is considered ($\Delta \mathcal{G} = \left\{ 30 \%, 60 \%\right\}$). As the stenosis is a smooth variation of the cross-sectional area at rest $A_0$ and of the arterial wall rigidity $K$, discontinuous variations of the arterial wall's geometrical and mechanical properties occur at each cell interface and the amplitude of these variations decreases with the number of cells $N$.  Hence, for $\Delta \mathcal{G}=\left\{30\%,60\%\right\}$ and $N=100$, the local discontinuous variations of the artery's properties are large enough for HR to be inaccurate. On the contrary, for $N=1600$, the local discontinuous variation of the artery's geometrical and mechanical properties are sufficiently small for HR to be as accurate as HR-LS and HR-S. \\
  
We have studied the wave capturing behavior of HR, HR-LS and HR-S. We showed that for arbitrary large smooth or discontinuous variations of the artery's cross-sectional area at rest $A_0$ and arterial wall rigidity $K$, both HR-LS and HR-S are able to compute the expected reflected and transmitted waves. On the contrary, HR is unable to correctly compute reflected and transmitted waves when large discontinuous variations of the artery's properties are considered. In particular, HR overestimates the reflected wave and underestimates the transmitted wave. Therefore, HR-LS and HR-S are good choices to compute wave reflections and transmissions in low-Shapiro flow regimes. In the following subsection, we will analyze the behavior of the different well-balanced methods in large network computations, where multiple effects come into play.

\section{A realistic example: stenosis of the iliac artery in a 55 arteries network}
\label{sec:Ex-55}

We study the response at the systemic level of a model network to the presence of a pathology. Indeed, the observed pressure and flow waveforms in the systemic network are the superposition of multiple reflected and transmitted waves, generated at each arterial bifurcation and dampened and diffused by the viscosity of the blood and the arterial wall. The presence of a pathology creates additional reflected and transmitted waves that change the reflection pattern and therefore the shape and amplitude of the observed waveforms. When such pathologies are modeled in the network, a well-balanced method is required to take into account the geometrical and mechanical source term induced by the local variations of the cross-sectional area and arterial wall rigidity representing the pathology. \\

\begin{figure}[!h]
\makebox[1.\textwidth][c]{
\begin{minipage}[t]{.1\textwidth}
  \centering
  \includegraphics[scale=0.45,angle=0,trim={40cm 0 40cm 0}]{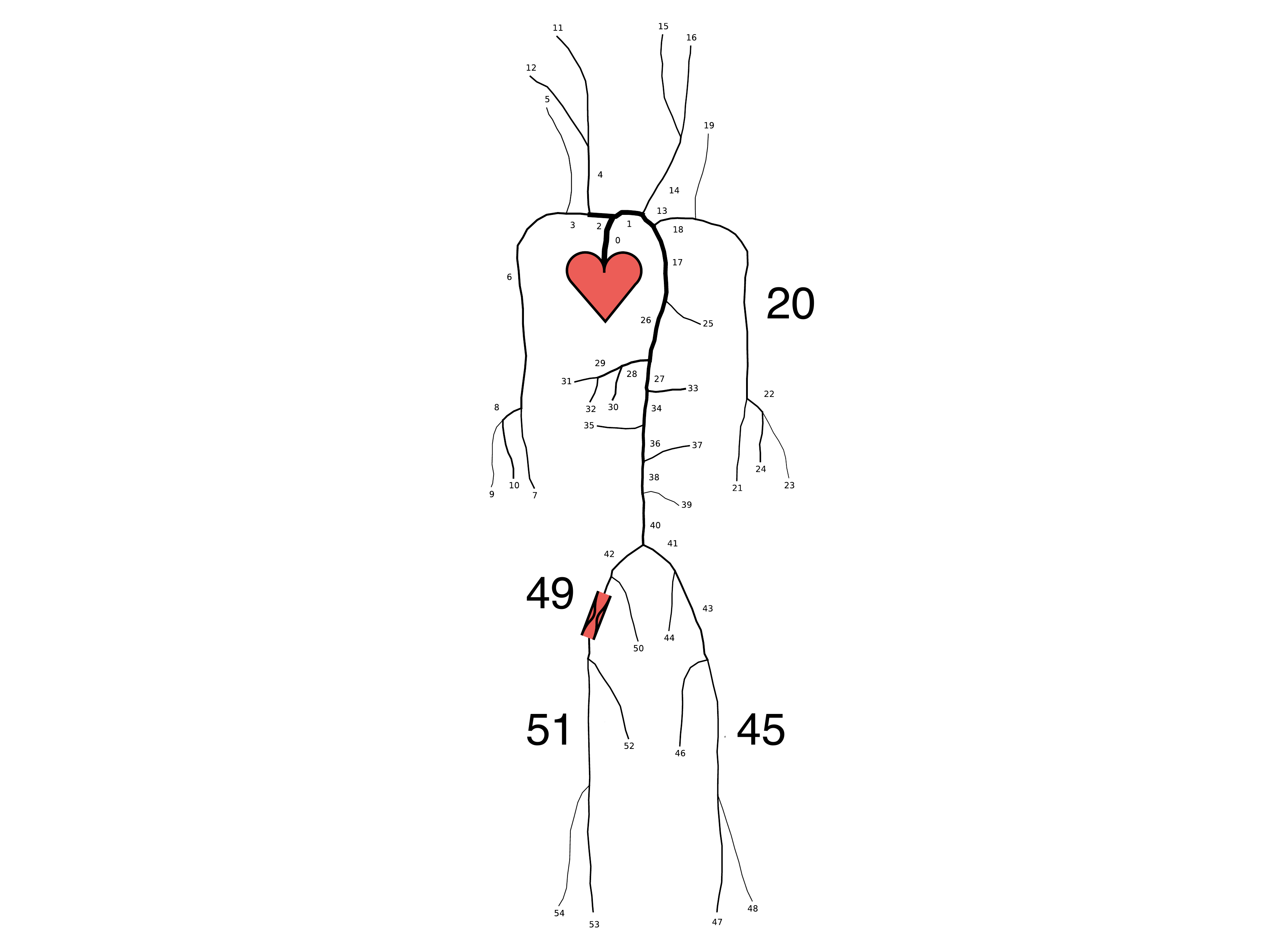}\\
\end{minipage}%
}
\caption{ Scheme of the 55 arteries network proposed in \cite{Sherwin2003-2} and used in this article. The numbered segments represent arteries described in the model. Each of the 55 arteries is characterized by a constant cross-sectional area at rest $A_0$, an constant arterial wall rigidity $K$ and a length $L$. At the end of each terminal segment, a constant reflection coefficient is imposed to model the resistive behavior of the distal network that is not taken into account in the modeled network. Therefore, a pulse wave propagates in the network starting from the heart and is reflected at each arterial bifurcation and terminal artery. The stenosis is represented in red in artery 49.}
\label{fig:Network-55}
\end{figure}

In the purpose of performing large network blood flow simulations, we use the arterial network proposed by Sherwin in \cite{Sherwin2003-2} which was adapted from Westerhof \cite{Westerhof1970}, describing 55 of the great arteries of the systemic network (human arterial network). This model was more recently used by Wang \cite{Wang2015} to perform viscoelastic blood flow simulations using different numerical methods. The network under consideration is represented in figure \ref{fig:Network-55}. The parameters of the model were obtained using physiological data and in each artery the geometrical and mechanical parameters do not vary with the axial position $x$. Therefore, in the absence of an arterial pathology, a well-balanced method is not required to compute blood flow in the considered network. The details of the parameters of the model are not recalled here and we refer the reader to the cited publications.

The pathology considered is a stenosis of the right iliac artery (artery 49 in figure \ref{fig:Network-55}). We consider two  possible shapes for the stenosis. The first corresponds to a succession of an increasing and a decreasing step and will be referred to as the square stenosis. It is defined by the following radius at rest $R_0$ and arterial wall rigidity $K$ 
\begin{equation}
\left\{
\begin{split}
& R_0\left( x \right) = &	\left\{
	\begin{split}
	&R_{in} & \: \: \: \text{ if } & x < x_s \\
	&R_{in} \left( 1 - \Delta \mathcal{G} \right) & \: \: \: \text{ if } & x_s < x < x_e\\
	&R_{in} & \: \: \: \text{ if } &  x \geq x_e\\
	\end{split}
	\right.\\
& K \left( x \right) = &\left\{
	\begin{split}
	&K_{in} & \: \: \: \text{ if } & x < x_s \\
	&K_{in} \left( 1 + \Delta \mathcal{G} \right) & \: \: \: \text{ if } & x_s < x < x_e\\
	&K_{in} & \: \: \: \text{ if } &  x \geq x_e .\\
	\end{split}
	\right.\\
\end{split}
\right.
\label{eq:Ex-Stenosis-Square-Geom-55}
\end{equation}
We choose $x_s = 6.25$ cm and $x_f =  8.25$  cm. The second geometry is the stenosis \eqref{eq:Ex-Stenosis-Geom} presented in subsections \ref{subsec:Steady} and \ref{subsec:Wave-Stenosis}. Its radius at rest $R_0$ and arterial wall rigidity $K$ vary as 
\begin{equation}
\left\{
\begin{split}
& R_0\left( x \right) = &	\left\{
	\begin{split}
	&R_{in} & \: \: \: \text{ if } & x < x_s \text{ or } x > x_f \\
	&R_{in} \left( 1 - \frac{\Delta \mathcal{G}}{2} \left[ 1 + \cos \left( \pi +2 \pi \frac{x-x_s}{x_f -x_s} \right) \right] \right) & \: \: \: \text{ if } &  x_s \leq x \geq x_f \\
	\end{split}
	\right.\\
& K \left( x \right) = &\left\{
	\begin{split}
	&K_{in} & \: \: \: \text{ if } & x < x_s \text{ or } x > x_f \\
	&K_{in} \left( 1 + \frac{\Delta \mathcal{G}}{2} \left[ 1 + \cos \left( \pi +2 \pi \frac{x-x_s}{x_f -x_s} \right) \right] \right) & \: \: \: \text{ if } &  x_s \leq x \geq x_f . \\
	\end{split}
	\right.\\
\end{split}
\right.
\label{eq:Ex-Stenosis-Cos-Geom-55}
\end{equation}
We choose $x_s = 5.5$ cm and $x_f = 9.5$ cm. We will refer to this stenosis as the cos stenosis. The cos stenosis is twice as long as the square stenosis to match the deformation area of the square stenosis. However, the maximal amplitudes of both configuration are the same and are proportional to $\Delta \mathcal{G}$.\\

In \cite{Ghigo2015}, the authors studied a similar pathological network and showed that the presence of the stenosis has a noticeable impact on the global hemodynamics for large values of $\Delta \mathcal{G}$. To that effect, we choose $\Delta \mathcal{G} = 65 \%$.\\

The results presented in this section are obtained using a time step $\Delta t  = 5 \times 10^{-5}$ and mesh size $\Delta x = 0.2$ cm and are compared to results obtained with the 55 arteries network without the pathology. This network will be referred to as the "Sane" network and does not require the use of any well-balanced method. We focus on four measurement points corresponding to typical measurement points used by medical practitioners during surgery. These points are situated in the middle of the following arteries, and the numbers indicated correspond to the numbering of the arteries in figure \ref{fig:Network-55}: the Left Subclavian II (20), Left Femoral (45), Right Femoral (51) and Right External Iliac (49), before the stenosis. Furthermore, as the pressure $P$ is the most common and simple variable to measure in vivo, we present only pressure waveforms in the following.\\

\begin{figure}[!h]
\makebox[1.\textwidth][c]{
\begin{minipage}[t]{.5\textwidth}
  \centering
  \includegraphics[scale=0.40,angle=0]{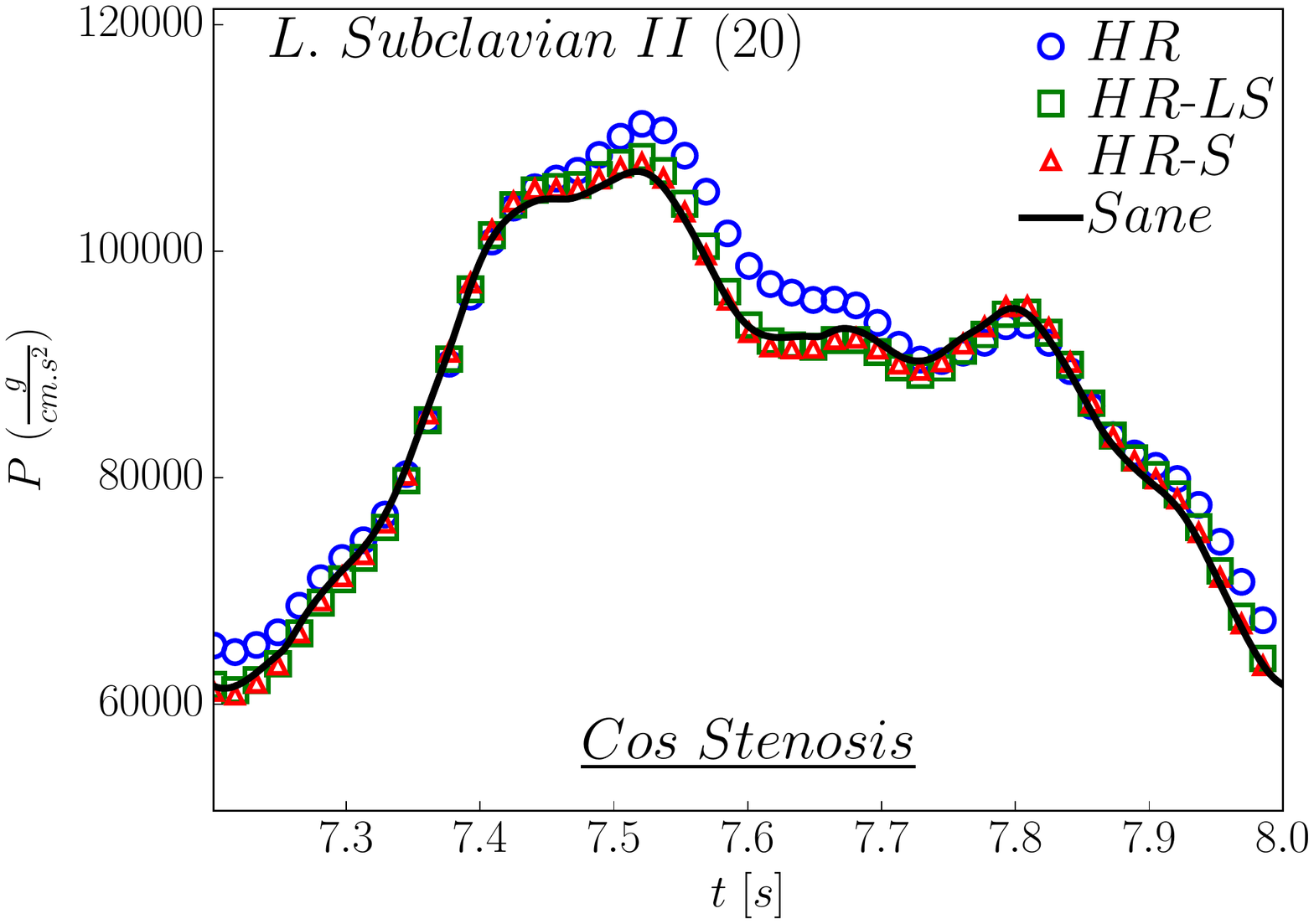}\\
\end{minipage}%
\begin{minipage}[t]{.5\textwidth}
  \centering
  \includegraphics[scale=0.40,angle=0]{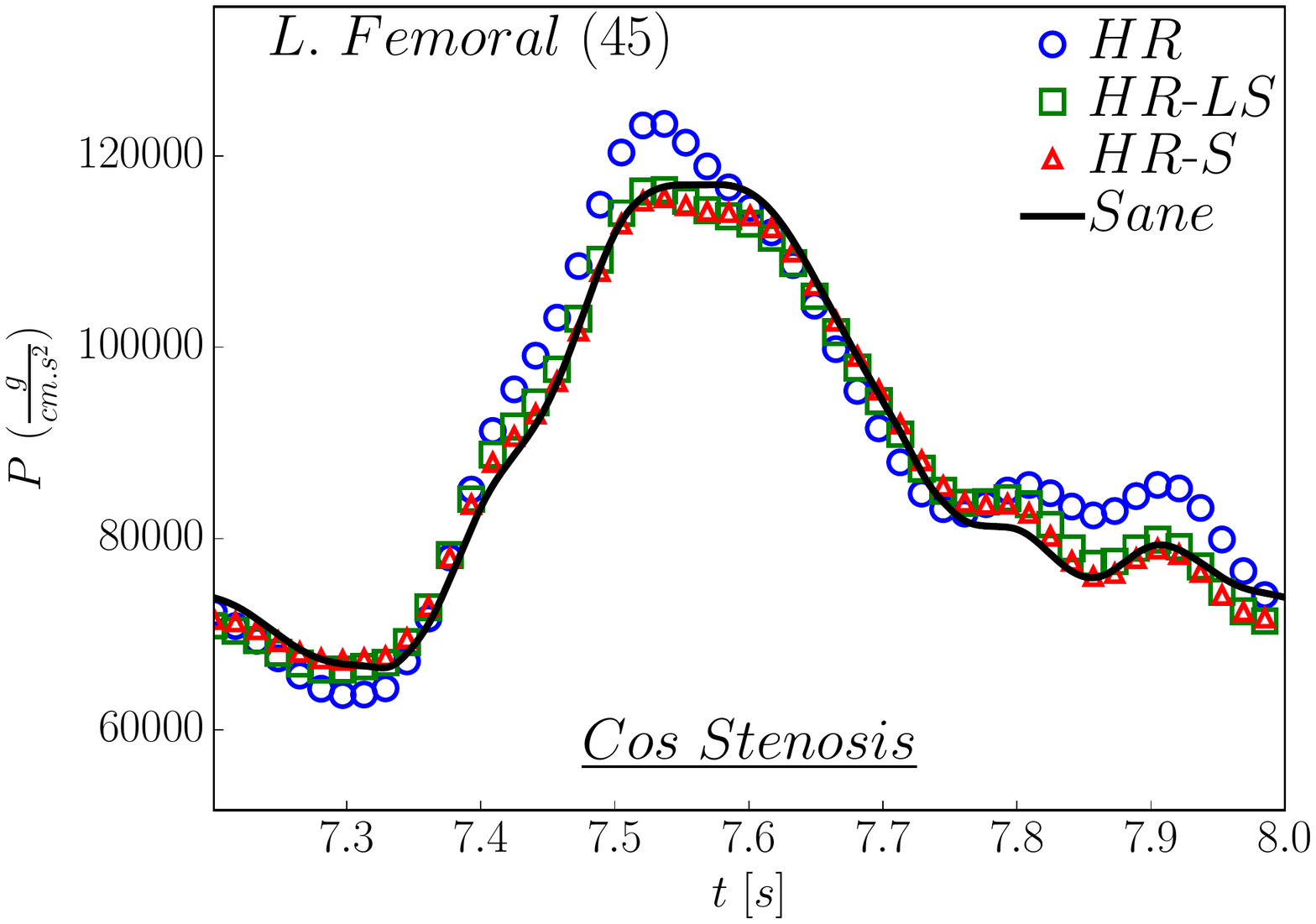}\\
\end{minipage}%
}
\makebox[1.\textwidth][c]{
\begin{minipage}[t]{.5\textwidth}
  \centering
  \includegraphics[scale=0.40,angle=0]{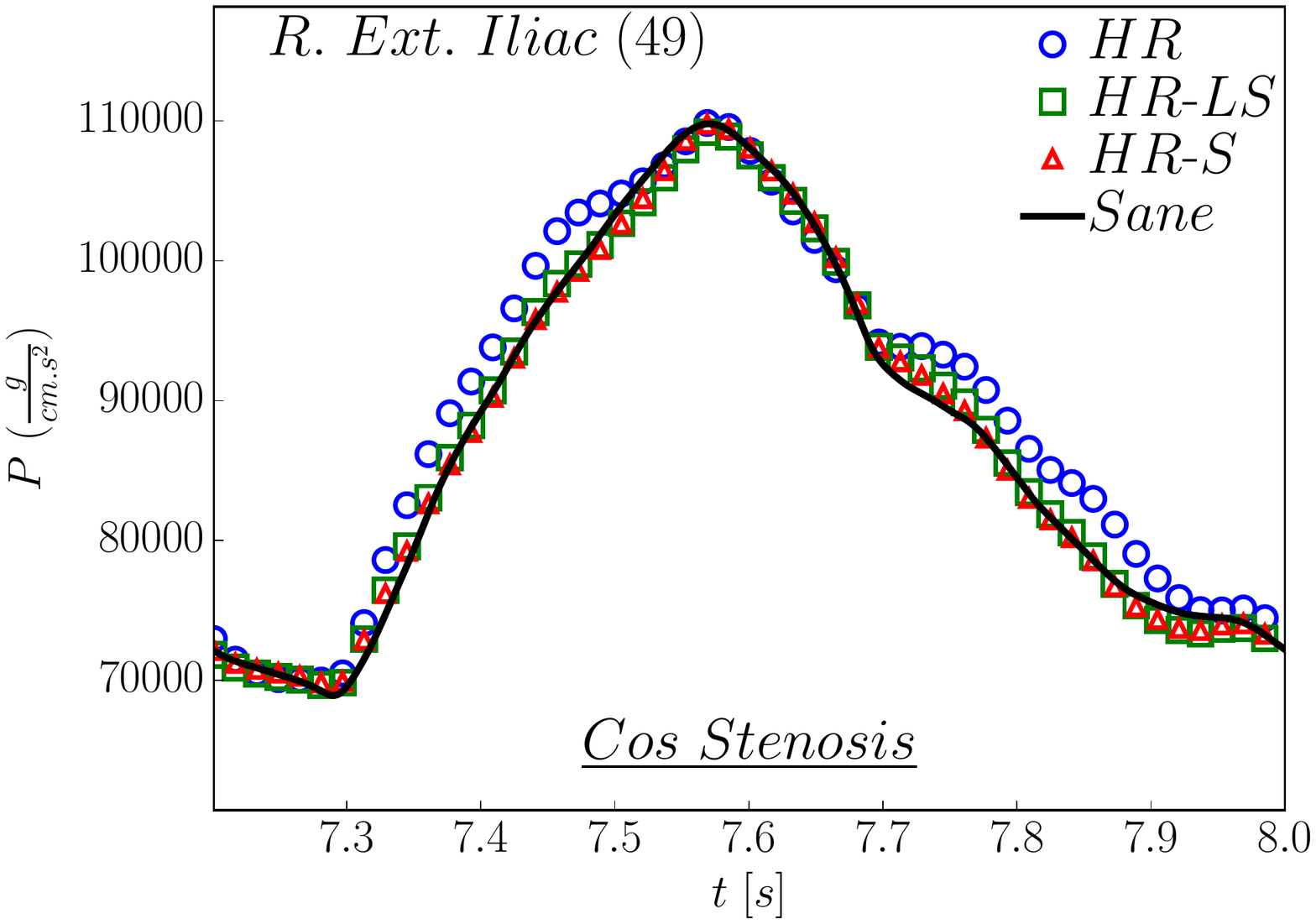}\\
\end{minipage}%
\begin{minipage}[t]{.5\textwidth}
  \centering
  \includegraphics[scale=0.40,angle=0]{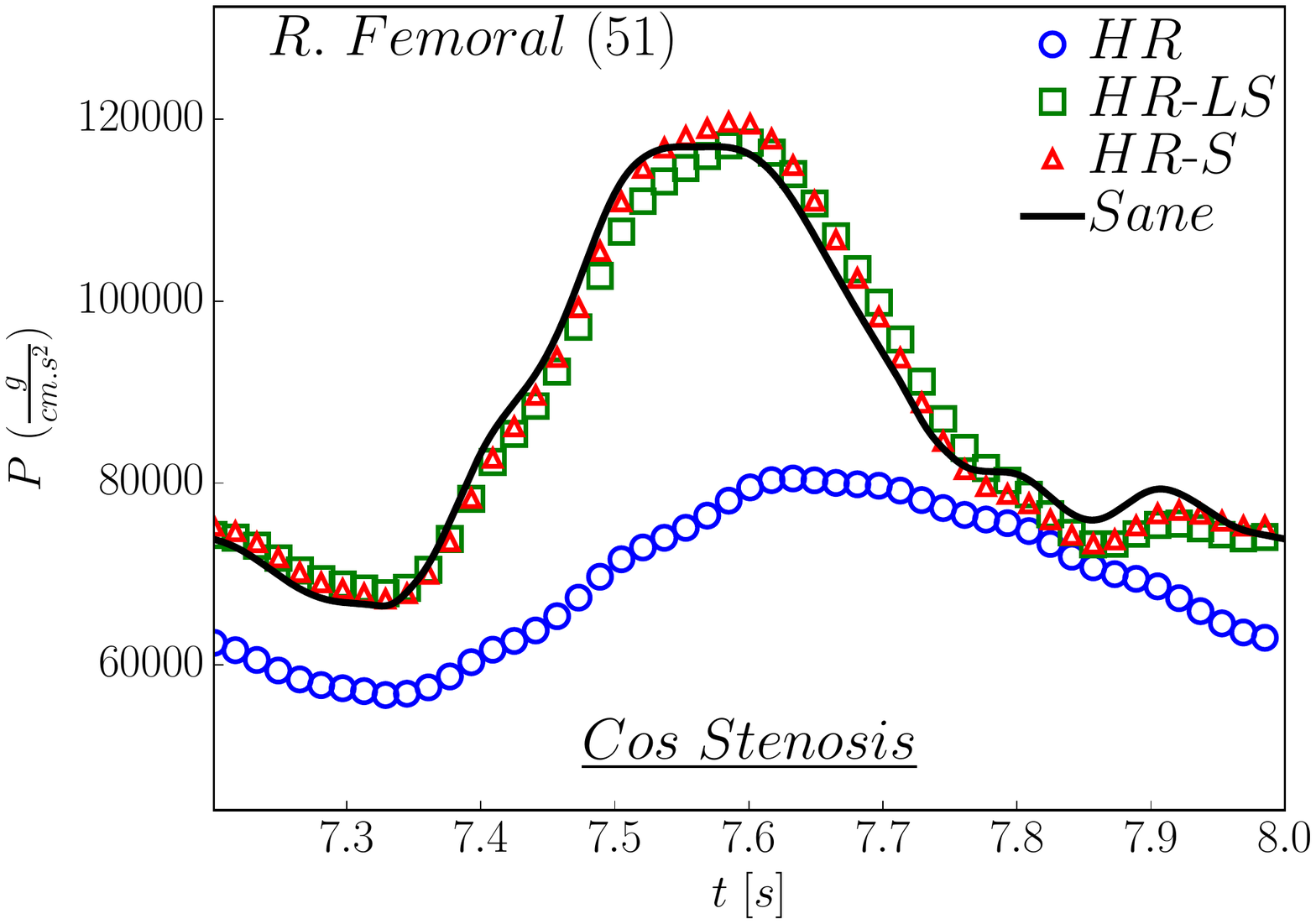}\\
\end{minipage}%
}
\caption{ Pressure $P$ in the middle of different arteries of the model network (20, 45, 49, 51) using an inviscid fluid and an elastic wall model. Comparison between the sane case (black line) and the cos stenosis using HR (blue circle), HR-LS (green square) and HR-S (red triangle). HR-LS and HR-S are different only in artery 51. HR is different from HR-LS and HR-S.}
\label{fig:55Arteries-Inviscid}
\end{figure}

In figure \ref{fig:55Arteries-Inviscid}, we compare the pressure waveforms obtained using HR, HR-LS and HR-S for the cos stenosis. The results obtained using HR-LS and HR-S are almost identical in the arteries 20 and 45, suited far from the stenosis, but also in the artery 49, located before the stenosis. In artery 51, situated after the stenosis, small differences exist between the results obtained with HR-LS and HR-S, especially during diastole ($t=7.5$ s and $t=8$ s). Moreover, the results obtained with HR-LS and HR-S in arteries 20, 45 and 49 are very close to those obtained with the Sane network, indicating that in these arteries, the resistive behavior of the stenosis is negligible compared to the global resistance of the network. However, in artery 51, the results obtained with HR-LS and HR-S slightly differ from those obtained with the Sane network, indicating that the presence of the stenosis has a local effect, especially during diastole ($t=7.5$ s to $t=8$ s). HR produces significantly different results from HR-LS and HR-S in each artery considered. In arteries 20, 45 and 49, HR overestimates the amplitude of the pressure waveform, whereas in artery 51 it underestimates it. These results are in good accord with the observations made in figures \ref{fig:Step-Fr-10m2-dR-60} and \ref{fig:Stenosis-Fr-10m2-dR-60} in subsections \ref{subsec:Wave-Step} and \ref{subsec:Wave-Stenosis}. \\	

\begin{figure}[!h]
\makebox[1.\textwidth][c]{
\begin{minipage}[t]{.5\textwidth}
  \centering
   \includegraphics[scale=0.40,angle=0]{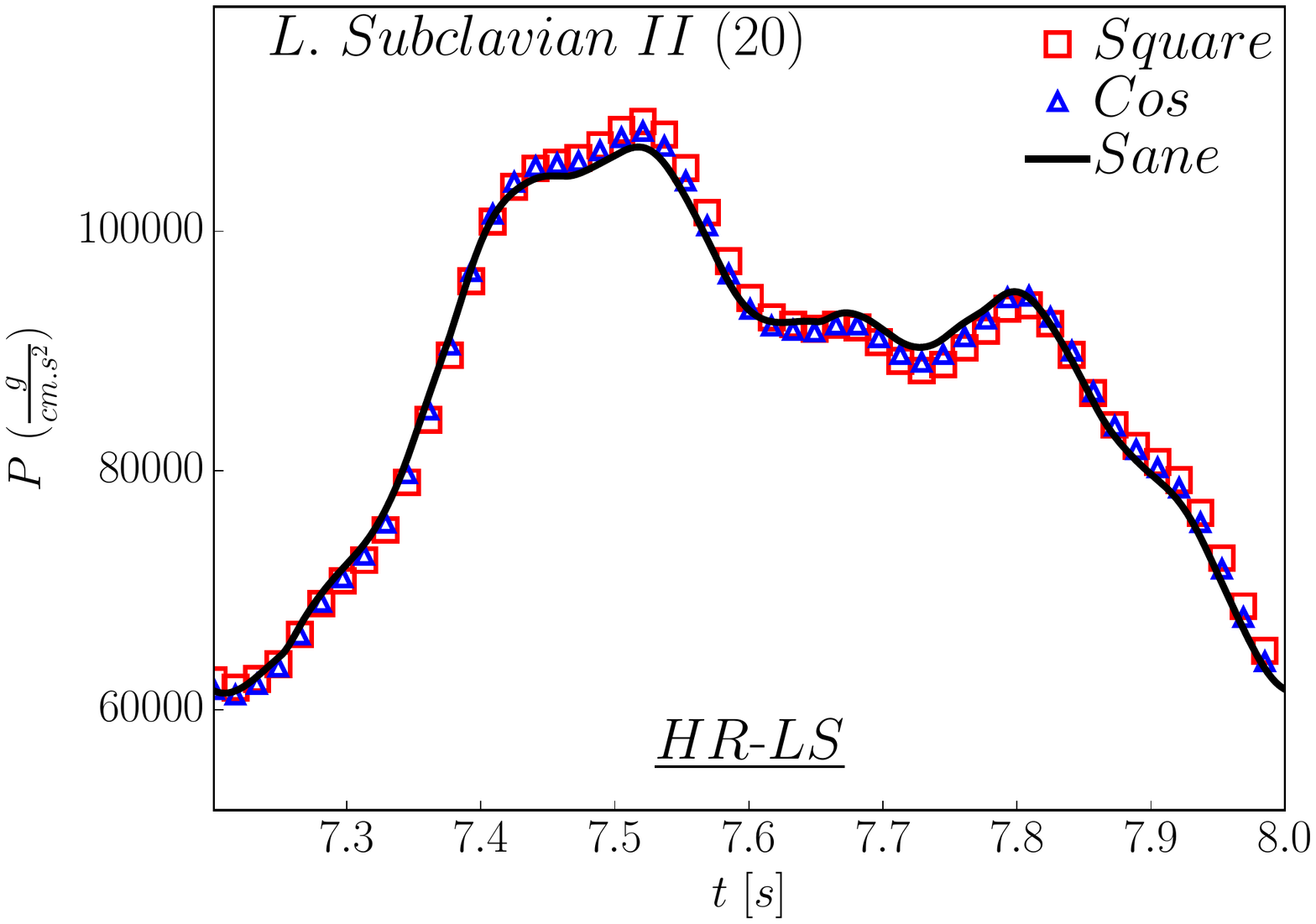}\\
\end{minipage}
\begin{minipage}[t]{.5\textwidth}
  \centering
   \includegraphics[scale=0.40,angle=0]{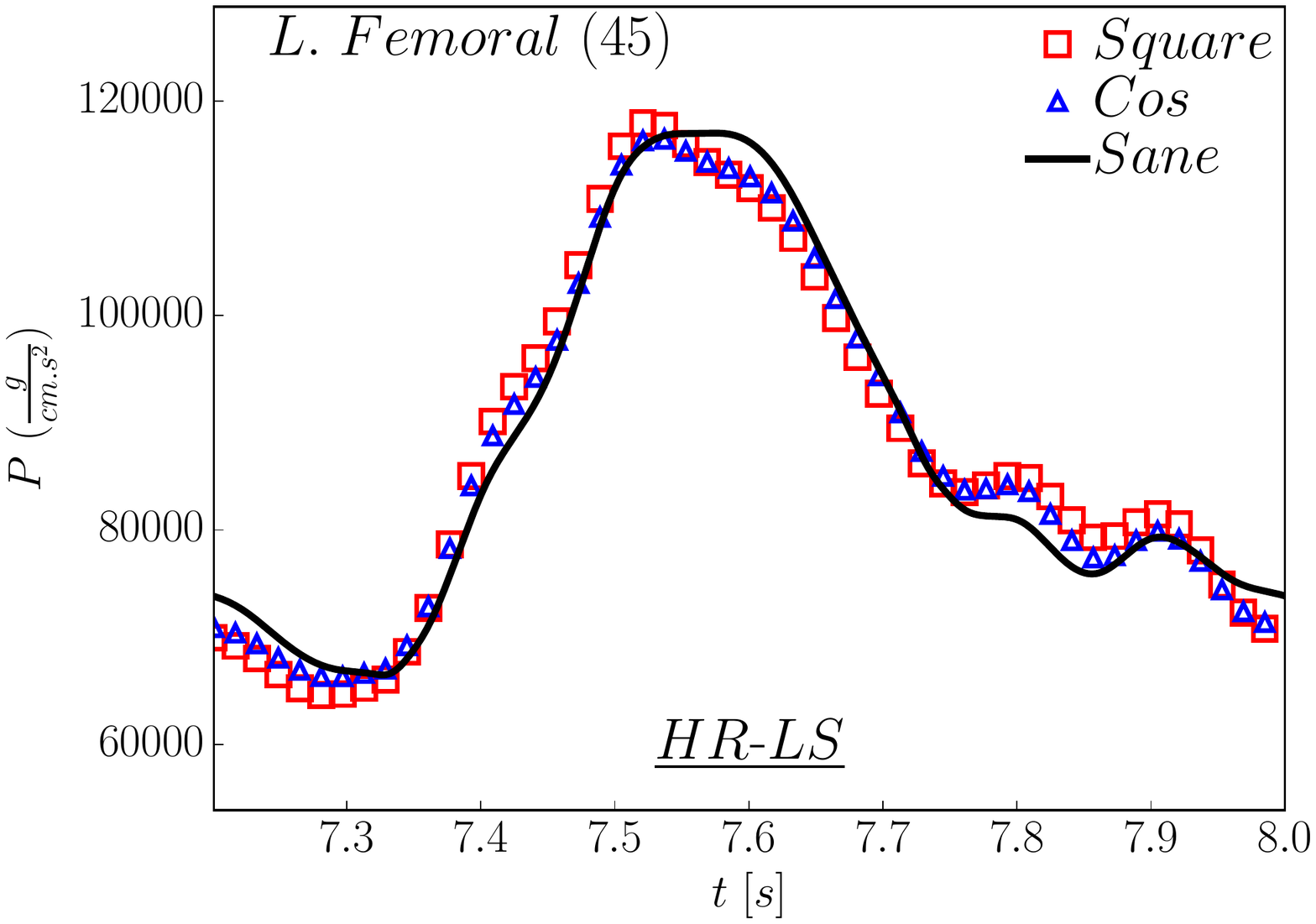}\\
\end{minipage}
}
\makebox[1.\textwidth][c]{
\begin{minipage}[t]{.5\textwidth}
  \centering
   \includegraphics[scale=0.40,angle=0]{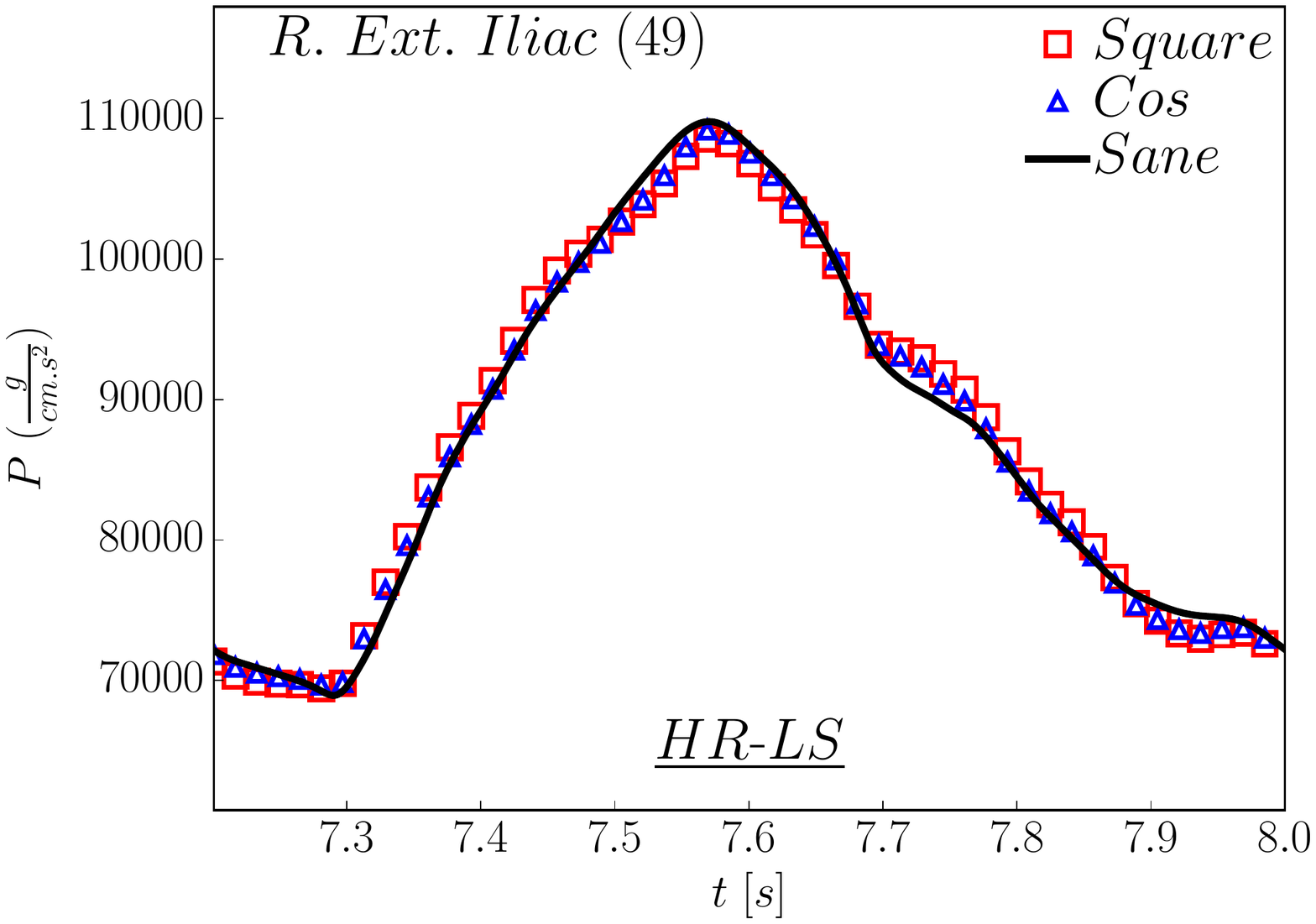}\\
\end{minipage}
\begin{minipage}[t]{.5\textwidth}
  \centering
   \includegraphics[scale=0.40,angle=0]{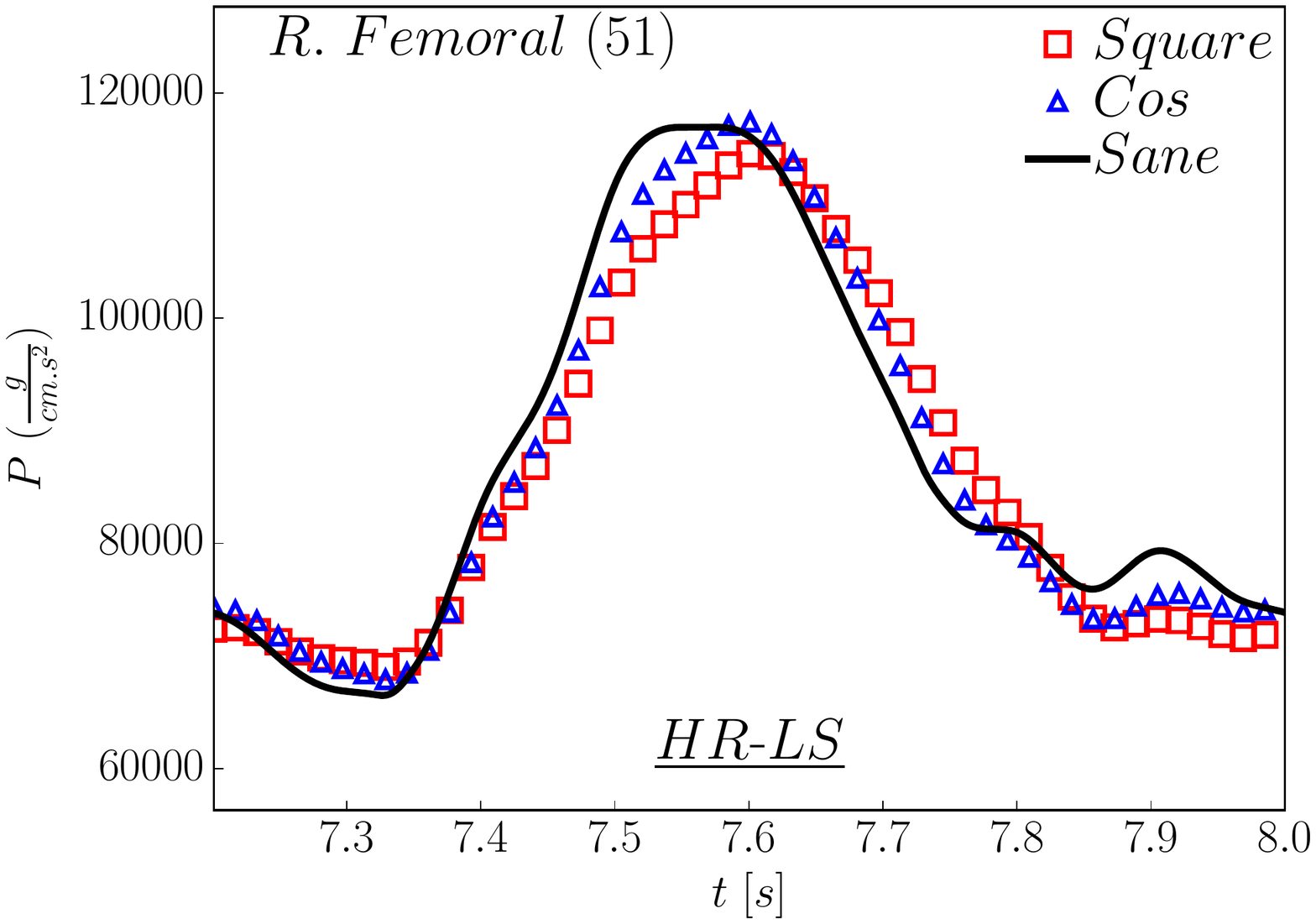}\\
\end{minipage}
}
\caption{ Pressure $P$ in the middle of different arteries of the model network (20, 45, 49, 51) using an inviscid fluid and an elastic wall model. Comparison between the sane case (black line), the square stenosis (red square) and the cos stenosis (blue triangle) using HR-LS. Changes with the shape of the pathology are visible only in artery $51$.}
\label{fig:55Arteries-Inviscid-Shape}
\end{figure}

In figure \ref{fig:55Arteries-Inviscid-Shape}, we compare the pressure waveforms obtained using HR-LS for the cos and the square stenosis. The results indicate that in arteries 20, 45 and 49, there are no significant differences when using either the cos or the square stenosis. Only in artery 51 is the influence of the shape noticeable.\\

\begin{figure}[!h]
\makebox[1.\textwidth][c]{
\begin{minipage}[t]{.5\textwidth}
  \centering
  \includegraphics[scale=0.40,angle=0]{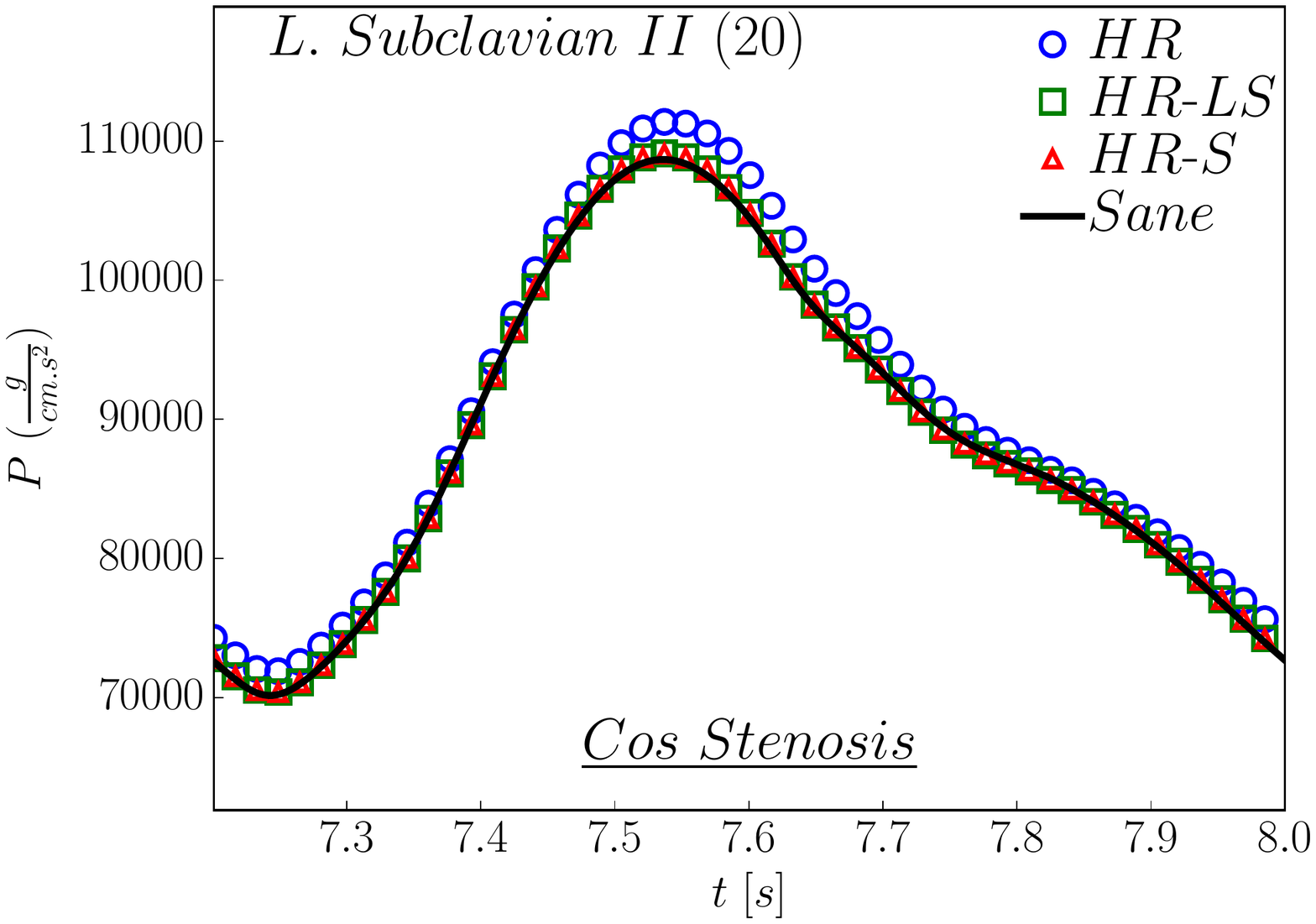}\\
\end{minipage}%
\begin{minipage}[t]{.5\textwidth}
  \centering
  \includegraphics[scale=0.40,angle=0]{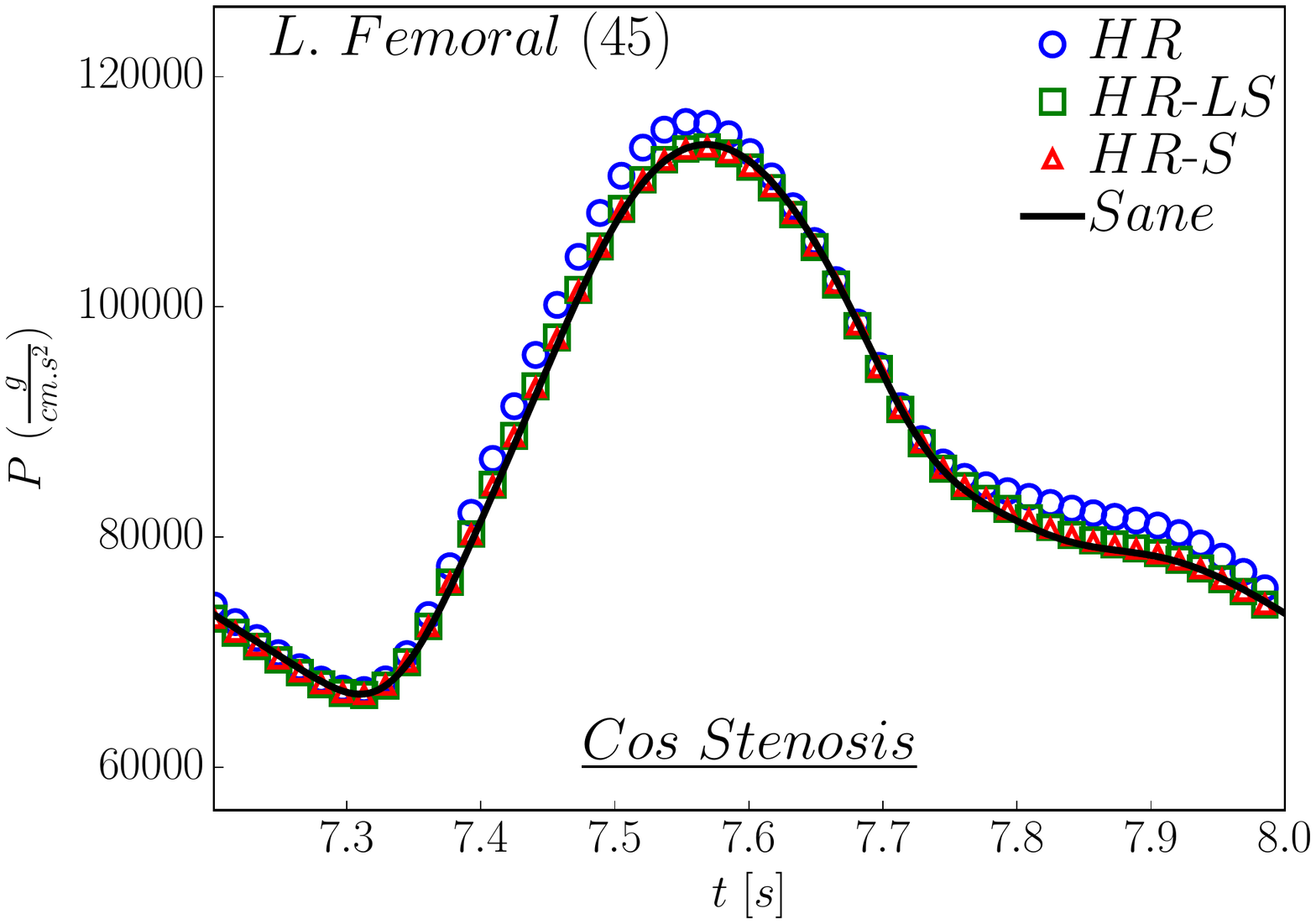}\\
\end{minipage}%
}
\makebox[1.\textwidth][c]{
\begin{minipage}[t]{.5\textwidth}
  \centering
  \includegraphics[scale=0.40,angle=0]{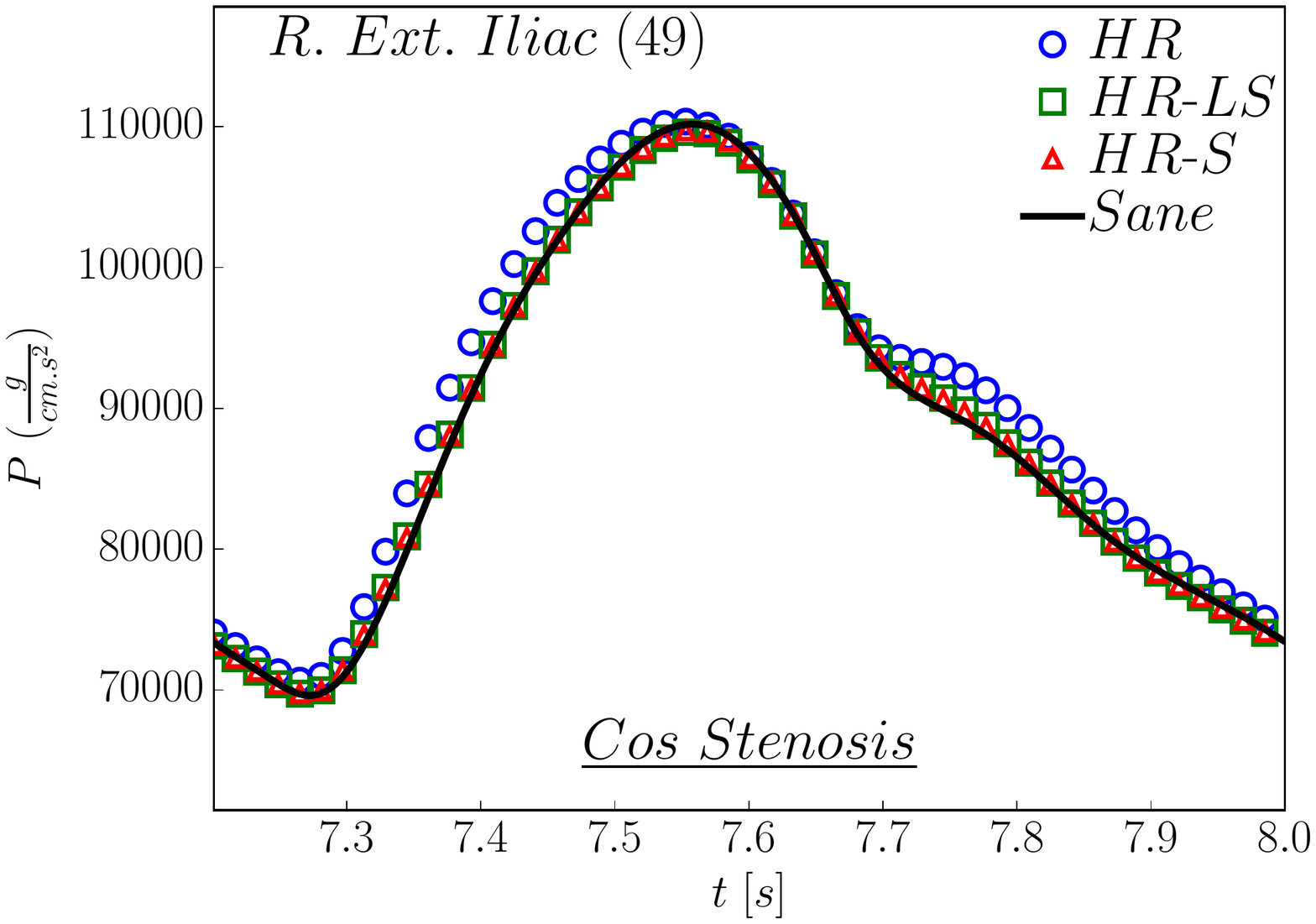}\\
\end{minipage}%
\begin{minipage}[t]{.5\textwidth}
  \centering
  \includegraphics[scale=0.40,angle=0]{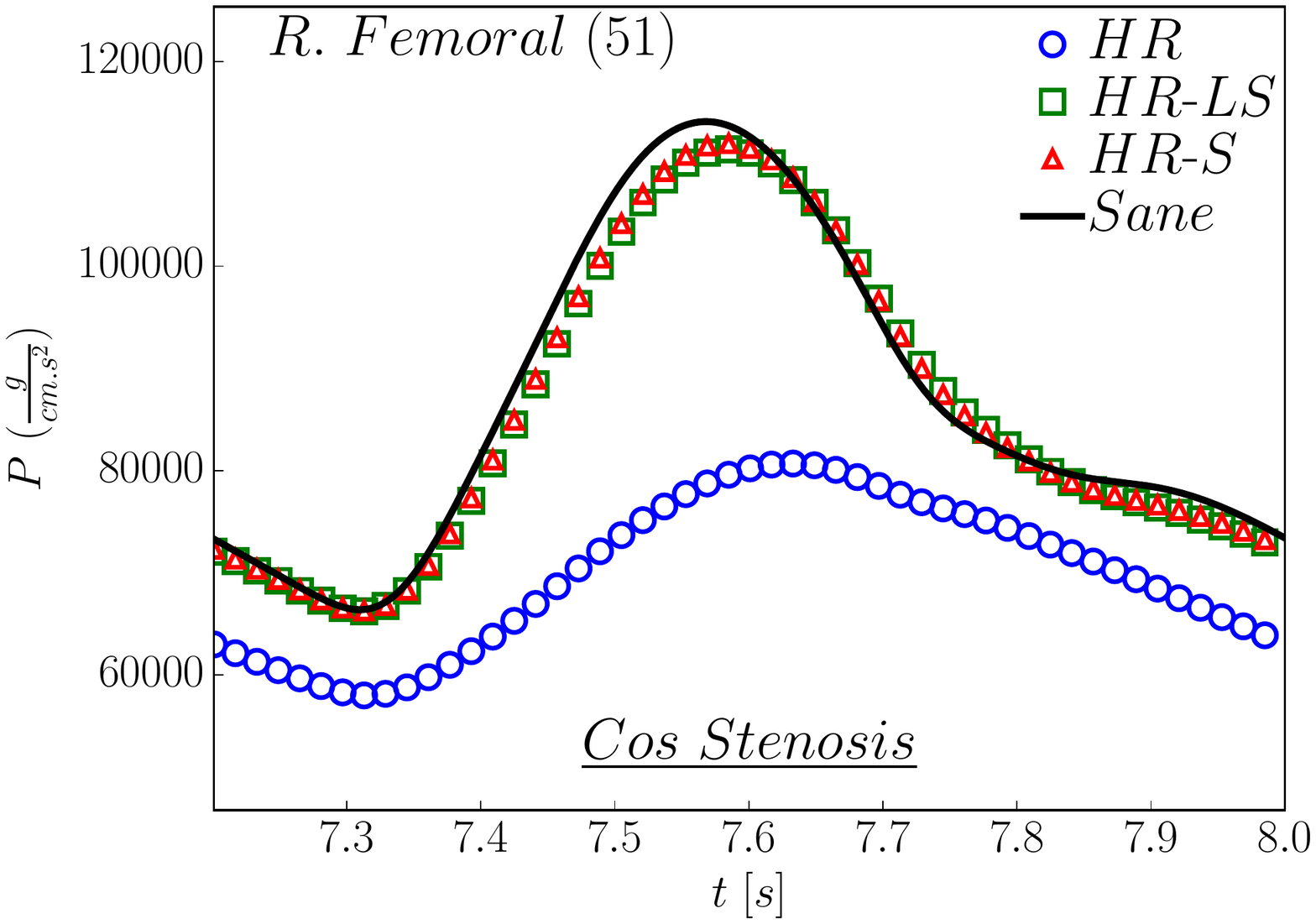}\\
\end{minipage}%
}
\caption{Pressure $P$ in the middle of different arteries of the model network (20, 45, 49, 51) using a viscous fluid and a viscoelastic wall model. Comparison between the sane case (black line) and the cos stenosis using HR (blue circle), HR-LS (green square) and HR-S (red triangle). Viscosity and viscoelasticity erase the differences between HR-LS and HR-S, which are now identical even in artery 51. HR is still different from HR-LS and HR-S.}
\label{fig:55Arteries-Viscoelastic}
\end{figure}

In figure \ref{fig:55Arteries-Inviscid}, the effects of the flow viscosity and the wall viscoelasticity are neglected. However, they play an important role in the global hemodynamics and need to be taken into account to obtain an accurate description of pressure and flow waves in a network. In figure \ref{fig:55Arteries-Viscoelastic} , we present similar results to those obtained in figure \ref{fig:55Arteries-Inviscid}, but now viscous and viscoelastic effects are taken into account. For the implementation of the viscous and viscoelastic terms, we refer the reader to \cite{Wang2015}. The results indicate that viscosity and viscoelasticity have a dissipative and diffusive effect and in the presence of such effects, the results obtained with HR-LS and HR-S overlap, even in artery 51. However, HR still overestimates the amplitude of the pressure waves in arteries 20, 45 and 49 and underestimates it in artery 51.\\

The results presented in this section indicate that the pressure waveform is sensitive to the choice of the well-balanced method. Even though HR-LS, HR-S have comparable behaviors, the results obtained with these methods are very different from those obtained with HR. On the contrary, changing the shape of the pathology has little effect on the shape and amplitude of the pressure waveforms. The small differences between the results are erased when blood viscosity and wall viscoelasticity are taken into account, due to the damping and diffusion behavior of the fluid and wall viscosities. Overall, HR-LS and HR-S behave similarly and produce satisfying results.\\

\section{Conclusion}

We introduced two well-balanced hydrostatic reconstruction techniques for blood flow in large arteries with varying geometrical and mechanical properties. The low-Shapiro hydrostatic reconstruction (HR-LS) is a simple and efficient well-balanced reconstruction technique, inspired from the hydrostatic reconstruction technique (HR) proposed in \cite{Audusse2004,Delestre2012}. It accurately preserves low-Shapiro number (equivalent of the Froude number for shallow water equations and the Mach number for compressible Euler equations) steady states that may occur in large network simulations and are the appropriate conserved properties at discontinuities of the geometrical and mechanical properties of the artery. The subsonic hydrostatic reconstruction (HR-S), introduced in \cite{Bouchut2010} and adapted here to blood flow, exactly preserves all subcritical steady states. We performed a series a numerical computations to compare the properties of HR, HR-LS and HR-S. In all numerical computations, HR was the least accurate method and was unable to correctly compute wave reflection and transmission when large variations of the artery's geometrical and mechanical properties were considered. HR-S proved to be exactly well-balanced for all low-Shapiro number steady states and the most accurate reconstruction technique. We showed that HR-LS is well-balanced only for steady states at rest, but provides satisfactory approximations of low-Shapiro steady states. HR-LS is also able to capture wave reflections and transmissions for arbitrary large variations of the artery's geometrical and mechanical properties, which is an essential property to compute realistic flow and pressure waveforms. We have also evaluated the sensitivity of the model to well-balanced methods and to the shape of the pathology in an 55 arteries network simulation. We showed that the model is not sensitive to the geometry of the pathology. However, important differences were observed between HR and the other well-balanced methods, namely HR-LS and HR-S, due to the fact that HR is unable to capture wave reflection and transmission. Finally, we observed that the small differences between HR-LS and HR-S are erased when adding viscous and viscoelastic effects, which are required to obtain realistic pressure and flow waveforms. This analysis allows us to conclude that both HR-LS and HR-S are adequate well-balanced methods to compute blood flow in large arteries with varying cross-sectional area at rest and arterial wall rigidity. However, in large networks where many arteries present variations of their geometrical and mechanical properties, the extra iterations required by HR-S increase the computational cost compared to HR-LS. We therefore recommended using HR-LS in this case, as it is a good compromise between simplicity, numerical accuracy and efficiency. In future works, we will investigate further the properties of HR-LS and propose an extension of the method to higher order.\\

\section{Acknowledgments}

The authors are grateful to thank F. Bouchut and E. Audusse for their helpful remarks and comments.


\clearpage
\pagebreak
\newpage
\fancyfoot{}
\fancyhead{}
\renewcommand{\headrulewidth}{0pt}
\renewcommand{\footrulewidth}{0pt}

\bibliography{\myreferences}


\end{document}